\documentclass[12pt]{article}
\usepackage[round,sort,authoryear]{natbib}

\usepackage{amsmath,amssymb,amsthm,amsfonts, fullpage, setspace}

\usepackage[colorlinks]{hyperref}
\usepackage{hyperref}
\hypersetup{colorlinks=red,citecolor=blue,pdfstartview=FitH, pdfpagemode=None}

\usepackage{eucal}
\usepackage{subeqnarray,bm}
 \usepackage[utf8]{inputenc}
\usepackage{graphicx}[final]
\usepackage{psfrag} 
\usepackage{color,soul, xcolor}

\usepackage{pdfpages}
\usepackage{epstopdf}

\usepackage{epic}%
\usepackage{color}%

\usepackage{multirow}

\usepackage{ifthen} 
\usepackage{sectsty}
\usepackage{subfig}
\usepackage{float}
\usepackage[english]{babel}
\usepackage[nottoc]{tocbibind}
 \DeclareGraphicsExtensions{.eps,.png,.pdf}
\usepackage{mwe}
\makeatletter
\def\maxwidth{ %
  \ifdim\Gin@nat@width>\linewidth
    \linewidth
  \else
    \Gin@nat@width
  \fi
}
\makeatother

\definecolor{fgcolor}{rgb}{0.345, 0.345, 0.345}

\allsectionsfont{\normalsize\bfseries}

\usepackage{framed}
\makeatletter
 {\par\unskip\endMakeFramed%
 \at@end@of@kframe}
\makeatother

\definecolor{shadecolor}{rgb}{.97, .97, .97}
\definecolor{messagecolor}{rgb}{0, 0, 0}
\definecolor{warningcolor}{rgb}{1, 0, 1}
\definecolor{errorcolor}{rgb}{1, 0, 0}

\newtheorem{thm}{Theorem}

\newtheorem{defn}[thm]{Definition}

\newcommand{\ds}{\displaystyle}

\newcommand{\norm}[1]{\left\Vert#1\right\Vert}
\newcommand{\abs}[1]{\left\vert#1\right\vert}

\newcommand{\R}{\mathbb{R}}

\numberwithin{equation}{section}

\usepackage{calrsfs}

\allsectionsfont{\normalsize\bfseries}

\numberwithin{equation}{section}
\allowdisplaybreaks

\makeatletter
\def\@xfootnote[#1]{%
  \protected@xdef\@thefnmark{#1}%
  \@footnotemark\@footnotetext}
\makeatother

\begin{document}
\title{Analysis of EEG data using complex geometric  structurization}
\author{E. A. Kwessi\footnote{Corresponding author, Trinity University, Department of Mathematics, Visiting the  University of Alabama at Birmingham, Department of Biostatistics, ekwessi@trinity.edu, ekwessi@trinity.edu}\hspace{2cm}
L. J. Edwards\footnote{Department of Biostatistics, University of Alabama at Birmingham, ljedward@uab.edu}
 }

\date{}
\maketitle

\begin{abstract}
Electroencephalogram (EEG) is a common tool  used to understand brain activities. The data  are typically obtained by placing electrodes at the surface of the scalp and recording the oscillations of currents passing through the electrodes. These oscillations can sometimes  lead to various interpretations, depending on the subject's health condition, the experiment carried out, the sensitivity of the tools used, human manipulations etc. The data obtained over time can be considered  a time series. There is evidence in the literature that epilepsy EEG data may be chaotic. Either way, the embedding theory in  dynamical systems  suggests that time series from a complex system could be used to reconstruct its phase space under proper conditions. In this paper, we propose an analysis of epilepsy  electroencephalogram time series data based on a novel approach dubbed complex geometric  structurization.  Complex geometric structurization stems from the construction of strange attractors using embedding theory from dynamical systems. The complex  geometric structures  are themselves  obtained  using a geometry tool, namely the  $\alpha$-shapes from shape analysis. Initial analyses show a proof of concept in that  these complex structures capture the expected changes brain in lobes under consideration. Further, a deeper analysis suggests that  these complex structures can be  used as biomarkers for seizure changes. 
\end{abstract}
  \vspace{0.25cm} 

\section{Introduction}

In neuroscience, to understand brain functions and brain related diseases, it is very common to place electrodes on a subject's head and record the electrical activities that result. These activities, when represented,  are often oscillations that  change in shape, frequency, and range. They can be analyzed to see if meaningful information can be  extracted  that gives a clue about the brain state of the subject at a certain time, or in a certain region of the brain. The accuracy of the recording highly depends on the instrument used, the region of interest (ROI), the experimenter's experience, the time of the day, the subject's discipline during the recording and many other factors. This is to say that uncertainty in the accuracy of the information recorded is always present. Even when the information is being recorded on the same subject, on the same day, on the same ROI, but at different intervals, changes are bound to occur. Electroencephalogram or EEG, a term coined by \cite{Hberger}, are electrical activities recorded on humans or animals that display prominent oscillatory behavior subject to important changes during various behavioral states. These changes show a high degree of nonlinearity in the signals that may be important. Indeed, in the field of biomedical signal processing (analysis of heart rate variability, electrocardiogram, hand tremor, EEG), the presence or absence of nonlinearity often conveys information about the health condition of a subject. In particular, EEG signals are often examined using nonlinearity analysis techniques or by comparing signals that are recorded during different physiological brain states (e.g. epileptic seizure). The differences observed during these analyses can either be due to genuine differences in dynamical properties of the brain or due to differences in recording parameters. The EEG are often analyzed as times series and there are many methods for analysis of  times series in the literature. The methods can be grouped into two categories: univariate measures and multivariate measures. A good review on the topic can be found in \cite{Carney2011a}. 

Among the methods that have been touted as more efficient at providing an insight into the real dynamics of EEG  is the famous Embedding Theorem of \cite{Takens1981a}. This Embedding Theorem has been instrumental to understanding how to reconstruct the true dynamics of systems  based on the times series measured on these systems. Essentially, this reconstruction theory, in layman's terms, suggests that  a times series measured over sufficiently long period of times contains enough information to reconstruct the phase space in which the associated system  normally evolves. This allows also to show that there are other intricate subunits  that influence the changes observed in the measured variables that are represented in  the time series. This theorem was used for example by \cite{Grassberger1983} to propose a measure called Correlation Dimension which was in turn  instrumental  to  \cite{Lehnertz1998}  for the prediction of epilepsy seizures. Despite all the methods proposed in the literature, there is no agreement on  which method constitutes the best tool at extracting the most meaningful information that could be useful for the physician for the prediction of seizure-like diseases. Moreover, with the increasing use of the concepts of chaos and complexity in health sciences, it is becoming more and more  difficult to distinguish their  adequate application. For example, there have been evidence of chaos in EEG data, see \cite{Destexhe1986, Destexhe1992, Destexhe1998}, and since  chaotic systems are  inherently complicated, they may  look complex. Likewise, complex systems may also look chaotic.  Distinguishing these two notions is important in applications, especially in health sciences. Henceforth, we will adopt a  terminology along the lines of \cite{Rickles2007} for understanding complexity and chaos. 

In this paper, we propose a new  method for analyzing EEG times series data, which we call Complex Geometric Structurization (CGS). The complex nature of the method stems  from the fact that  we have multiple subunits interacting together resulting in a rich collective behavior feeding back into the behavior of individual parts. The method is inspired by the Embedding Theorem of \cite{Takens1981a} for the construction of a geometric structure whose volume can be evaluated from shape analysis technique. The volume of this  geometric  structure behaves as a key statistic akin to  a biomarker for the phenomenon or ROI of interest. Using data driven approaches to study brain pathologies is a very active field of study nowadays due to improvement in life expectancy across the world with its cohorts of problems such as brain disorders as illustrated in \cite{Zheng2019}, \cite{David2020} and the references therein. Moreover, the push to use data and methodological driven approaches to brain pathologies is also evidenced by the numerous grants offered by the National Institute of Health and private  foundations such as the Michael J. Fox Foundation for Parkinson Disease, the Bill and Melinda Gates Foundations, just to name some.

 The remainder of the paper is organized as follows: In Section \ref{UCA}, we make a brief review on how to use the Embedding Theorem to construct strange  attractors;  in Section \ref{sect:statsMorp},  we review important notions of statistical morphometry; in Section \ref{CS}, we introduce the complex geometric structurization method; in Section \ref{Appl}, we discuss some applications of the CGS method on real data; in Section \ref{Discussion}, we discuss the pros and cons of the proposed method in different contexts. 
\section{Understanding Embedding Theory} \label{UCA}
\begin{figure}[H] 
   \resizebox{1\textwidth}{!}{\begin{minipage}{1.1\textwidth}
   \centering
   \includegraphics[scale=0.5]{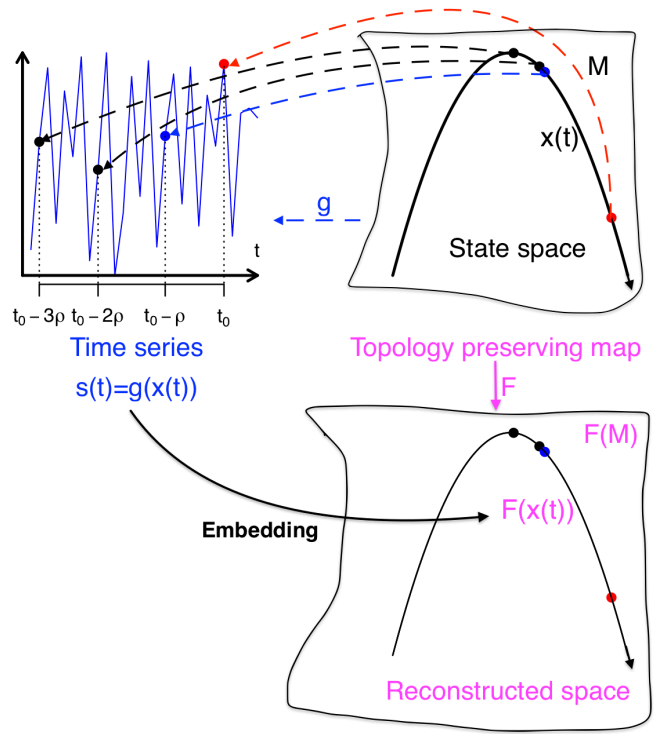} 
   \end{minipage}}
   \caption{An illustration of the embedding mechanism.}
   \label{fig:embedding}
\end{figure}

As its name suggests, a dynamical system is a system  whose variables evolve over time. Its phase space is a geometric representation of the trajectory of its variables over time. The values taken by the system's variables at an instant describe the system's states. To understand how to reconstruct the phase space of a dynamical system based on observations (times series) of one of  its variables, we need to revisit the Embedding Theory of \cite{Takens1981a}, which is essentially a high dimension transformation of the time series.  Consider the $n$-dimensional space $\mathbb{R}^n$. We  recall that a manifold $ M$ in the space  $\mathbb{R}^n$ is a topological space that locally looks like a Euclidian space near each point. Topology here means that bending is  ignored. For example, the surface of the globe is a  topological manifold in the space $\mathbb{R}^3$. Now, consider a dynamical system with a system's state $x(t)$ lying on a manifold $M$ of  $\mathbb{R}^n$. Let $\rho$ be a sampling interval and let the time series $s(t)=g(x(t))$ be given as a one-dimensional observation of the system dynamics through an observation function $g$.
\cite{Takens1981a} embedding theory states that for almost every smooth function $g$, the delay coordinate map defined as  $\mbox{F}: \mathbb{R}^n\to \mathbb{R}^m$ with   $\mbox{F}(x(t))=[s(t),s(t-\rho), \cdots, s(t-(m-1)\rho)]^T$ is an embedding, that is, it is a one-to-one immersion of the state space attractor with dimension $d$ when $m>2d$. In other words, the result states that $\mbox{F}(x(t))$ is a representative of $x(t)$, even if the true state space $M$ has not been observed, see Figure \ref{fig:embedding}. The quantity $m$ is  referred to in the literature as the embedding dimension and $\rho$ as the time delay (or lag).
\noindent The embedding theory  is predicated on the observation that a time series observed over a long period of time may show an internal structure. In fact, considering a time series $s(t)=g(x(t))$, we only observe an incomplete picture of $x(t)$ since $s(t)$ is a scalar.  However, if we observe it for a long period of time, a more precise description will emerge, which will help understand its dynamics. In practice, most of the focus is given into how to find appropriate  values for the time delay $\rho$ and the embedding dimension $m$, see Appendix \ref{sect:est}. We observe that this type of reconstruction technique has been applied successfully in the case of epilepsy EEG data before, see for instance \cite{Destexhe1986}, \cite{Destexhe1992}, and  \cite{Destexhe1998}.

\subsection{Examples}
\begin{figure}[H] 
   \centering
   \resizebox{0.75\textwidth}{!}{\begin{minipage}{1.3\textwidth}
 \centering   \includegraphics[scale=0.65]{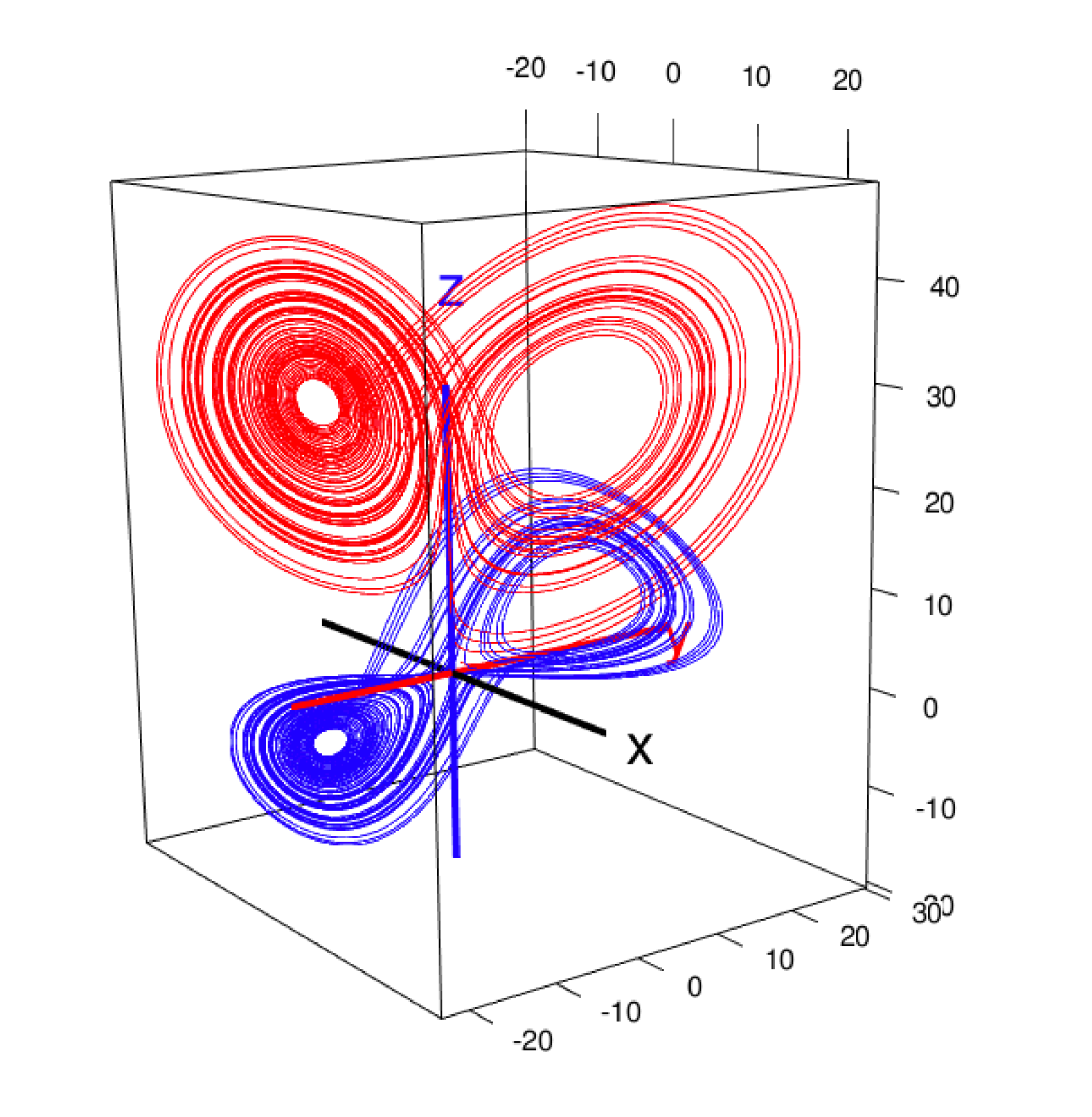} 
      \end{minipage}}
   \caption{ Lorenz  attractor (red) with its reconstructed counterparts (blue) and plotted in the same coordinates system. We can observe the topological equivalence between the original phase space and its reconstructed counterpart.}
   \label{figLorenzRossler}
\end{figure}

\begin{center}
\begin{figure}[H]
  \centering  \includegraphics[scale=1]{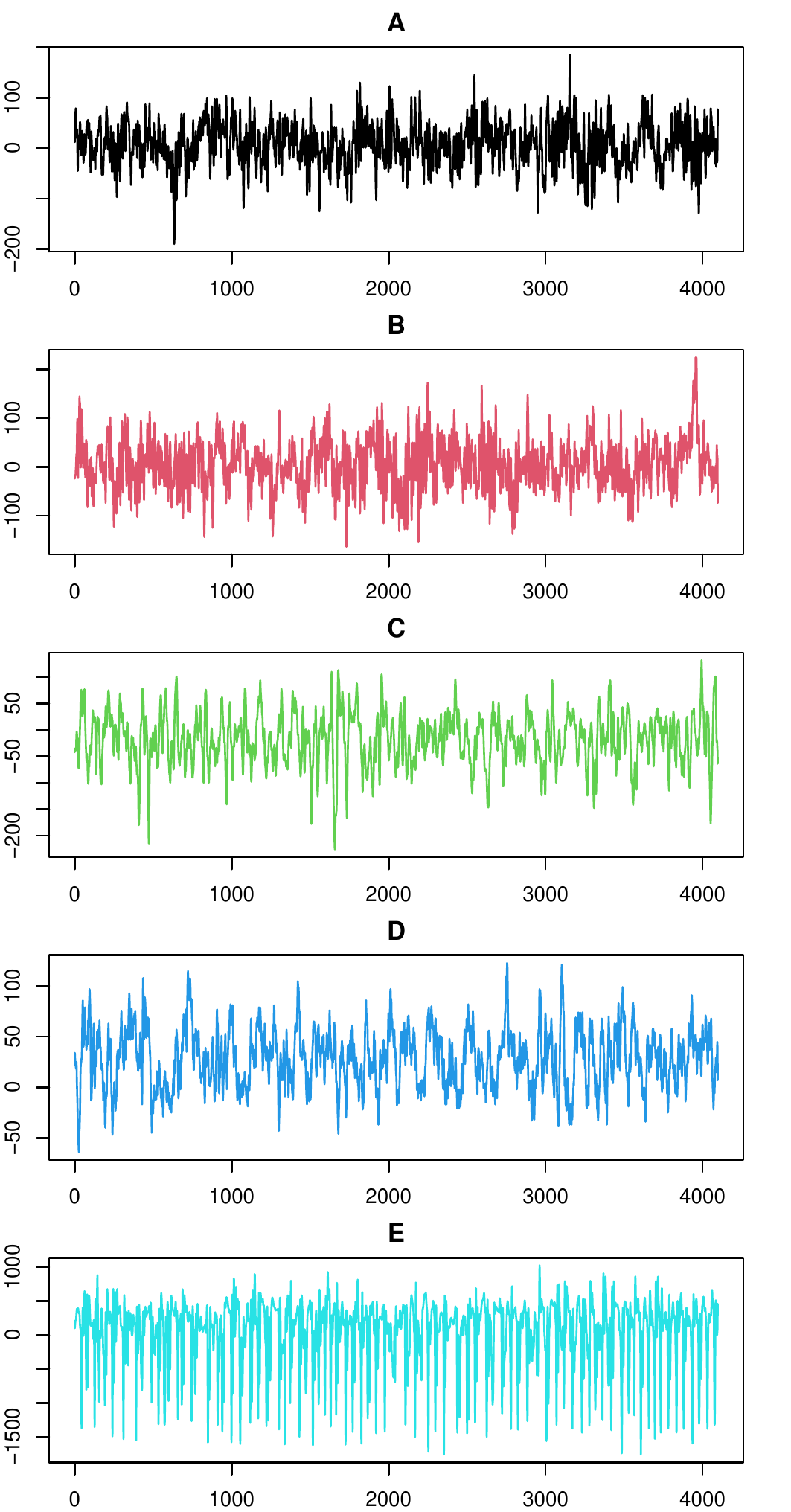} 
      \caption{This figure represents one time series selected at random from each set A--E. We observe that the amplitude is much more pronounced in the set E, which represents the seizure prone patients. }
      \label{figRealData0}
\end{figure}
\end{center}

{\bf Dynamical Systems:} The Takens' embedding theory can be used to reconstruct the Lorenz,  a famous attractor often mentioned in dynamical systems.
The \cite{Lorenz1963}  
systems of differential equations is  given   as 
\begin{equation*}
\begin{cases}
\dot{x}=s(y-x)\\
\dot{y}=rx-y-xz\\
\dot{z}=xy-bz
\end{cases}\;.
\end{equation*}
It is known for example that the Lorenz system is chaotic for $s=10, r=28$, and $b=8/3$. 
Figure \ref{figLorenzRossler} below is a depiction of this  attractor for these parameter values plotted  in the space $x=x(t), y=x(t-\rho)$, and $z=x(t-2\rho)$. The time step used is $\Delta t=0.005$ for an interval time of [0, 75] for the  Lorenz system. We also note that the time lag or delay is  $\rho=31 \Delta t$, and was  found using either the autocorrelation (ACF) or average mutual information (AMI), see for instance Section \ref{sect:est} below.
\noindent {\bf Real Data:} There have been many applications of Takens' embedding theory with real data.  For example, in  \cite{Fisher2009a},\cite{Carney2011a}, an attractor is constructed from a publicly available (at \url{http://www.meb.unibonn.de/epileptologie/science/physik/eegdata.html}) epilepsy data set called here EDATA for simplicity. The data consist of five sets  A, B, C, D, and E. Each containing 100 single-channel EEG segments of 23.6 seconds, each of which was selected  after visual inspection for artifacts (such as acoustic and electrical shielding, separation of earth ground for laboratory, interconnectivity of devices on the same phase and ground centrally and locally) and has passed a weak stationarity criterion. Sets A and B  were  obtained from surface EEG recordings of five healthy subjects  with eyes open and closed, respectively. Data  were obtained  in seizure-free intervals from five patients in the epileptogenic zone for set D and from the hippocampal formation of the opposite hemisphere of the brain for set C. Set E contains seizure activity, selected from all recording sites exhibiting ictal activity. Sets A and B have been recorded extracranially, whereas sets C, D, and E have been recorded intracranially. All EEG signals were recorded with the same 128-channel amplifier system, using an average common reference [omitting electrodes containing pathological activity (C,D, and E) or strong eye movement artifacts (A and B)]. After 12 bit analog-to-digital conversion, the data were written continuously onto the disk of a data acquisition computer system at a sampling rate of 173.61 Hz. Band-pass filter
settings were 0.53--40 Hz (12 dB/oct.), see \cite{Andrzejak2001}. Figure \ref{figRealData0} below is an illustration of  the dataset EDATA. Each row represents one time series from  sets A, B, C, D, and E respectively. Clearly, time series in the seizure set E has a much pronounced amplitude synonymous with more brain activities. 
In Figure \ref{figRealData1} below, we represent the reconstructed phase spaces based on the time selected times series from  set  A -- E . We note that the axes are $x=x(t), y=x(t-\rho), z=x(t-2\rho)$ where $\rho=1\Delta t$, with $\Delta t=\frac{1}{fs}=5.76$ ms.

  \begin{figure}[H] 
\resizebox{0.9\textwidth}{!}{\begin{minipage}{1.3\textwidth}
    \centering \begin{tabular}{cc}
  \bf (A) & \bf (B) \\
        \includegraphics[scale=0.51]{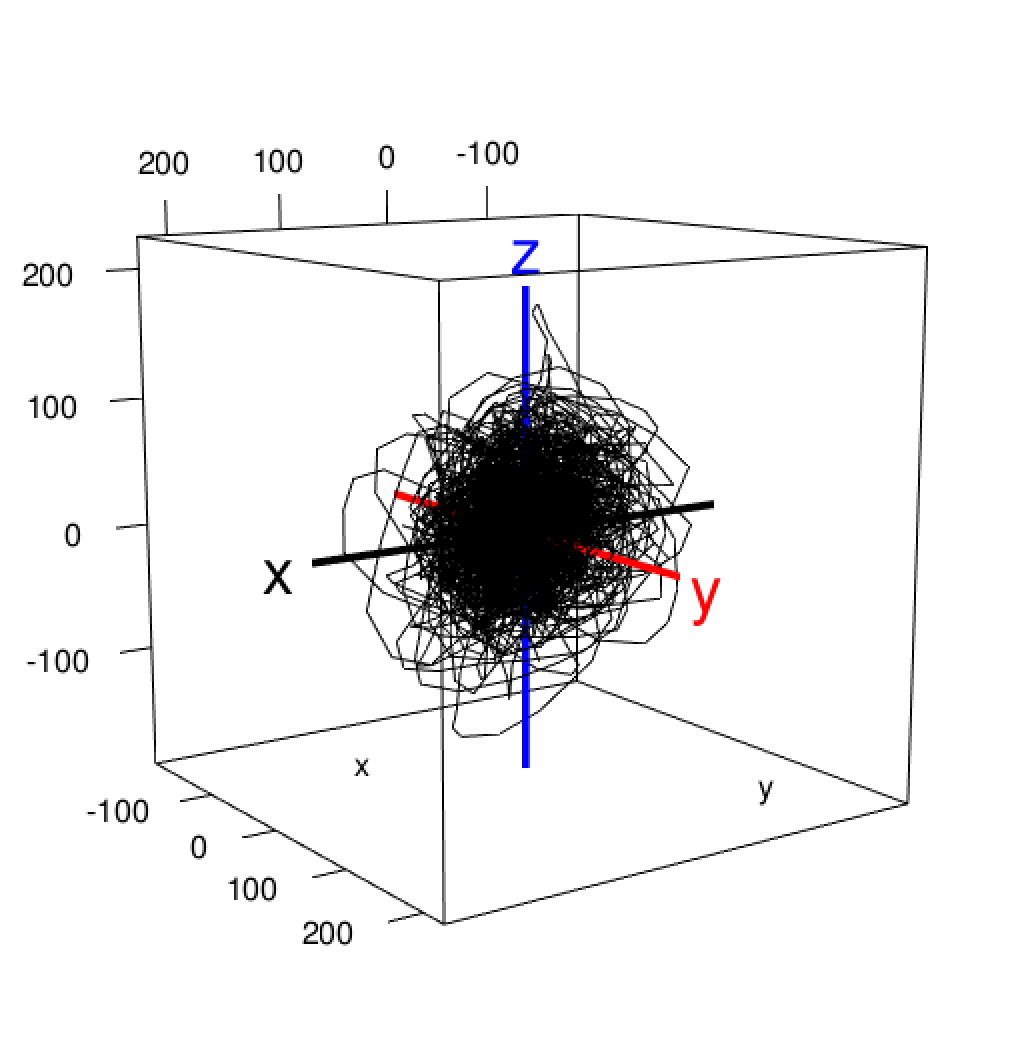}   &   \includegraphics[scale=0.51]{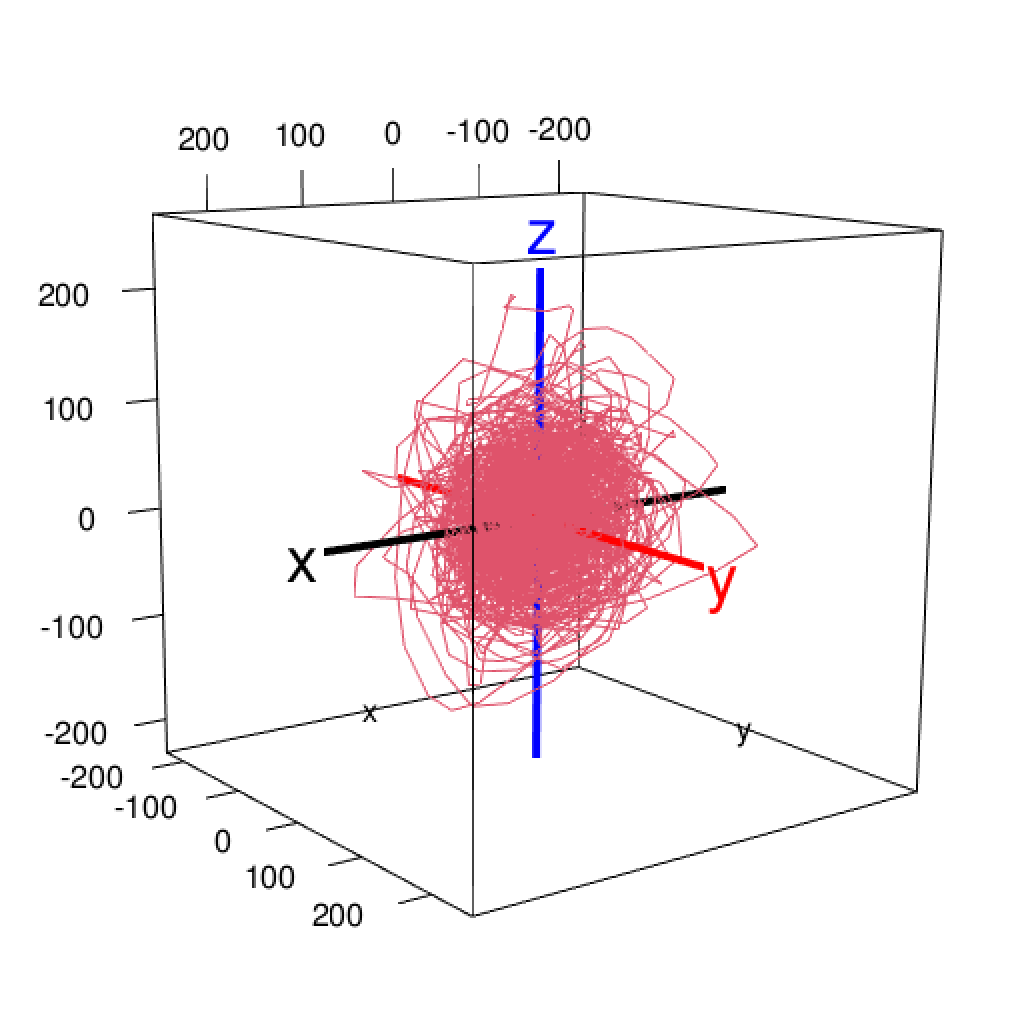} \\
 
  \bf (C) & \bf (D)   \\
            \includegraphics[scale=0.51]{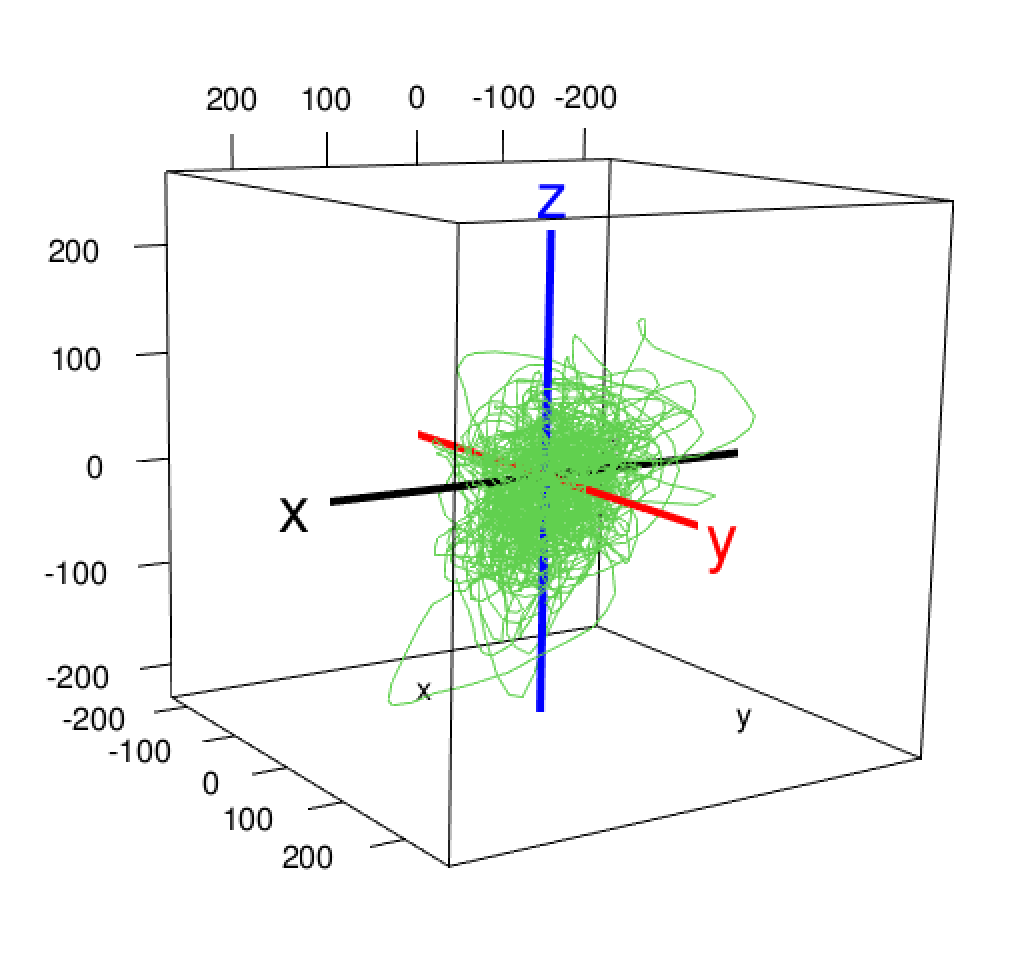}  &  \includegraphics[scale=0.51]{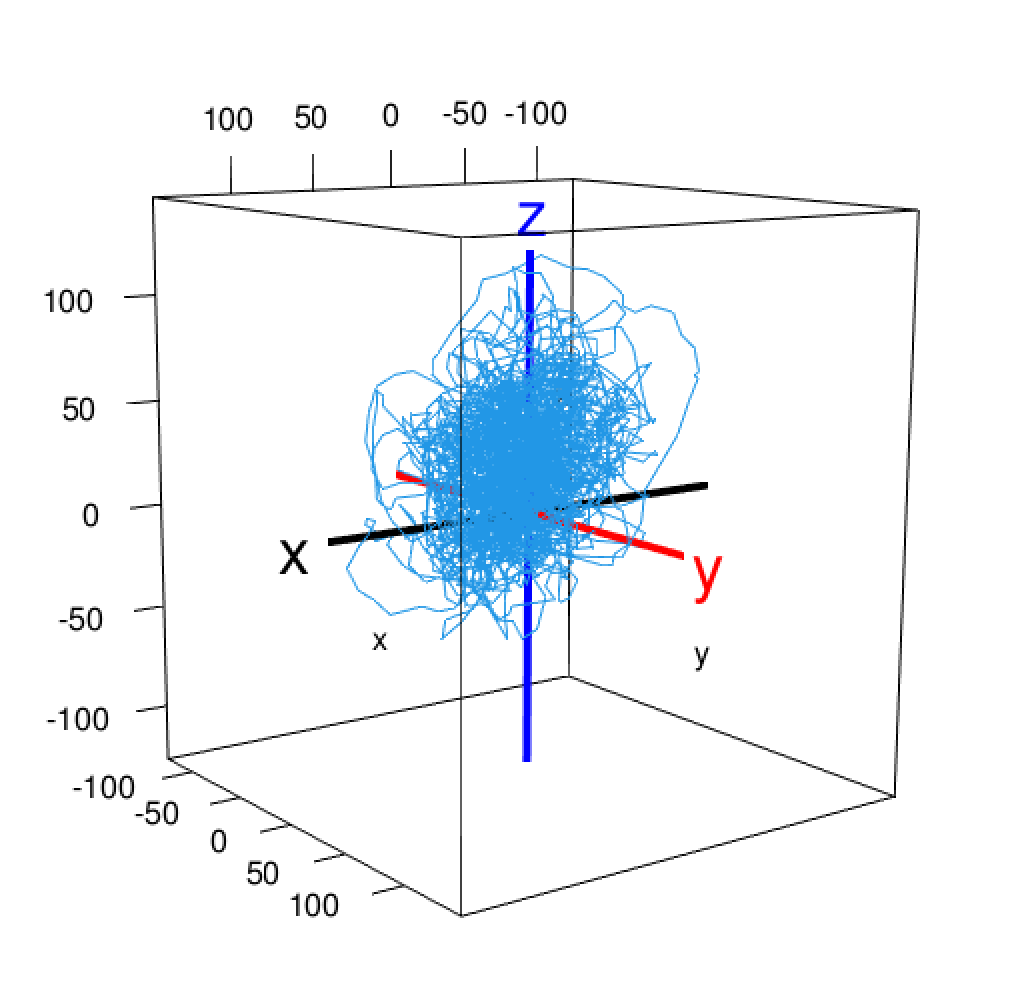}  \\
             \bf (E) &   \\
            \includegraphics[scale=0.51]{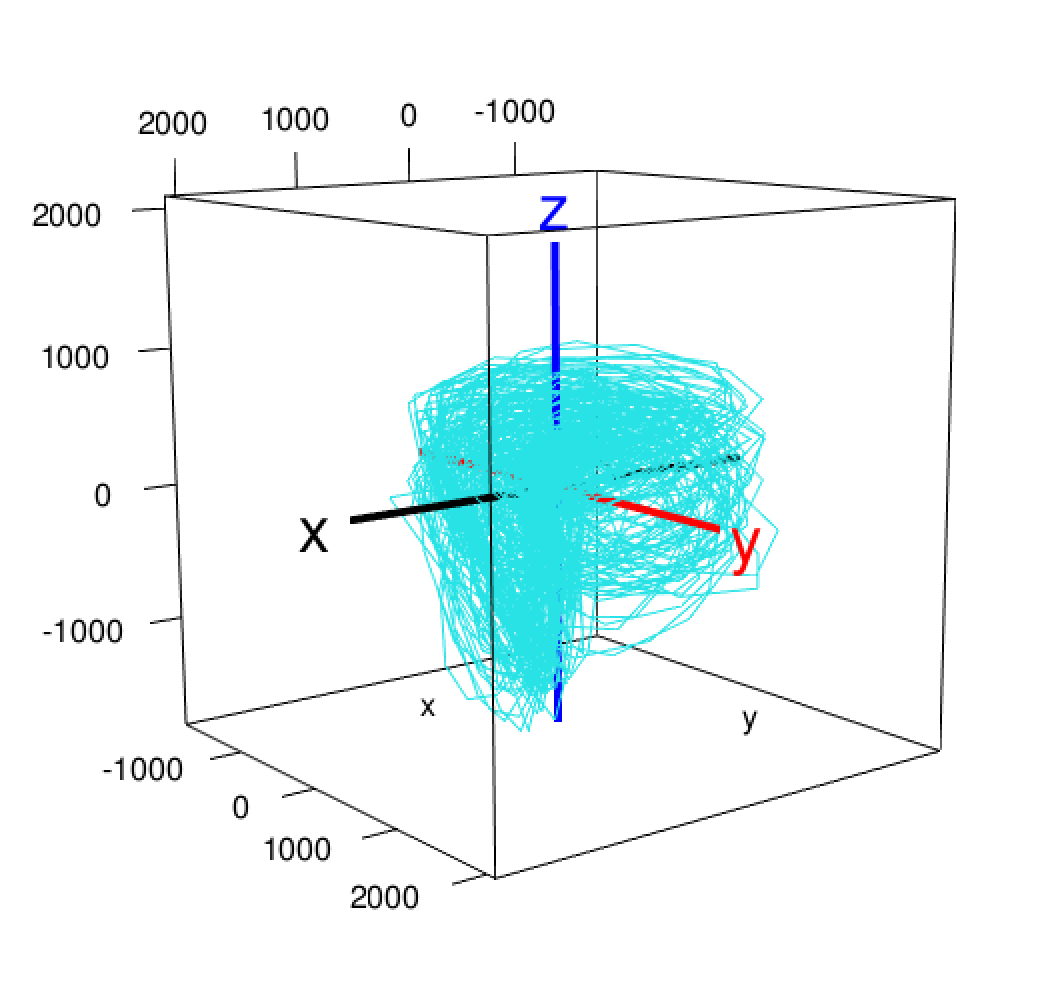}  &
        \end{tabular}
        \end{minipage}}
   \caption{Reconstructed phase space diagram, restricted to the space $xyz$.}
   \label{figRealData1}
\end{figure}

\section{Statistical Morphometry}\label{sect:statsMorp}
Given a set of points in a two or three-dimensional space, statistical morphology (or morphometrics) amounts to finding an appropriate geometric  characterization of the variability of the shape and size of the set of points. This characterization includes but is not limited to volumes and surface area, surface roughness, deviation from convexity, porosity, and permeability. We note that in general, the set of points may not be convex in the strictest sense, so there is a need for a method that relaxes  the convexity restriction.  The notions of $\alpha$-convexity  and $\alpha$-shape represent alternatives that relax the strict convexity assumption. 
\noindent These new $\alpha$-shaped objects are even more flexible than $\alpha$-convex objects in that $\alpha$ now controls the spatial scale of the estimator.  note that $\alpha$ is a unitless quantity. In fact, as $\alpha$ decreases, the $\alpha$-shape shrinks and more space appears among the sample points, whereas as $\alpha$ increases, the $\alpha$-shape object converges to the convex hull of the sample. In other words, in $\alpha$-shaped objects, $\alpha$ controls the amount of porosity between the sample points. These $\alpha$-shaped objects have been used in various fields for the characterization of biological systems, see for instance \cite{Lafarge2014} and the references therein,  or also \cite{Gardiner2018}. We  can  mention the pivotal work of \cite{Edelsbrunner1994} where the main algorithm for the construction of $\alpha$-shaped objects can be found.

\subsection*{Example}
In the example below, we illustrate the $\alpha$-shape construction respectively in two and three dimension, based on a random sample of data taken from the original object $S$.

We consider  $2500$ points in 3D, obtained from the object $S$ which is the object delimited by the curve  and $z=x^2+y^2$ where  $x$ and $y$ are random points selected in the interval $[-1,1]$.  In Figure \ref{fig:example2} below, we construct the $\alpha$-shape object (red) for  $\alpha=0$ (a), $\alpha=0.5$ (b), $\alpha=0.8$ (c), and $\alpha=2$ (d).

\begin{figure}[H] 
       \resizebox{1\textwidth}{!}{\begin{minipage}{1.1\textwidth}
   \centering \begin{tabular}{cccc}
   \bf (a) & \bf (b) & \bf (c) & \bf (d)\\
    \includegraphics[scale=0.15]{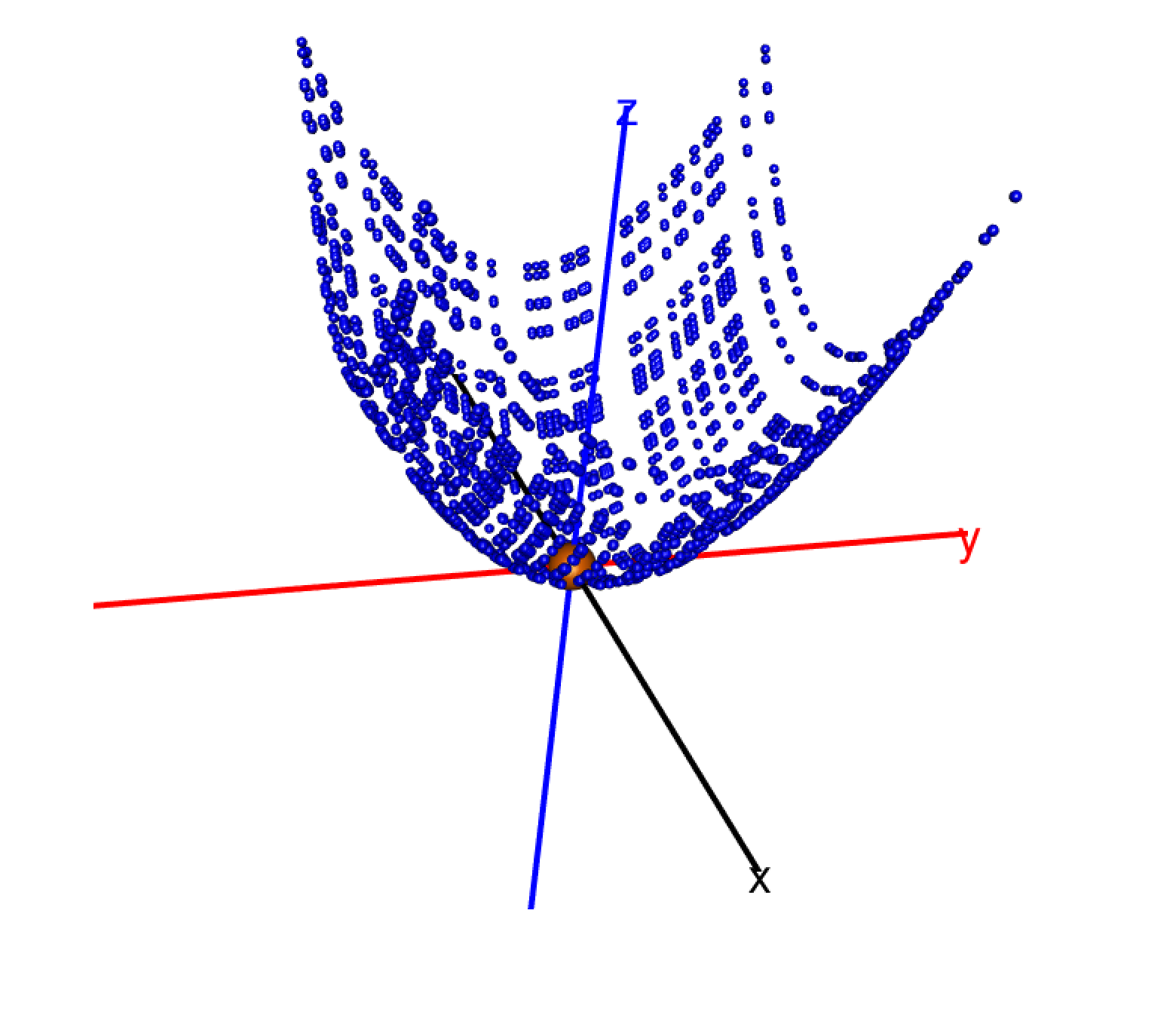}  &
   \includegraphics[scale=0.15]{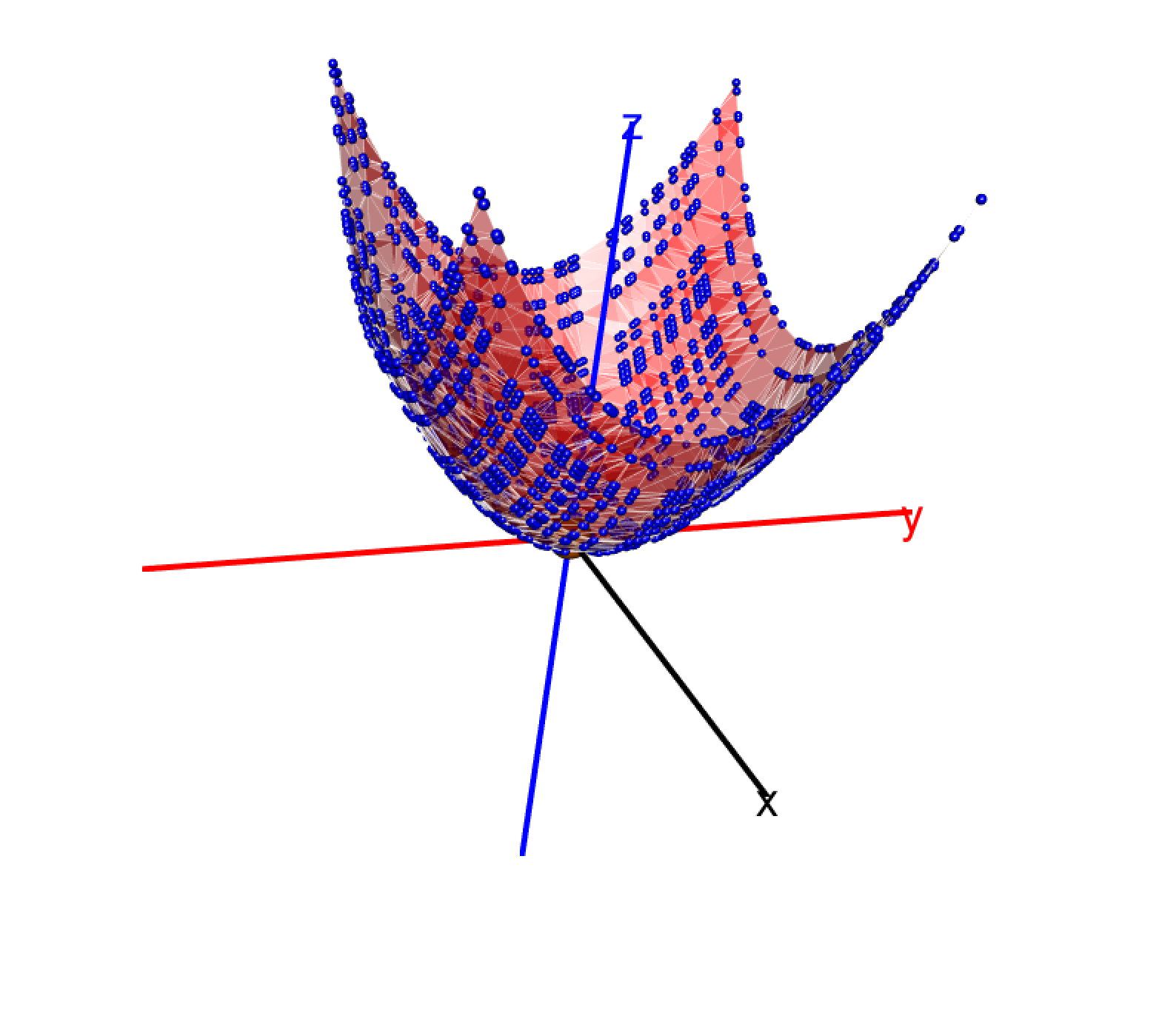}  &
     \includegraphics[scale=0.15]{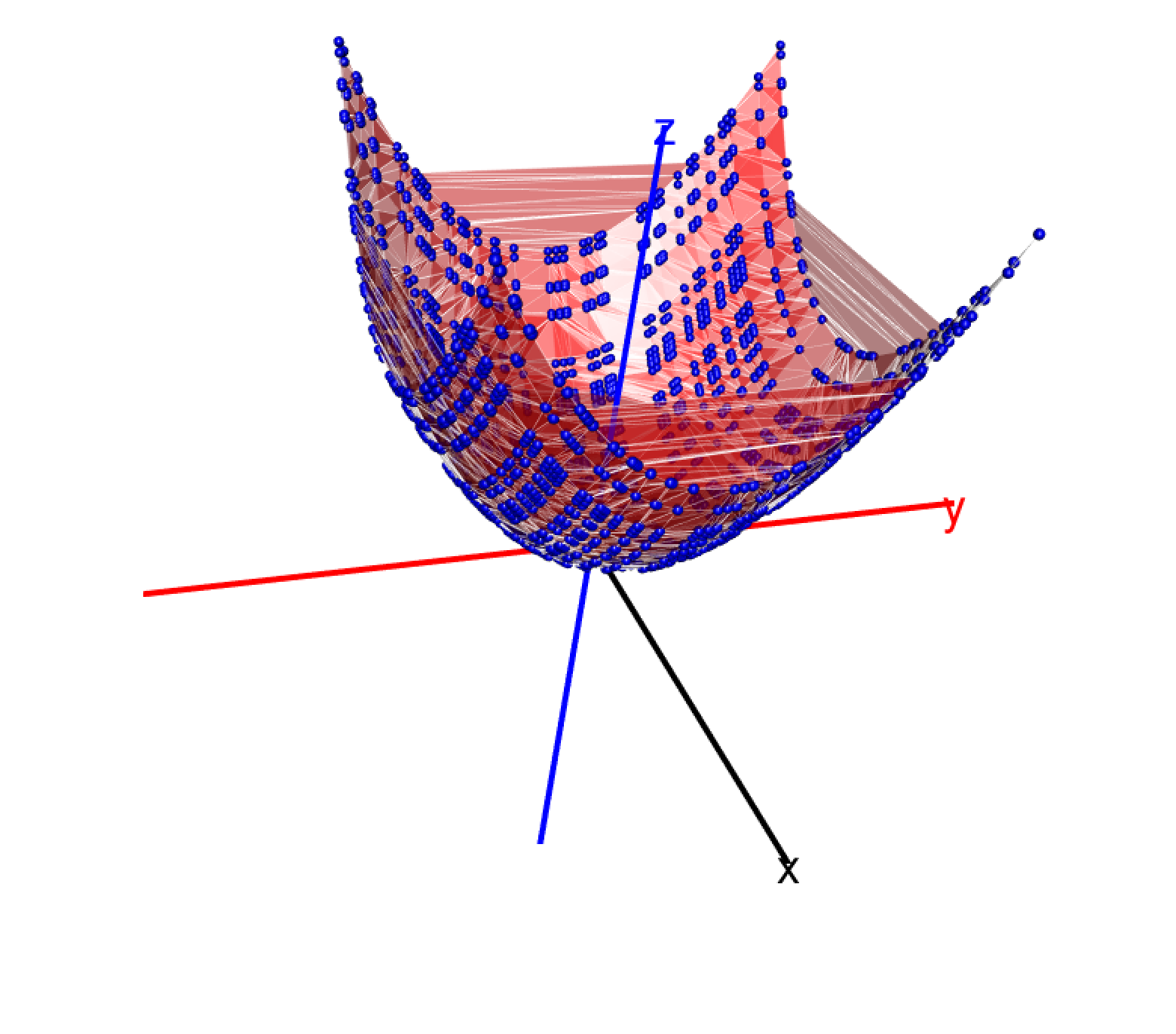}  &
         \includegraphics[scale=0.15]{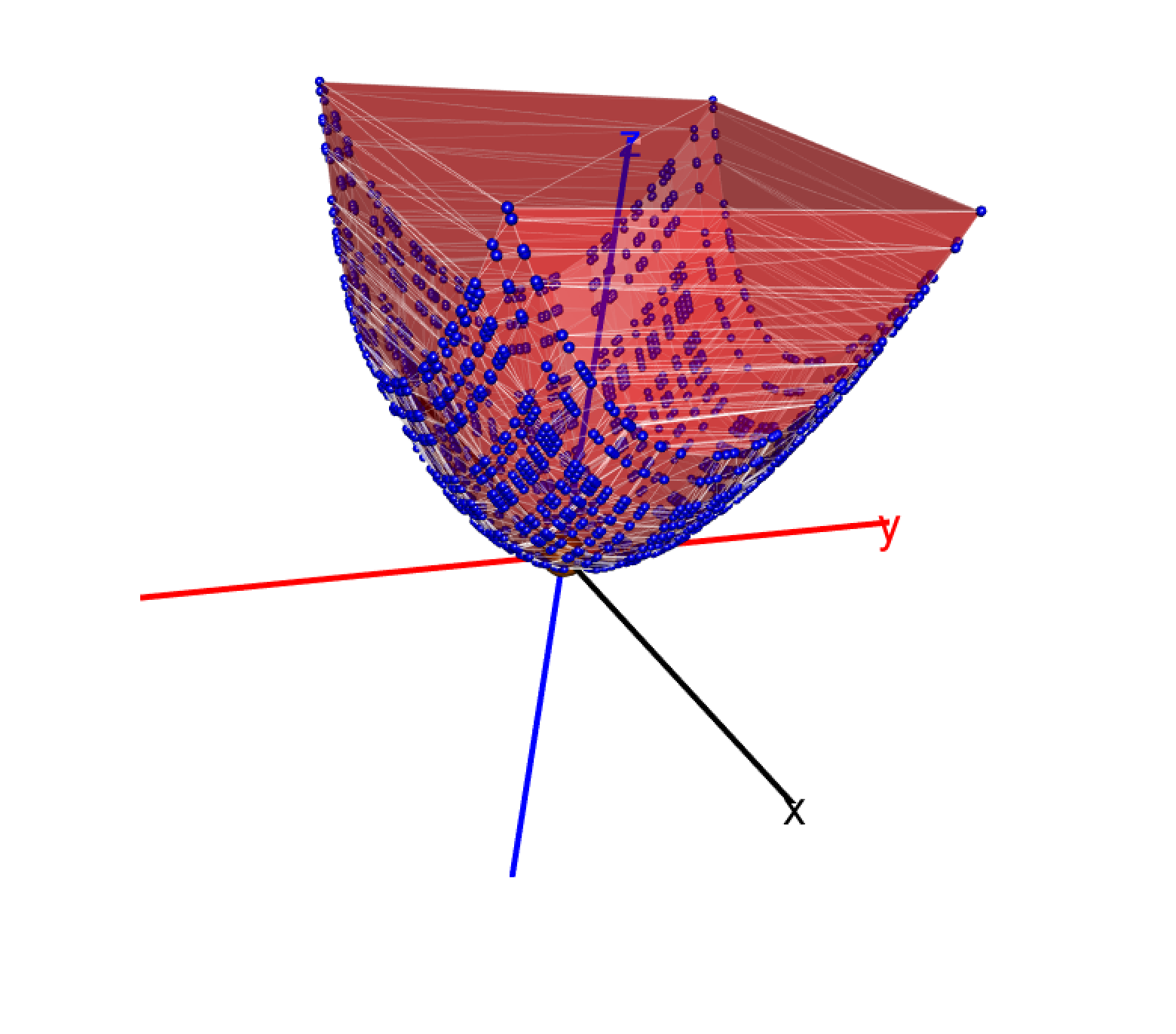}  
 \end{tabular}
  \end{minipage}}
   \caption{$\alpha$-shape object (red) of S ($z=x^2+y^2$) for  $\alpha=0$ (a), $\alpha=0.5$ (b), $\alpha=0.8$ (c), and $\alpha=2$ (d).}
   \label{fig:example2}
\end{figure}

\section{Complex Geometric Structurization}\label{CS}
In view of the apparent shape that can be observed from the reconstructed phase space above, the question of how to compare these complex structures arises. In other words, this question is related to the question of how to compare strange attractors. Among the methods proposed, we can mention the work of  \cite{Grassberger1983} and its many variants. The method  we propose is borne out of the observation that there seems to be a complex geometric structure whose shape changes from healthy patients to seizure-prone ones. So we will rely on the notion of $\alpha$-shape to construct the complex structure related to each situation. We will use the following algorithm to construct our ``complex structure" (CGS). The motivation comes from the fact that, given a time series, if it  has been observed long enough, it carries the signature of the original phase space diagram in which the true  system it originated from evolves. If we have many such time series from the same system, we should be able to capture enough information about the true phase space diagram. The large number of these time series should be enough to eventually eliminate noise or undesired artefacts from the reconstruction. We then expect each reconstructed phase diagram to be topologically equivalent to any other, therefore forming a structure compact enough to be the revolving center of all other reconstructed phase space diagram. In general, the the dimension of the space in which the true system evolves is greater than three, making it impossible to visualize with the naked-eye. Our empirical approach is to consider only a cross section of the reconstructed phase space in dimension three in this case. the reason for this choice is two-fold: one, we can actually visualize a part of the true phase space diagram.  Second, we can use existing method to evaluate the volume of the structure in lower dimensions. This is to say that the volume is preserved in reconstruction. In doing this, we are trying to find a measurable  identifier or markup for this group of time series that will vary from groups to groups and from individual to individual.
\subsection{Explanation of the method}
Given $N$ time series, we use the Takens reconstruction technique to obtain the embedding dimension $m_n$ (using ACF) and time delay (or sampling interval) $\rho_n$ for $n=1,2\cdots, N$ (using the method of false nearest neighbors). Let $m=\min\{m_1, \cdots, m_N\} $ and $\rho=\min \; \{\rho_n, n=1, 2, \cdots, N\}$.  If $m\geq 3$, then for each time series, we obtain the complex structure CGS$_{\alpha(n)}$. Let $\ds \alpha=\min\; \{ \alpha(n):  \mbox{Vol}_{3D} (\mbox{CGS}_{\alpha(n)}) ~\mbox{is maximized} \}$. We use the $\alpha$-shape technique to obtain the volume of the $CGS_{\alpha}$ in 3D. This step is crucial since we choose  to represent the complex structure only using the first three delay-coordinates $x_1=x (t), x_2=x(t-\rho), x_3=x(t-2\rho)$ even if the actual space has dimension $m>3$. The motivation for this selection  is that the volume of the CGS based on any  combination of three coordinates  would not significantly be different from any other volume obtained from any other combination of three coordinates.  The reasoning behind the choice of $\alpha$ is  that the volume of the complex structure is bounded by the volume of its convex hull,  as $\alpha$ increases, so we select the smallest alpha that maximizes the volume, see section \ref{sect:ChoiceOfDelayCoord}  below for an illustration. This leads to the following algorithm

\subsection{Algorithm}
 \begin{enumerate}
 \item Use all the times series collected on patients in different groups.  
 \item For each time series, reconstruct a 3D  ``strange attractor" from the first three delay-coordinates.
  \item Use the  $\alpha$-shape technique to construct a ``Complex Geometric Structure" (CGS) related to all  strange  attractors.
 \item Define a measure related to each CGS that that can be statistically analyzed. This could be the volume, the surface area, the center, etc.
 \item Repeat this procedure for all replicates (if there are any) and thus obtain new data. 
 \item 
 \begin{itemize}
 \item[(a)] For comparison:\\
 Use a statistical test to see if there is a significant difference between measures  of different groups. This could be a parametric or a nonparametric test.
 \item[(b)]  For Prediction:\\
 Use the data obtained for training in machine learning, see \cite{Kwessi2020} (preferred to standard generalized linear model (GLM) models because of the absence of a specific model) and test it on potential new data.
 \end{itemize}

 \end{enumerate}
 \subsection{Comments}
 \begin{enumerate}
 \item The CGS terminology stems from the fact the structures obtained after reconstruction do not have  of classical geometric shapes. Their shapes are  rather  ``complex" in nature.
 \item We propose to use all time series available  for a particular region of the brain, at a specific instant; for example before seizure, during seizure, or after seizure. It is common to use  one time series, see for instance \cite{Fisher2009a}. We note that repeated measures on the same subject do not yield the same values and thus in the reconstruction, there could be individual times series whose excursions in the phase space  are wider than others yielding a bigger geometric structure in size and volume. In this case, individual times series could be used and the subsequent large  volumes could be trimmed out if necessary. This is particularly important  if one is interested  for example in obtaining a richer data set to analyze.  
  \item The embedding dimension $m$ often obtained is greater than 3, so in the absence of a  visualization mechanism for  data  in spaces of  dimensions higher than 3, we propose to focus on only the first three delay coordinates.  

 \item Using the R-package \textit{Alphashape3d}, measures such as volumes or surface areas can be obtained for the $\alpha$-shape object under consideration.
 \item As we saw in the example above, the choice of $\alpha$ is critical in the $\alpha$-shape construction. We propose to select the minimum value of $\alpha$ for which the volume is maximized. 
 \item Finally, we note that in practice, the experimenter can used  data from chaotic or complex systems.  One can check for chaos in data by calculating Lyapunov exponents. If the data are  sensitive to initial conditions and  chaotic, the attractor would be referred to as  a ``strange attractor", see \cite{Celso1987}. Even if  on the other hand the data are sensitive to initial conditions but non chaotic, we would still keep the same ``strange attractor" denomination because in this case (the Lyapunov exponents are non-positive), the attractor obtained is still strange,  see for instance \cite{Celso1984, Paladin1987}. It also noteworthy to observe that  non chaotic systems that are sensitive to initial conditions are sometimes referred to in the literature as complex systems, see for instance the comparative review between chaotic and complex systems by  \cite{Rickles2007}. 
  \end{enumerate}

\subsection{Technical considerations}
In this section, we examine some technical considerations to keep in mind when implementing this method.
\subsubsection{Choice of $\alpha$}
As we mentioned above, the choice of $\alpha$ is critical in this process. In what follows, we show that the optimal value of $\alpha$ is  smallest that maximizes the volume of the CGS. Moreover, if one is interested in comparing  CGS among groups, it would be adequate to select a common value of $\alpha$, for example as  the largest $\alpha$ value, among the  values that maximize the CGS for each  group. For instance,  using the EDATA set, we obtained $m=10$ and $\rho=1\Delta t$, with $\Delta t=\frac{1}{fs}=5.76$ ms. In Figure \ref{figAlphaOptim} below, we observe that for a value of $\alpha\approx 200$, the volume is maximized for the subset A, whereas the maximum is reached for subset E at $\alpha\approx 580$. So, we will choose $\alpha_{\mbox{optim}}=580$ as the optimal value of $\alpha$ if we want to compare the two CGSs.

  \begin{figure}[H] 
  
 \resizebox{1\textwidth}{!}{\begin{minipage}{2.5\textwidth}
   \centering
    \includegraphics[scale=1]{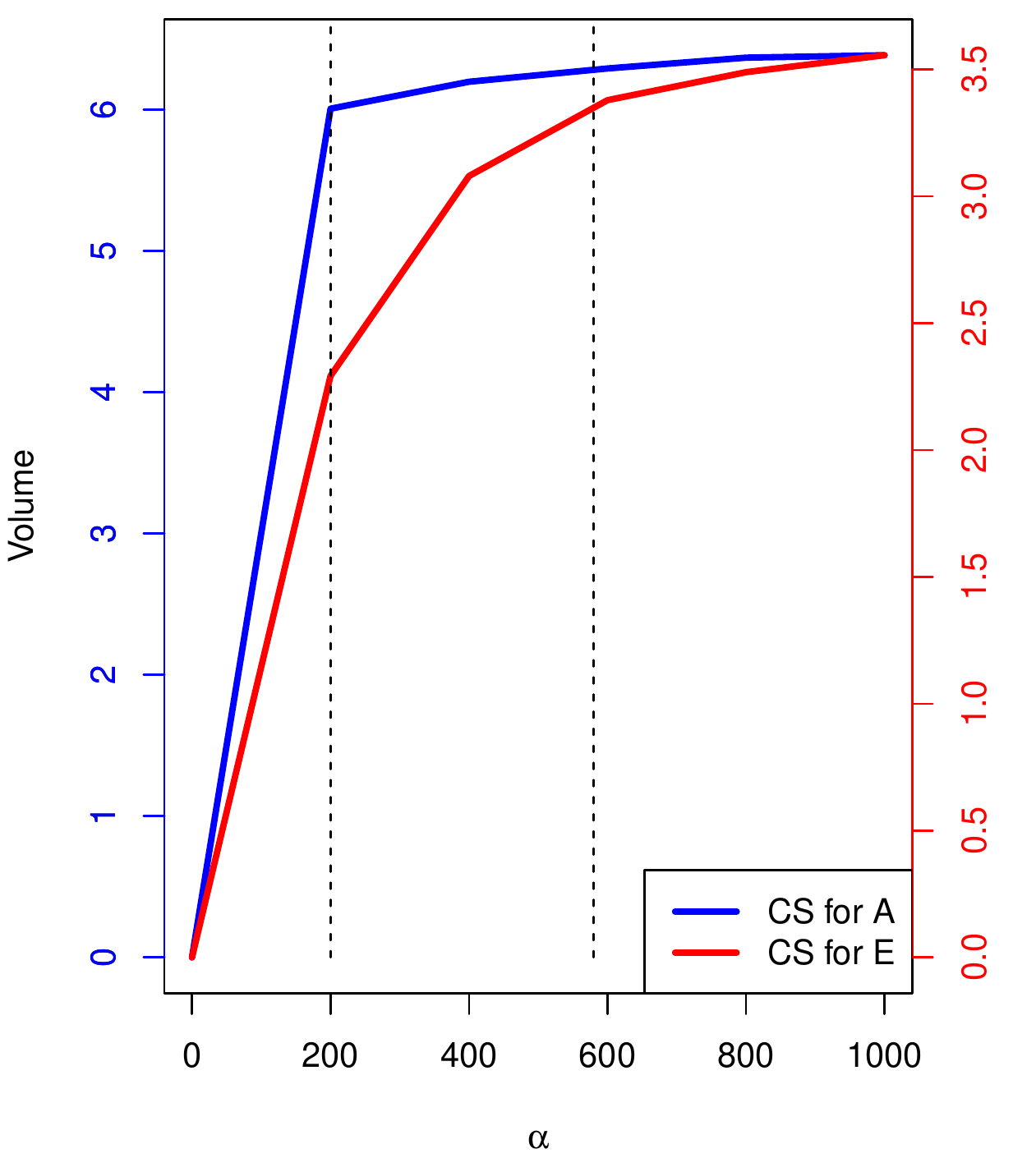}  
      \end{minipage}}
   \caption{Evolution of the volume  CGS as function of $\alpha$.}
   \label{figAlphaOptim}
\end{figure}
\subsubsection{Choice of the embedding dimension.}\label{sect:ChoiceEmbDim}

We note that the embedding dimension is closely related to the time lag. We are not going to discuss which should be estimated first. However, we note that  there are two methods for estimating the time lag and they do not always  yield the same embedding dimension. We propose to select the embedding dimension as the smallest value of  the two embedding dimensions when they are different.  

\subsubsection{Choice of the delay-coordinates}\label{sect:ChoiceOfDelayCoord}
\begin{figure}[H] 
   \resizebox{1\textwidth}{!}{\begin{minipage}{1.2\textwidth}
   \centering
\begin{tabular}{cc}
\bf (A) & \bf (E)\\
 \includegraphics[scale=0.4]{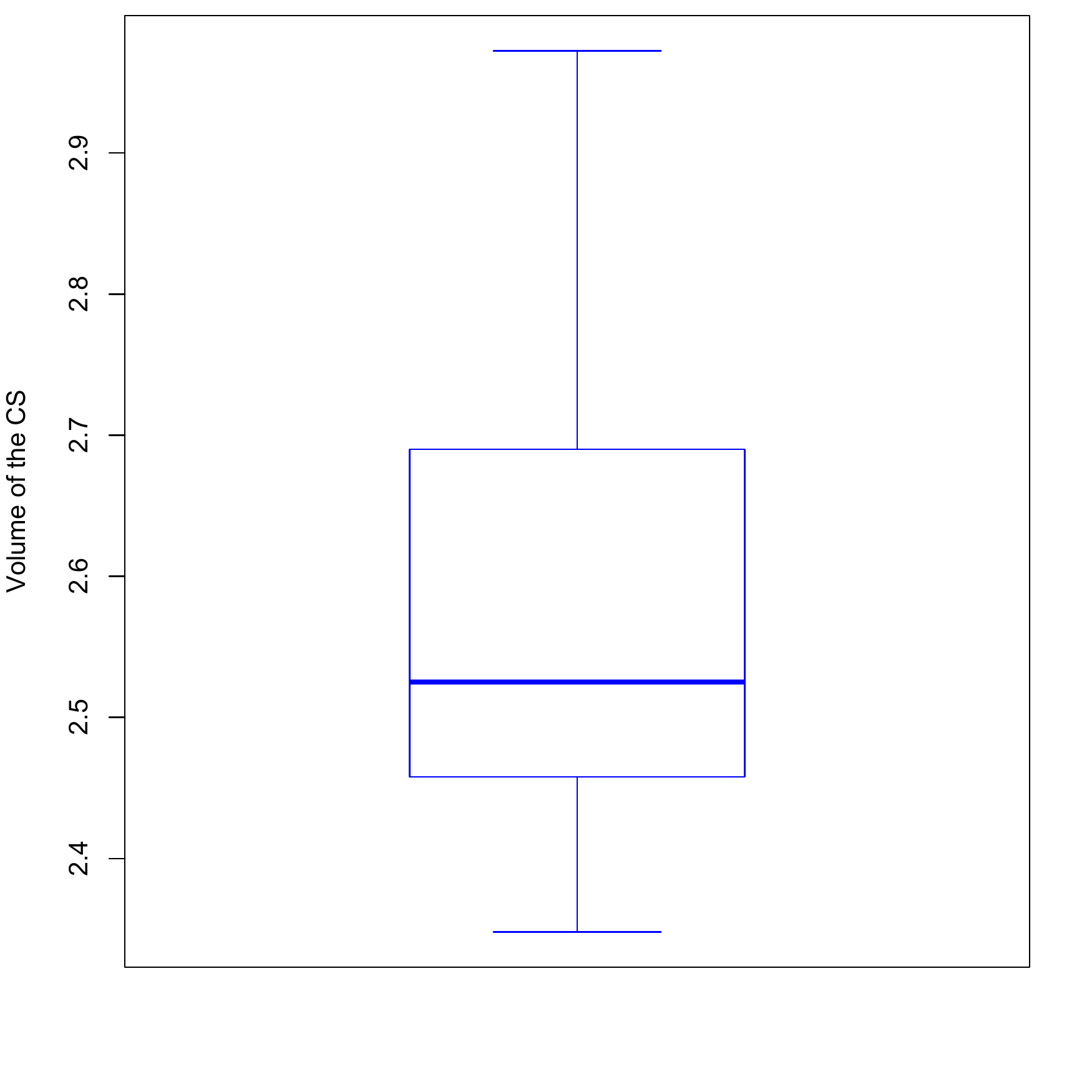}  &
    \includegraphics[scale=0.4]{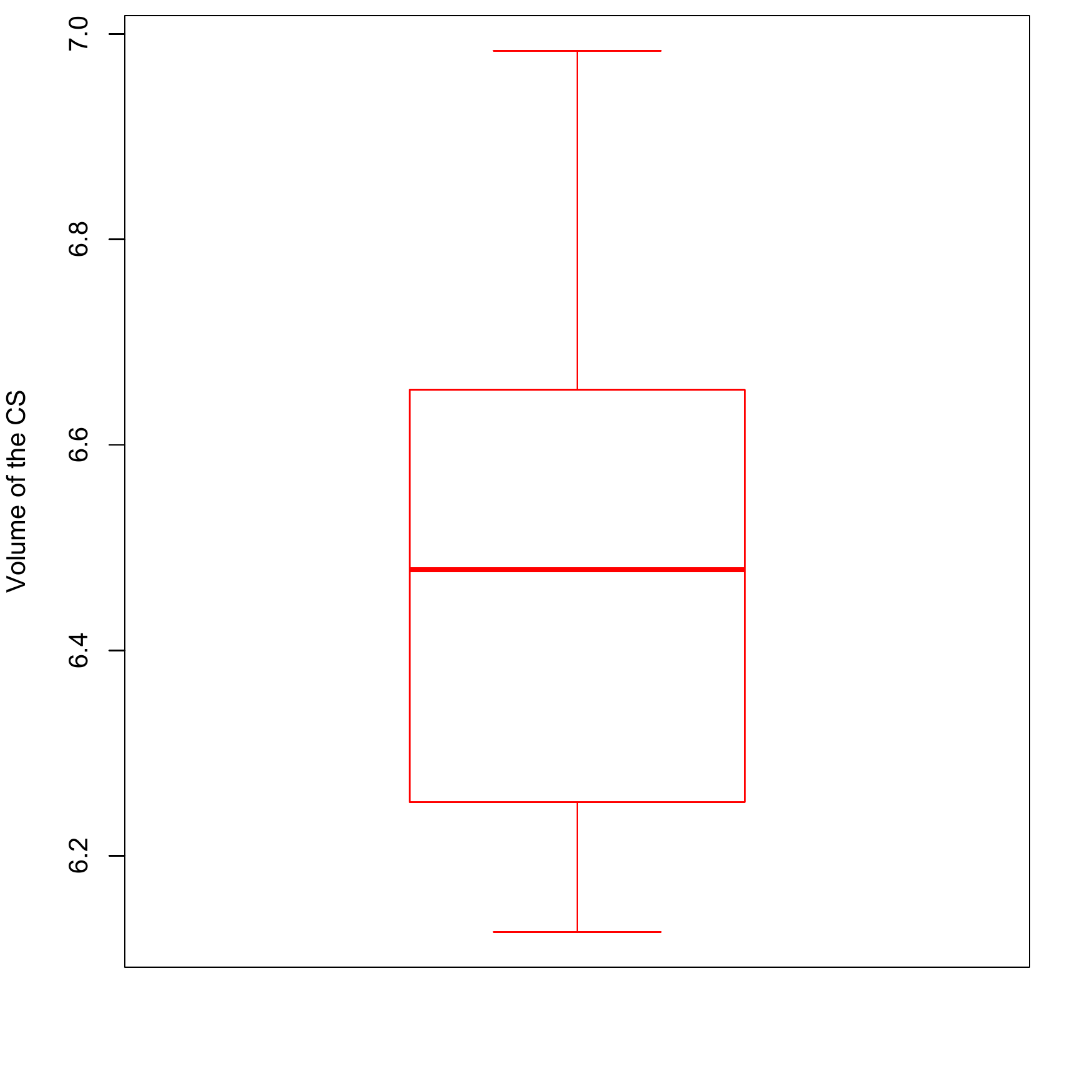}  
       
        \end{tabular}

  \end{minipage}}
   \caption{Evolution of the volume the CGS for the 120 different combinations of  3  delay-coordinates for subsets A and E.}
   \label{figDelayCoordCombo}
\end{figure}

In Section \ref{sect:ChoiceEmbDim}, we mentioned that we will select the first three delay-coordinates to construct the complex structure. However, it is worthwhile to consider the question of  whether the volume would change if a different combination of delay-coordinates is used. Our choice is based on the conjecture  that it does not matter which combination of delay-coordinates is selected. To emphasize that point, we will discuss the case of subsets A and E. In the case of subsets A and  E, the embedding dimension found is $m=10$. So there are ${10 \choose 3}=120$ possible different combinations of three delay-coordinates, selected among 10. So, for each combination, we will construct the corresponding CGS and assess whether the volume changes significantly  across all the different CGS's.  Figure \ref{figDelayCoordCombo} shows the volume of the CGS by combination of three delay-coordinates for subsets A and E. The boxplots suggest that  the distribution of volumes  for each set is reasonably concentrated around its median (thick red and blue lines) with a small range and no outliers. This is to say that taking the first three delay coordinates seems reasonable despite minor variations otherwise.

\noindent The main takeaway from the plots above  is that the variation of the volume as a function of the combination of delay-coordinates appear not to be significant. Even for seizure data like subset E, the variation in volume appears to be mild, which is our impetus for conjecturing that a similar observation could be made for other datasets. Obviously we cannot predict what will happen for all datasets, and ultimately that endeavor would require a mathematical proof that may go beyond the scope of the present manuscript.

\section{Applications}\label{Appl}

\subsection{Analysis of EDATA }\label{AnalysisEdata}
 
\begin{figure}[H] 
\resizebox{1\textwidth}{!}{\begin{minipage}{1.3\textwidth}
    \centering \begin{tabular}{cc}
  \bf (A1) & \bf (A2) \\

        \includegraphics[scale=0.43]{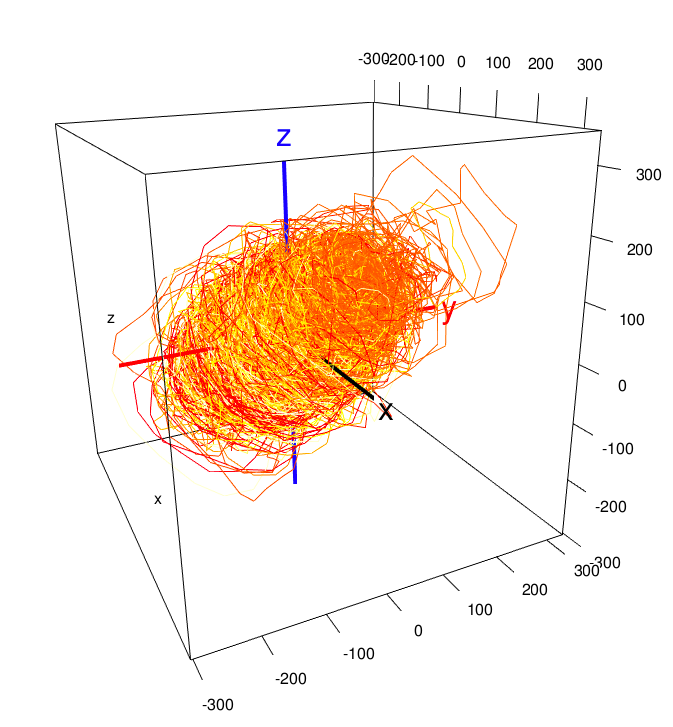}   &
    \includegraphics[scale=0.46]{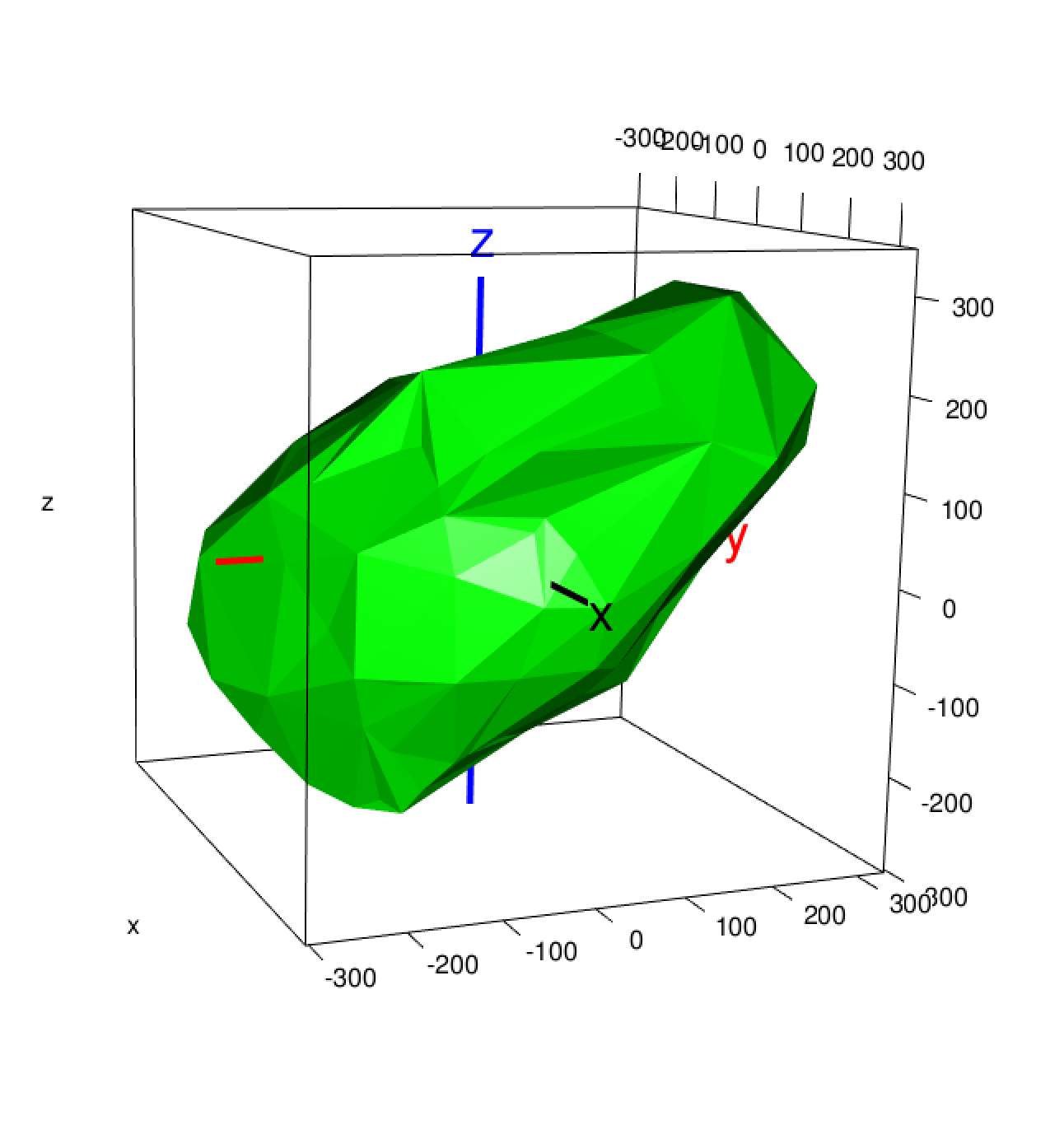}  
    
      \end{tabular}
        \end{minipage}}
          \caption{Complex geometric structure   for set A  in EDATA.}
   \label{figdataStructureA}
\end{figure}

\begin{figure}[H] 
\resizebox{1\textwidth}{!}{\begin{minipage}{1.3\textwidth}
    \centering \begin{tabular}{cc}
   \bf (B1) & \bf (B2)\\
   
      \includegraphics[scale=0.39]{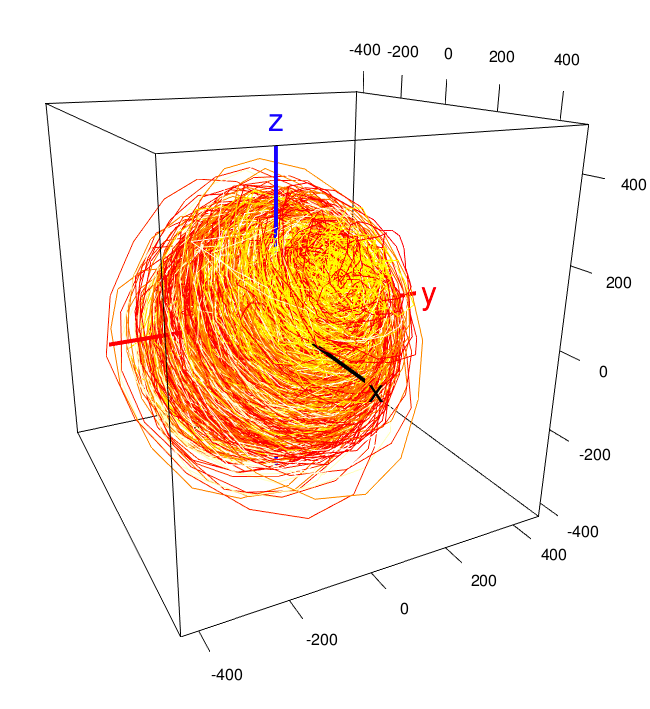} &
      \includegraphics[scale=0.46]{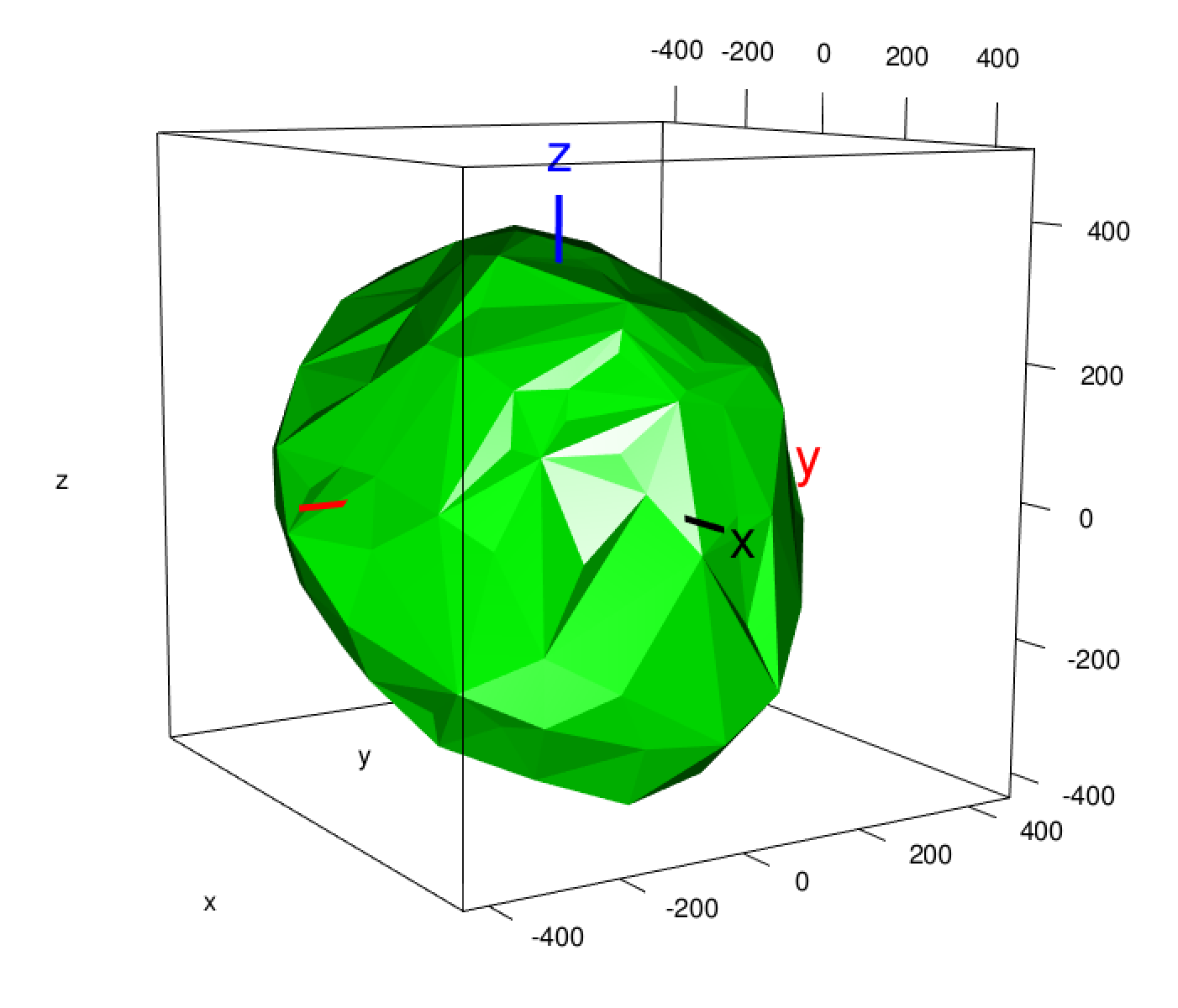}  
        \end{tabular}
        \end{minipage}}

   \caption{Complex geometric structure    for set  B in EDATA.}
   \label{figdataStructureB}
\end{figure}

\begin{figure}[H] 
\resizebox{1\textwidth}{!}{\begin{minipage}{1.3\textwidth}
   \centering\begin{tabular}{cc}
   \bf (C1) & \bf (C2) \\
        \includegraphics[scale=0.43]{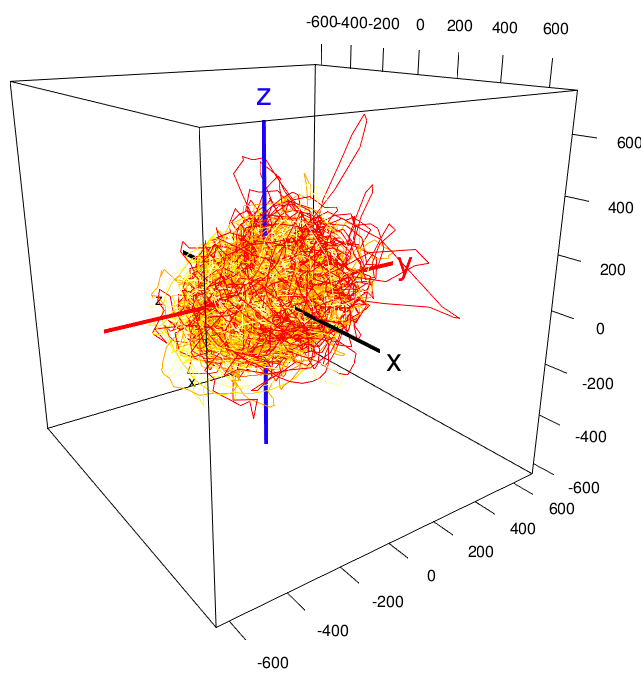}   &
    \includegraphics[scale=0.48]{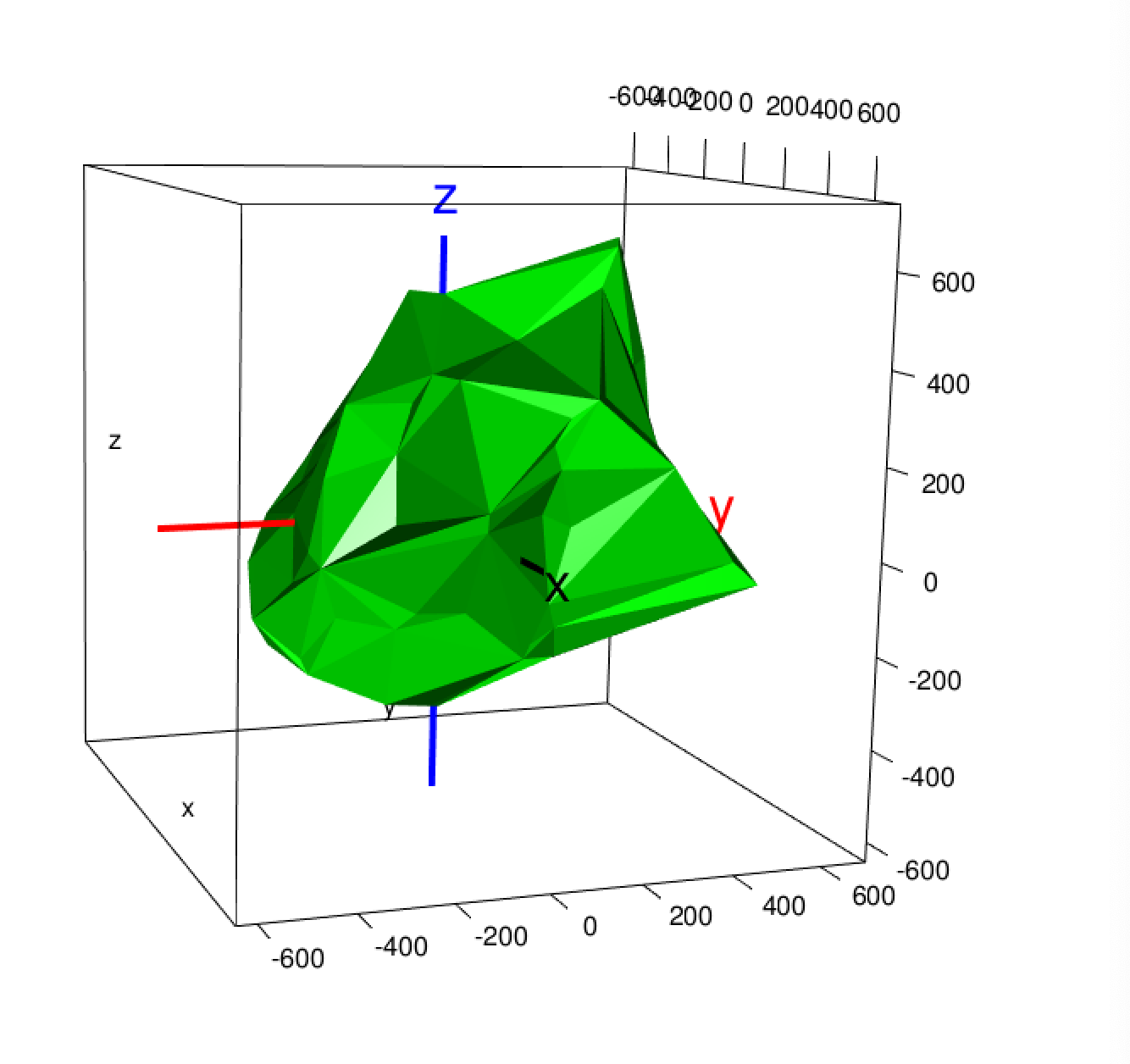}  
    
     \end{tabular}
        \end{minipage}}
           \caption{Complex geometric structure   for sets C in EDATA.}
   \label{figdataStructureC}
\end{figure}

\begin{figure}[H] 
\resizebox{1\textwidth}{!}{\begin{minipage}{1.3\textwidth}
   \centering\begin{tabular}{cc}    
\bf (D1) & \bf (D2)\\
      \includegraphics[scale=0.41]{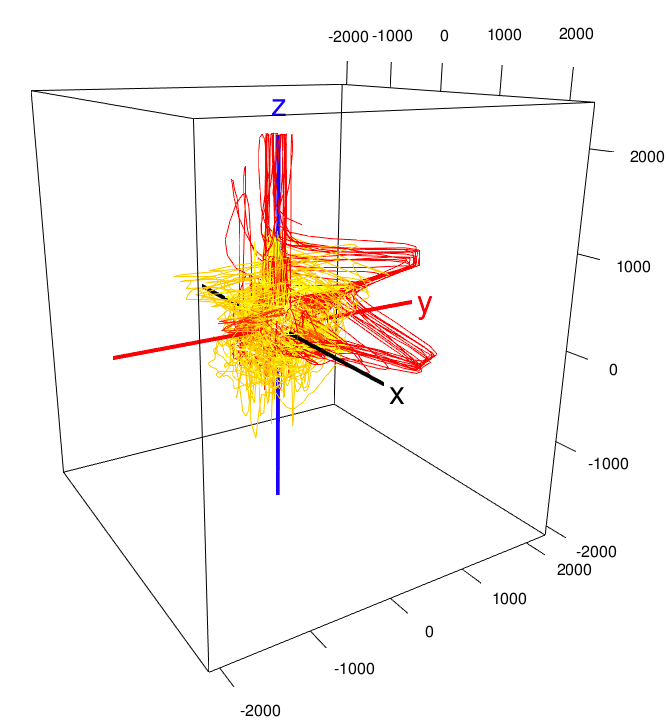} &
      \includegraphics[scale=0.46]{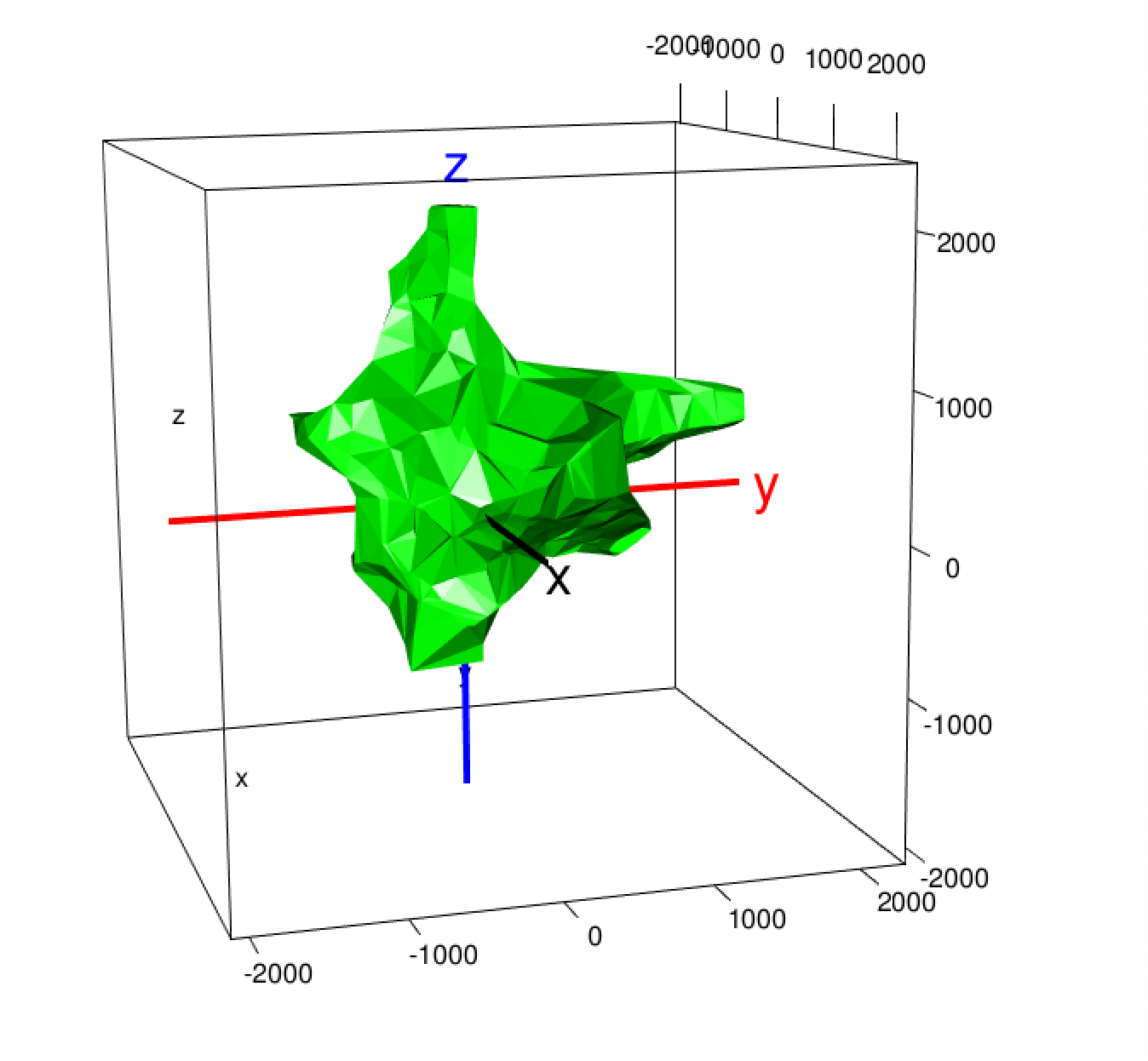}  
        \end{tabular}
        \end{minipage}}

   \caption{Complex geometric structure    for set D in EDATA.}
   \label{figdataStructureD}
\end{figure}

\begin{figure}[H] 
\resizebox{1\textwidth}{!}{\begin{minipage}{1.3\textwidth}
    \centering \begin{tabular}{cccc}
   \bf (E1) & \bf (E2) &  & \\
        \includegraphics[scale=0.4]{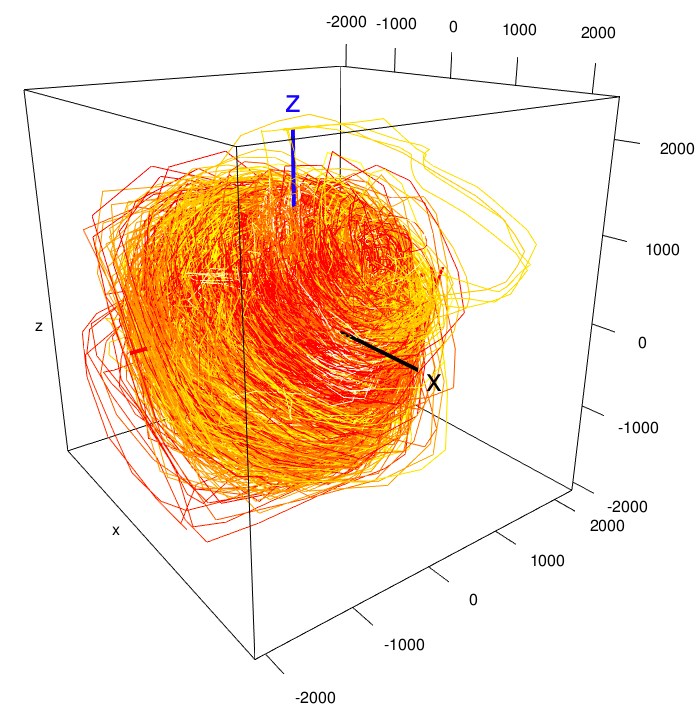}   &
    \includegraphics[scale=0.48]{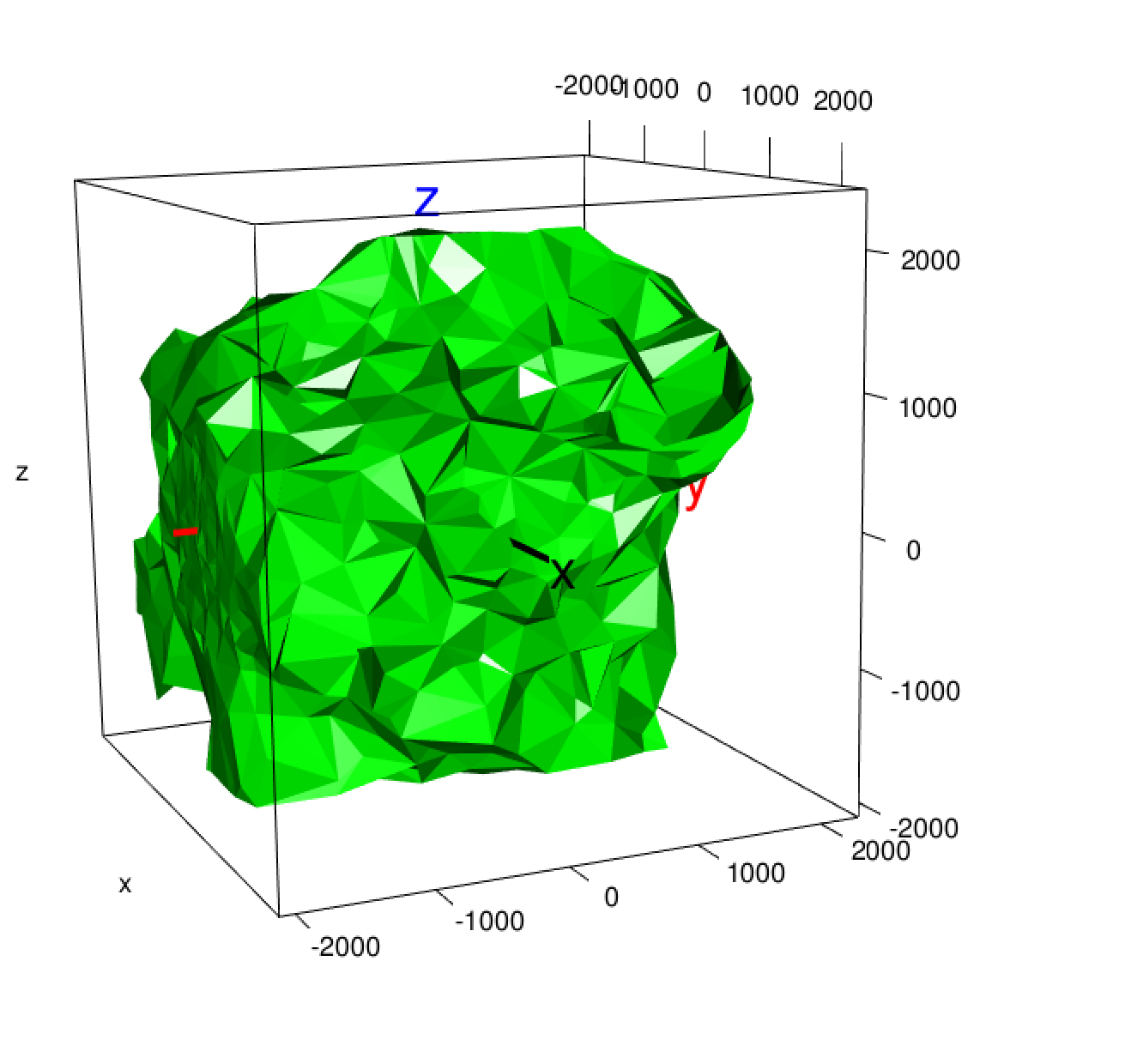}  & & 

        \end{tabular}
        \end{minipage}}

   \caption{Complex geometric structure for set E in  EDATA.}
   \label{figdataStructureE}
\end{figure}

In Figure \ref{figRealData1}, we have shown a representation of the reconstructed phase space diagram for one time series in the space $x=x(t), y=x(t-\rho), z=x(t-2\rho)$ and for each set A--E.  In Figures \ref{figdataStructureA}--\ref{figdataStructureE}, we construct the phase space diagram and the complex structures for the EDATA set,  for all the 100 time series in each subset. We observe that for each set, we obtain a compact structure whose volume we can now evaluate, see {(\bf A1)}--{\bf(E1)}.  We selected $\alpha=580$ for each case, because of all sets A-E, 580 is the value of alpha that maximizes the volume of the  complex structure of E, the largest of all of them (see  the selection criterion given above), see figure \ref{figAlphaOptim} above. The CGS structure  for each set is then obtained, see {\bf (A2)}--{\bf (E2)}. 
We  observe that the CGS for set E appears to be bigger than all other CGSs,  which is a sign of larger excursions in the phase space and therefore synonymous with a much  intense brain activity during seizure. 
  

\noindent Pooling the data may seem to inflate the volume of the CGS. So instead, we could construct the complex structure related to individual channel and evaluate their volumes. This will give a richer data to analyze, where outliers can be removed. Figure \ref{figRealData11} below shows the boxplot of the volume of the  complex structure for each  channel. The volume for subset E is very large compared to A-E and obscures the distribution for subsets A-D. As a result, in the 2nd panel in Figure \ref{figRealData11} we remove subset E so that the distributions of subsets A-D can be more clearly seen.  

 \begin{figure}[H] 
\resizebox{1\textwidth}{!}{\begin{minipage}{1.8\textwidth}
     \centering\includegraphics[scale=1]{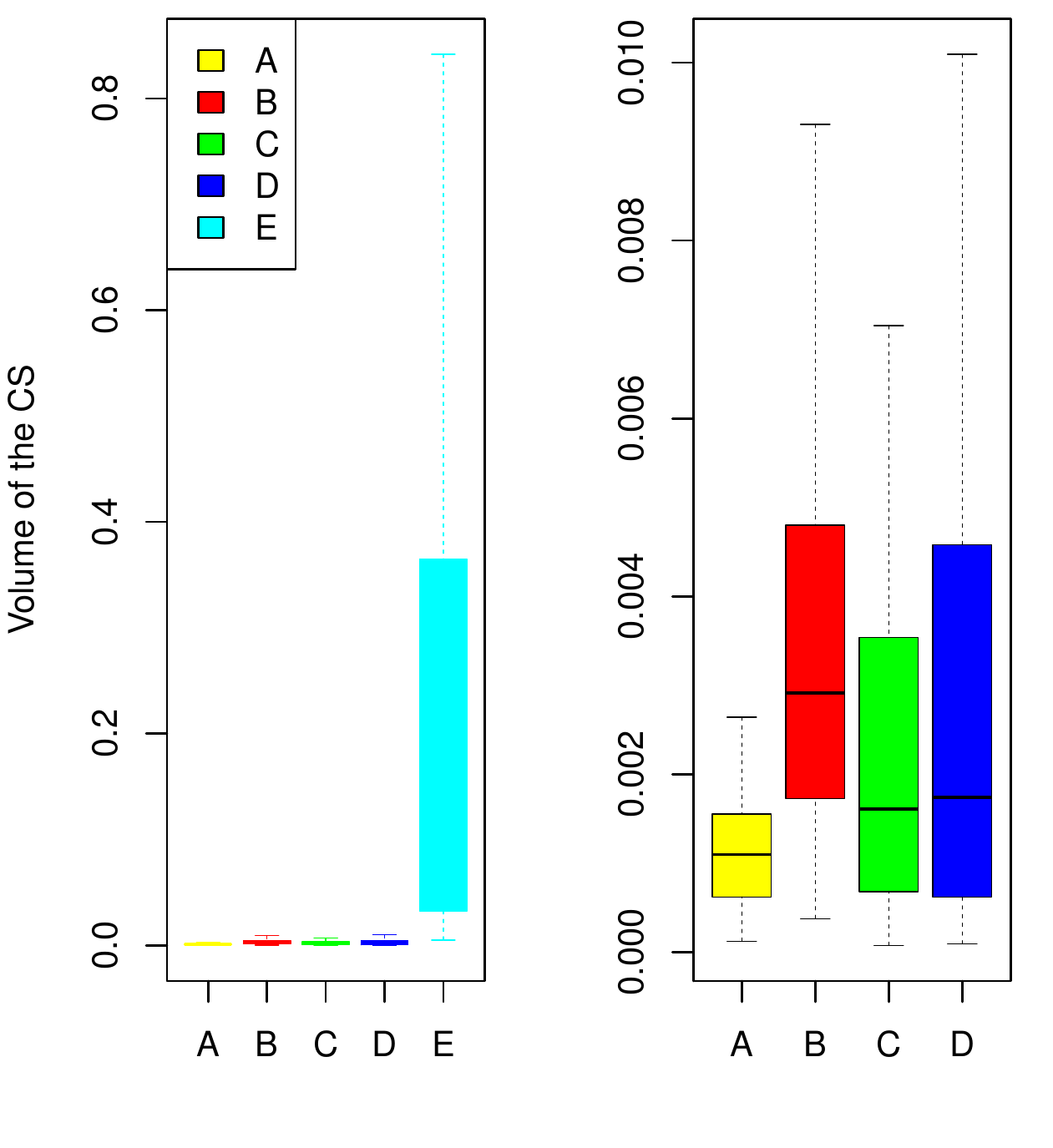}  
            \end{minipage}}
   \caption{Boxplots for the CGS   for subsets A--E in EDATA, obtained from individual times series.}
   \label{figRealData11}
\end{figure}
\noindent 
This Figure is further evidence of what already observable in the raw data in Figure \ref{figRealData0} and Figure \ref{figRealData1}. Sets A--D have comparable amplitude at the time series level confirmed by the fact that the volume of CGS are in similar range. The amplitude at the time series level of Set E is much higher than that of sets A--D, which is also confirmed by a larger volume at CGS level. This is further evidence that during seizure, these volumes increase substantially when compared to seizure-free intervals. More importantly, we can extract meaningful statistics from this data as suggested in Table \ref{table1} below:

 \begin{table}[H] 
\resizebox{1\textwidth}{!}{\begin{minipage}{1\textwidth}
\centering
\begin{tabular}{|c|c|c|c|c|c|c|c|}
\hline

&Min & $Q_1$ & Median & $Q_3$ & Max & Mean & SD\\ \hline
A & 12.30 & 63.63  &  115.87 & 154.17 &317.88 & 109.84 & 66.08\\ \hline
B & 37.89 & 63.63 & 173.50 & 291.31& 479.06  &364.25& 266.63 \\ \hline
C & 7.73 &  68.38 & 161.30 &  262.70 & 351.60 &  262.10  & 333.70 \\ \hline
D & 9.45&  62.78  & 174.10&  4569.00 & 17800.00  & 692.30&  206.54 \\ \hline
E& 516.3 & 3277.30 &  9279.40 &  36270.50 & 134300.00 & 26422.50 &33682.48  \\ \hline

\end{tabular}
\end{minipage}}
\caption{Table of volumes of CGS for each set A--E. The values are  of order $10^{-5}$.}
\label{table1}
\end{table}

\subsection{Analysis of auditory and visual cortex of the brain under auditory and  visual tasks,  and rest}\label{AnalysisVA}

In this example, we discuss data collected for the Brain Core Initiative at UAB. This data was collected on 20 individuals in two regions of their brain, namely the Auditory and Visual cortex, see Figure \ref{figBrain} below. The patients were subject to three ``tasks": Auditory, Visual, and Rest. Measurements of brain activity was obtained as EEG times series, after removing unnecessary artefacts. For this data set, the optimum value of $\alpha$ is $\alpha_{\mbox{optim}}=300$. We will consider the following subsets: Auditory cortex-Auditory task, Auditory cortex-Visual task, Auditory cortex-Rest, Visual cortex-Visual task, Visual cortex-Visual task, Visual cortex-Rest.

 \begin{figure}[H] 
   \centering
    \includegraphics[scale=0.35]{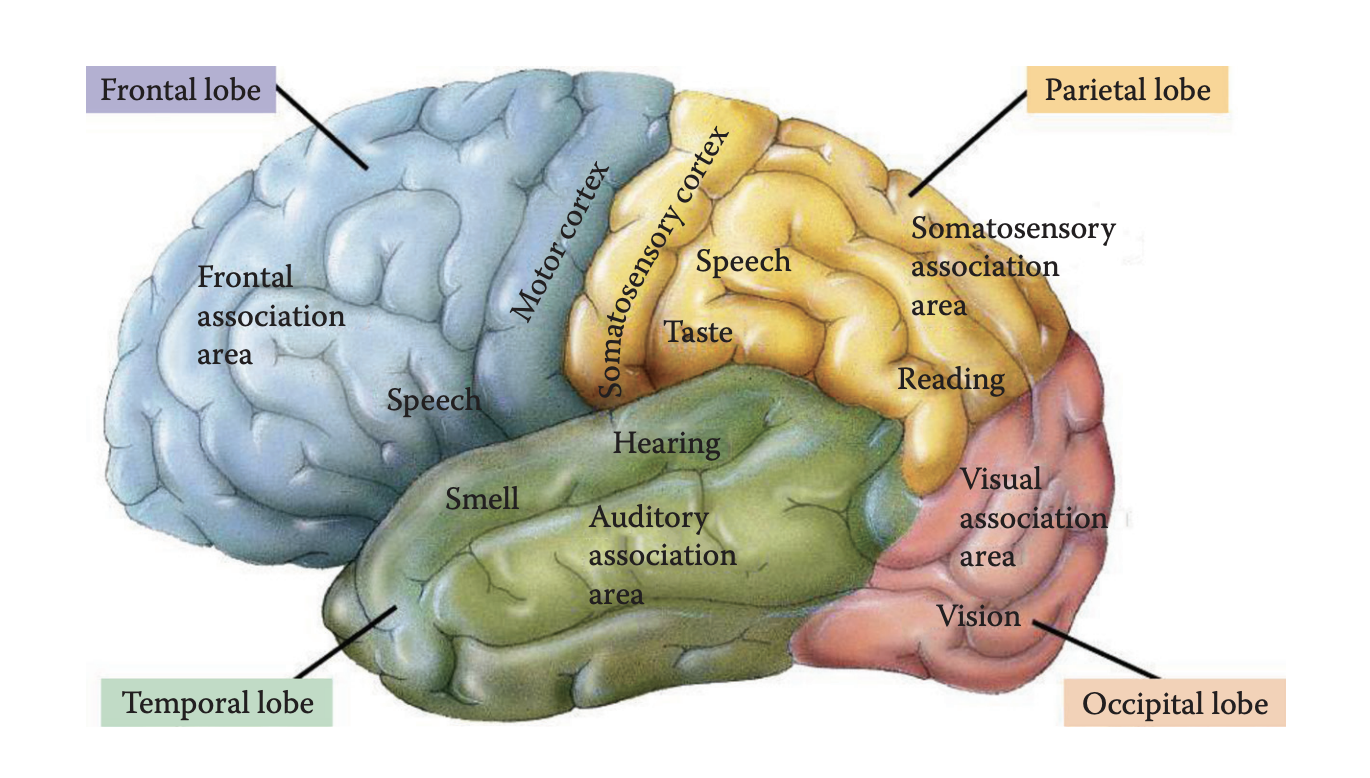}  
  
   \caption{Functional diagram of the Brain lobes. Image credit \cite{Chen}.}
   \label{figBrain}
\end{figure}

\subsubsection*{Macro-level analysis} 
\begin{figure}[H] 
  
  \resizebox{1\textwidth}{!}{\begin{minipage}{1.2\textwidth}
   \centering
\begin{tabular}{ccc}
\bf Auditory--Auditory Task & \bf Auditory--Visual task & \bf Auditory--Rest \\
\bf  Vol=7.8181 & \bf Vol=7.8911 & \bf Vol=11.449 \\
 \includegraphics[scale=0.26]{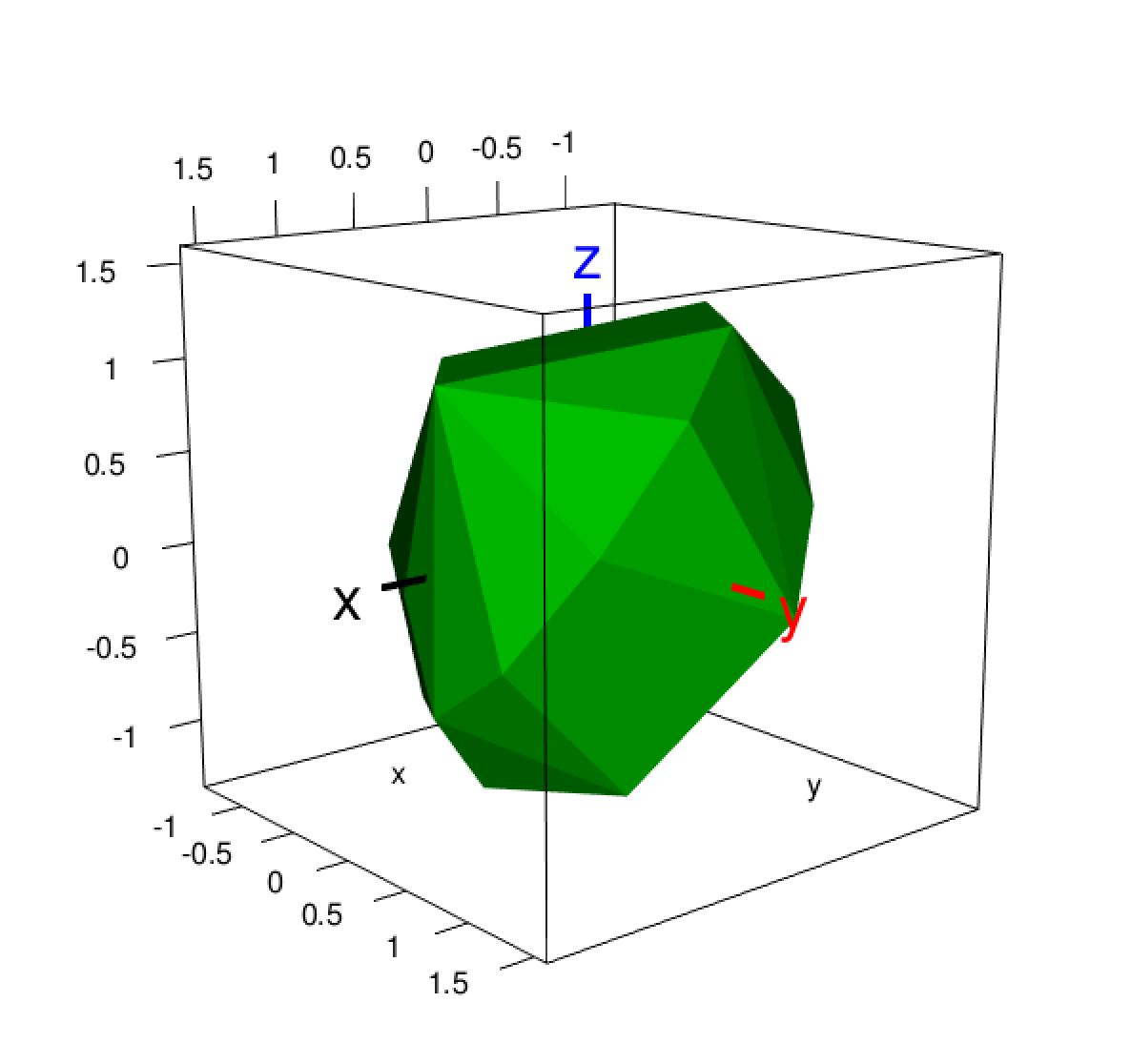}  &
        \includegraphics[scale=0.26]{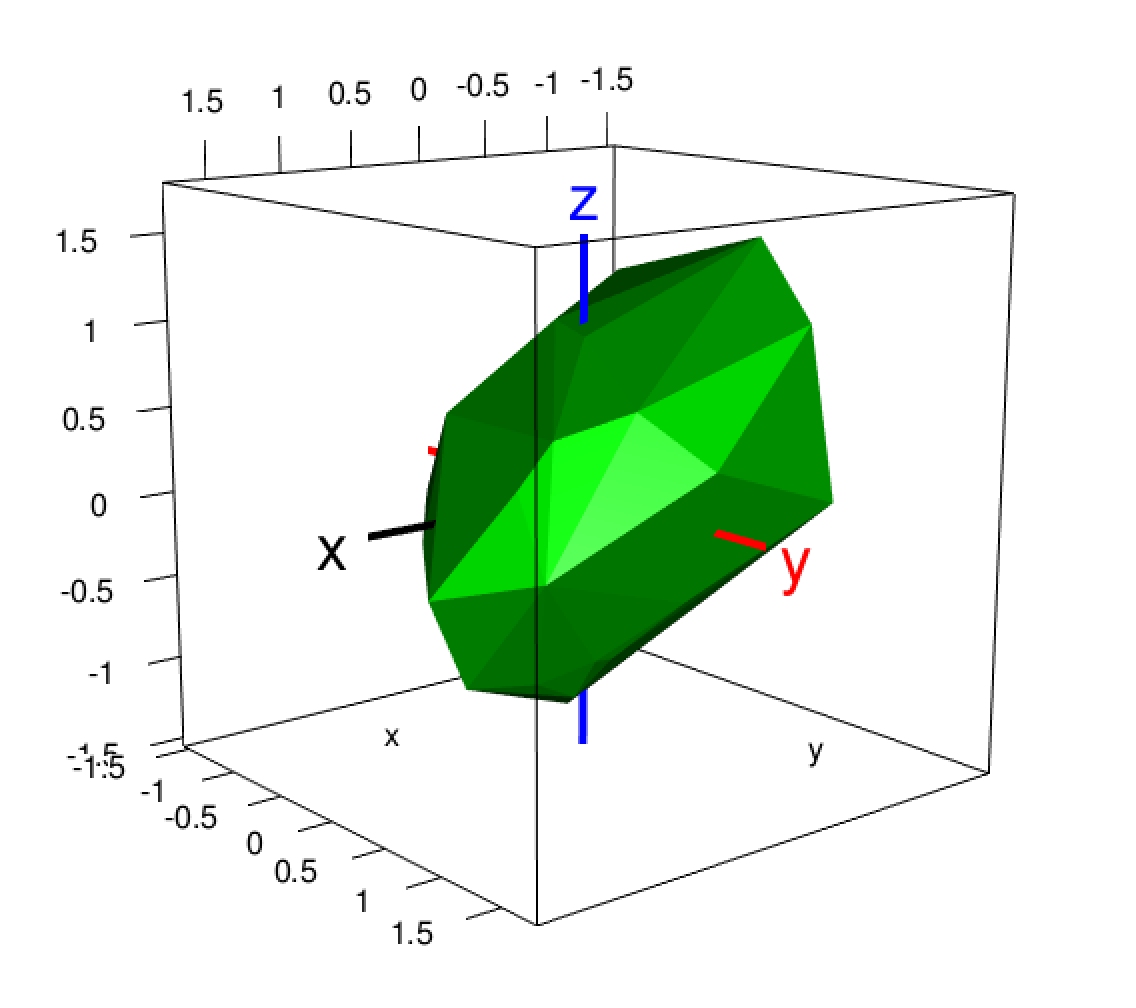}  &
   \includegraphics[scale=0.26]{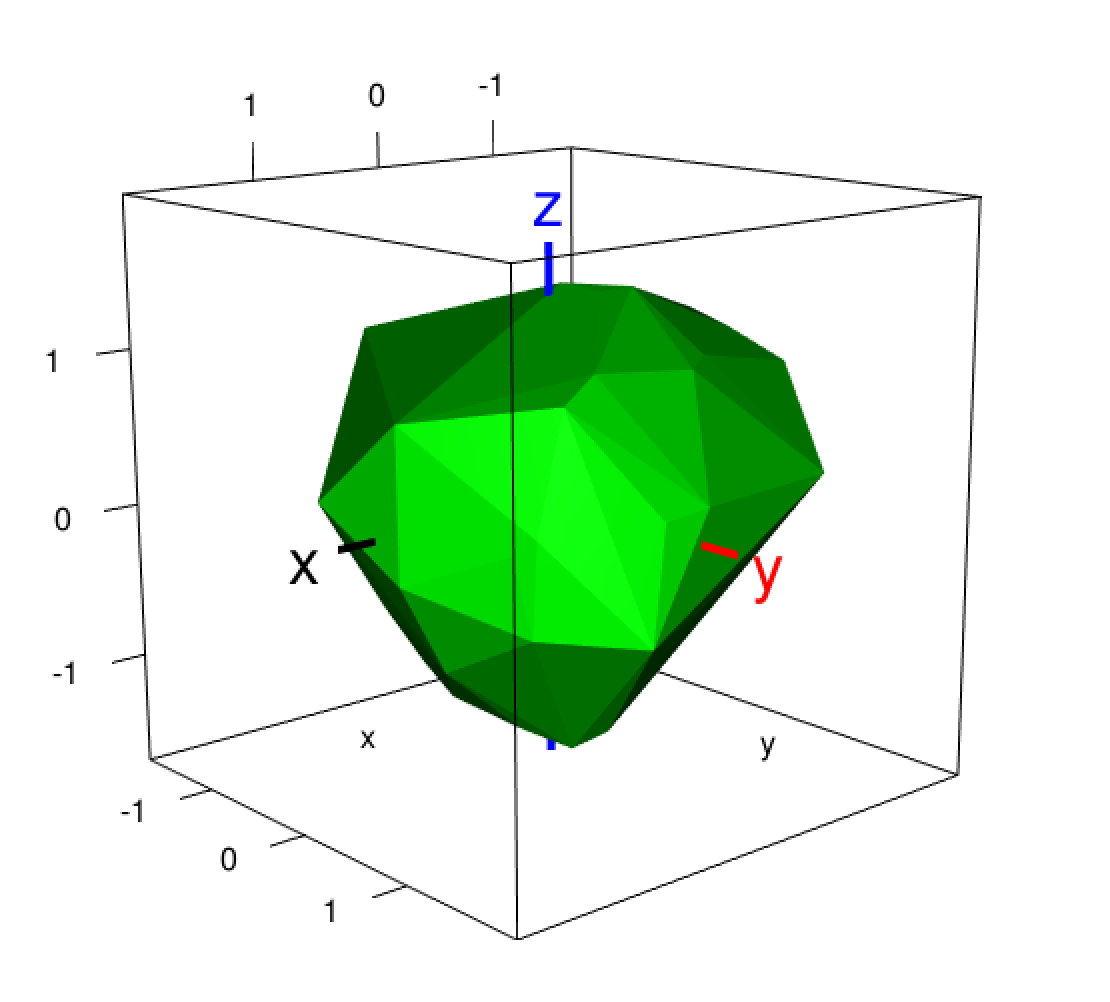} \\
   \bf Visual-Auditory Task & \bf Visual--Visual task & \bf Visual--Rest \\
    \bf Vol=8.2875 & \bf Vol=7.3373 & \bf Vol=14.745 \\
  \includegraphics[scale=0.26]{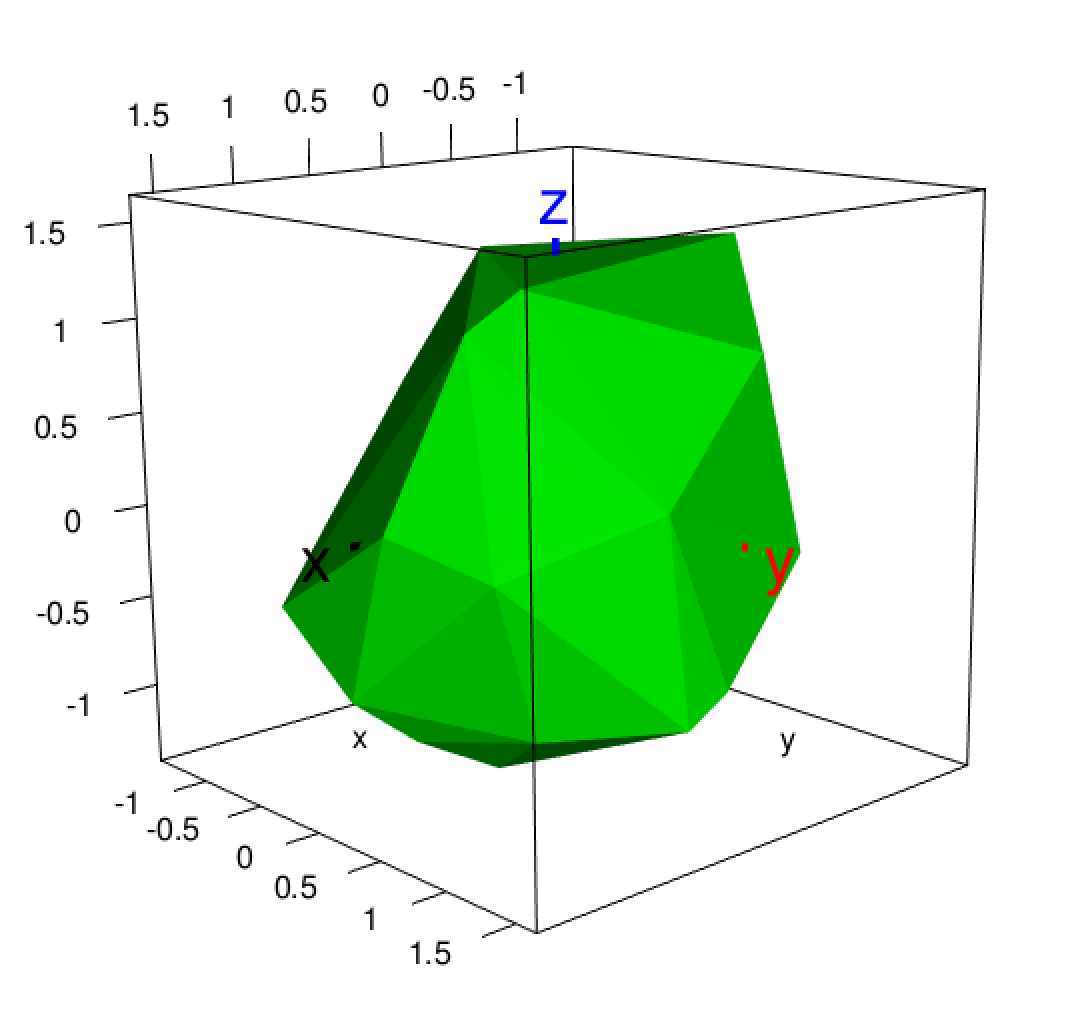}  &
    \includegraphics[scale=0.26]{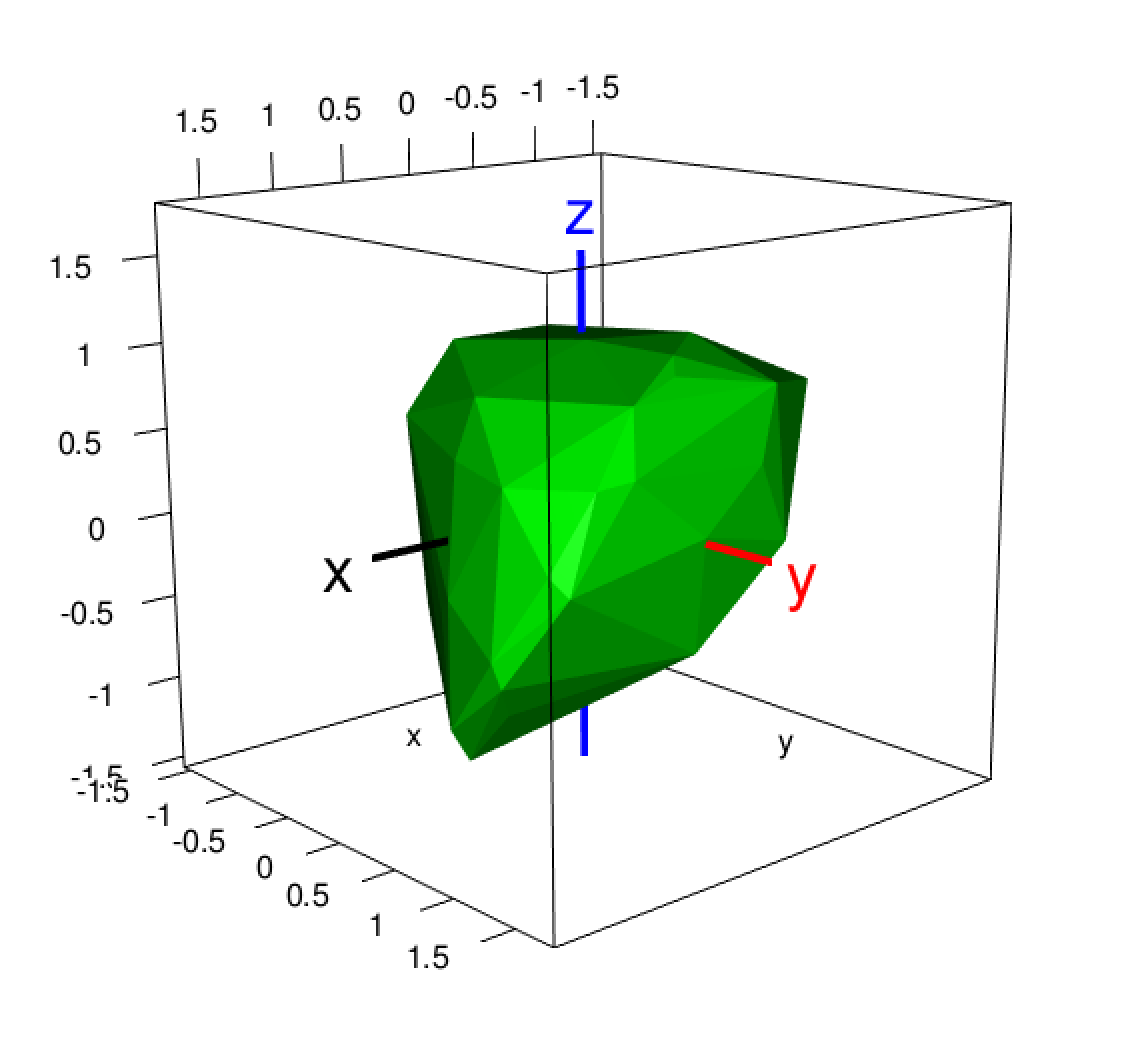} & 
   \includegraphics[scale=0.26]{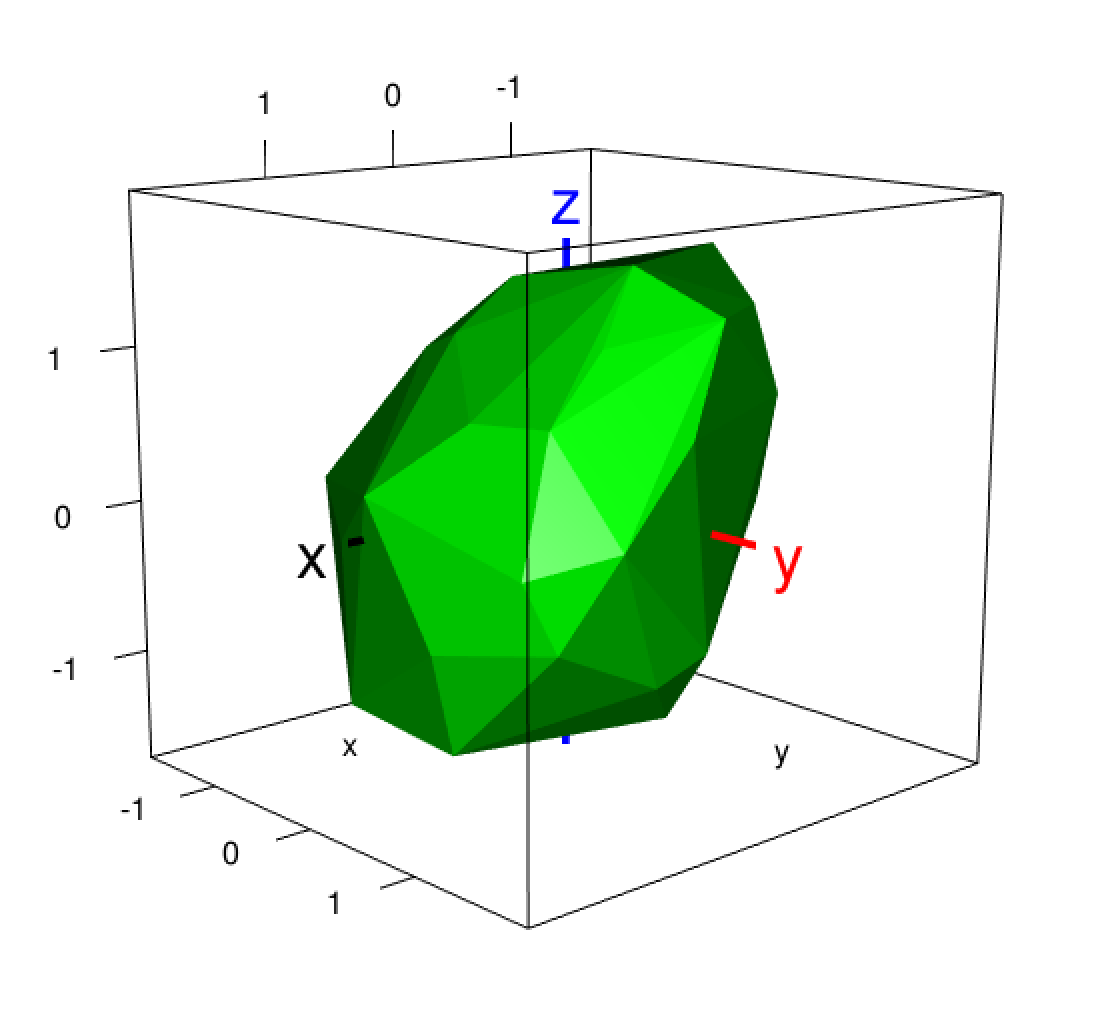} 
   \end{tabular}
   \end{minipage}}

   \caption{Complex geometric structures   for the above subsets.}
   \label{figAV}
\end{figure}

Here all the time series are used. Figure \ref{figAV} below represents the CGS for each of the subsets above. 
\noindent Comparing the first and second row, the volumes seem to differ by cortex. Comparing the first, second, and third column, the difference in volumes of tasks are respectively 0.449, 0.5538, and 3.296. These numbers show similarity between auditory and visual tasks but they both differ from rest.  The increase of volume during rest may not be  counter-intuitive. In fact, there is  vast literature linking resting brain activities to underlying high cognitive processes such as moral reasoning, self-consciousness, remembering past experiences, or planning for the future, see for instance \cite{Bruckner2008}. 

\subsubsection*{Micro-level analysis} 
However, if we use individual time series, since they represent individual patients, then we can  obtain their CGS by cortex and by task. This would enable having  a richer dataset and a more in-depth comparison. In Figure \ref{figComp} and \ref{figBoxPlotComp} below, we plot the density and boxplot of the volumes of the CGS by cortex and/or by task. Plots (A) represent the plots  of CGS by cortex, plots (B) the plots of  cortexes by task, plots (C) the plots by task, and plots (D) the plots tasks by cortex. At first glance, it seems as if a Beta or Lognormal distribution would  provide a good fit to these densities.
\begin{figure}[H] 
   \centering
    \includegraphics[scale=0.5]{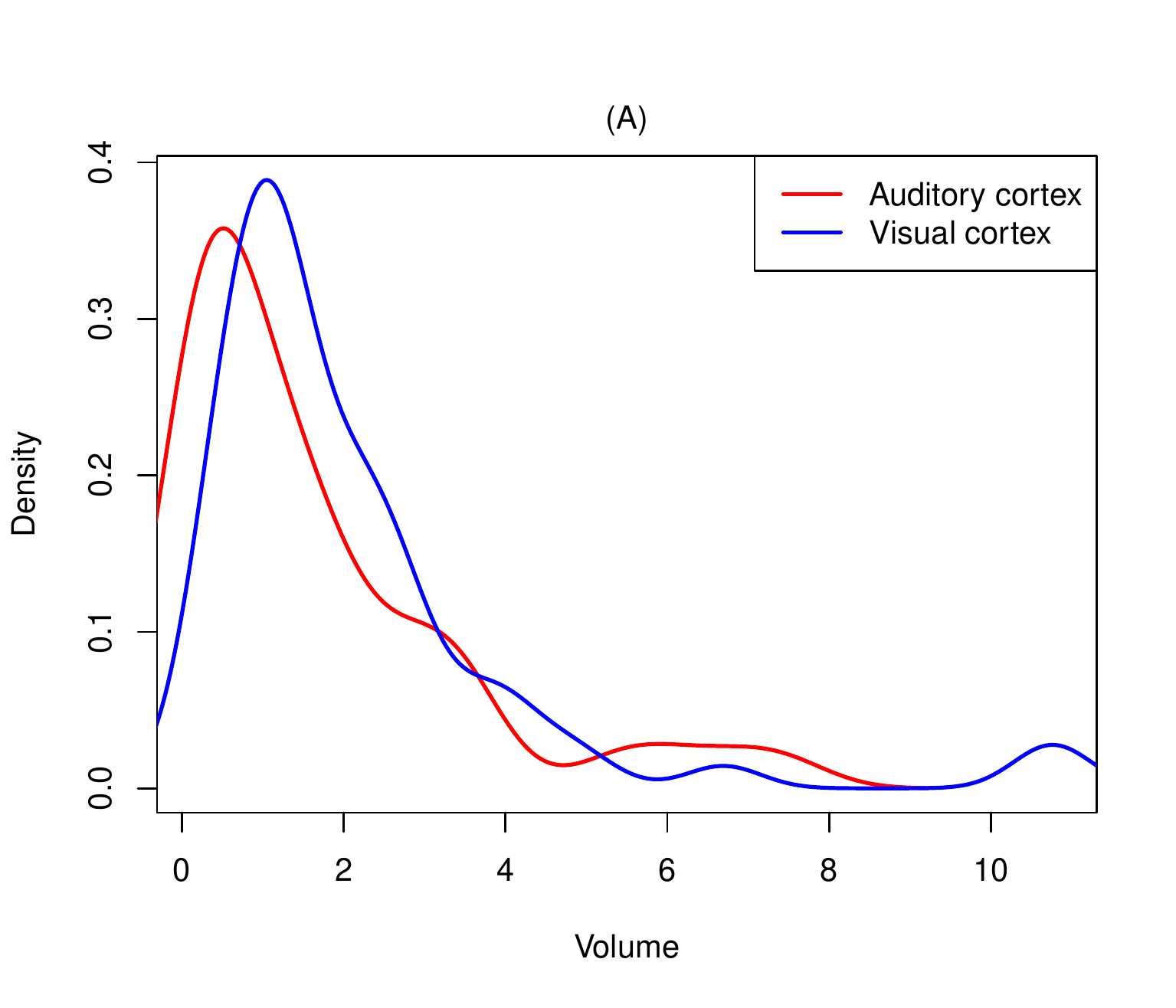}  
        \includegraphics[scale=.5]{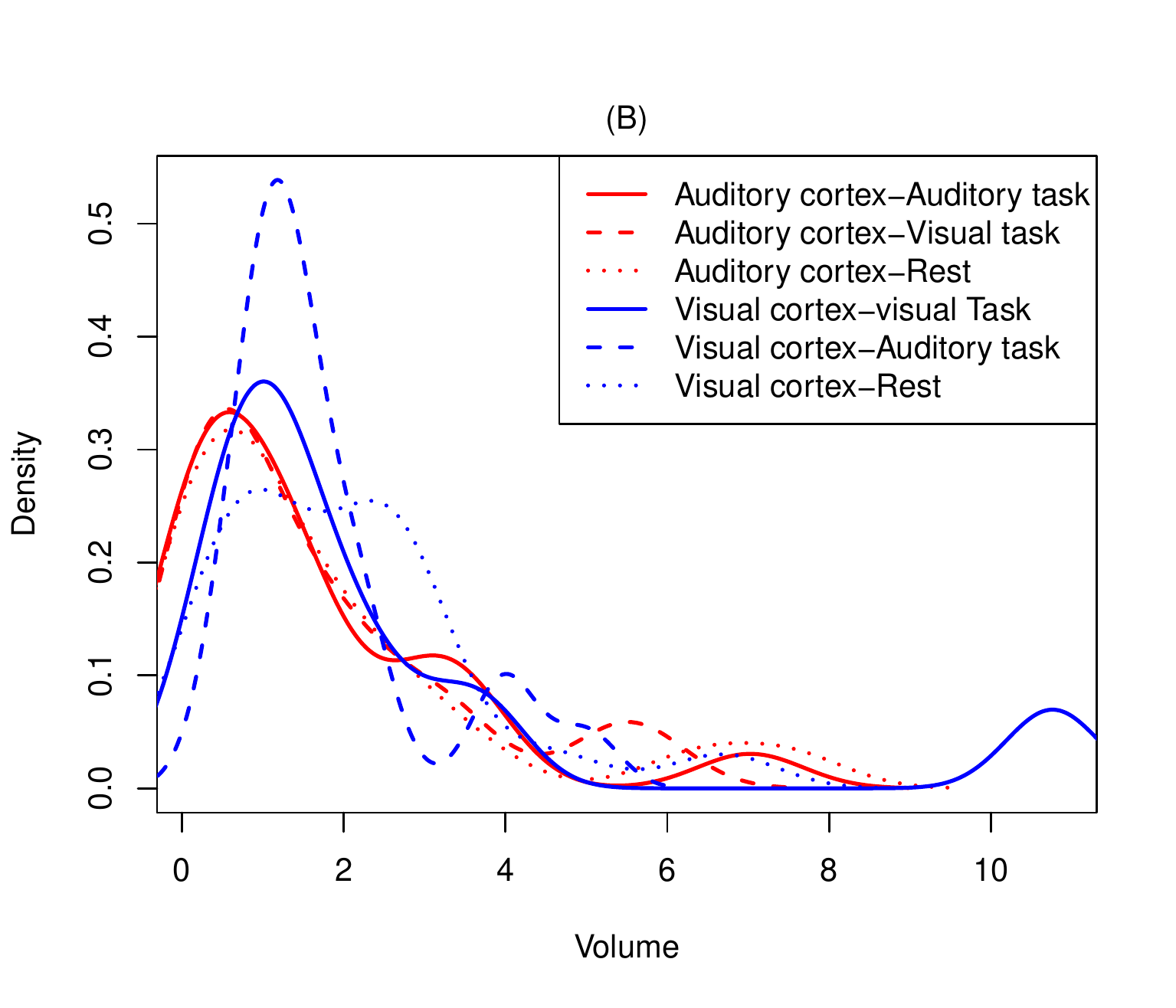}  
   \includegraphics[scale=0.5]{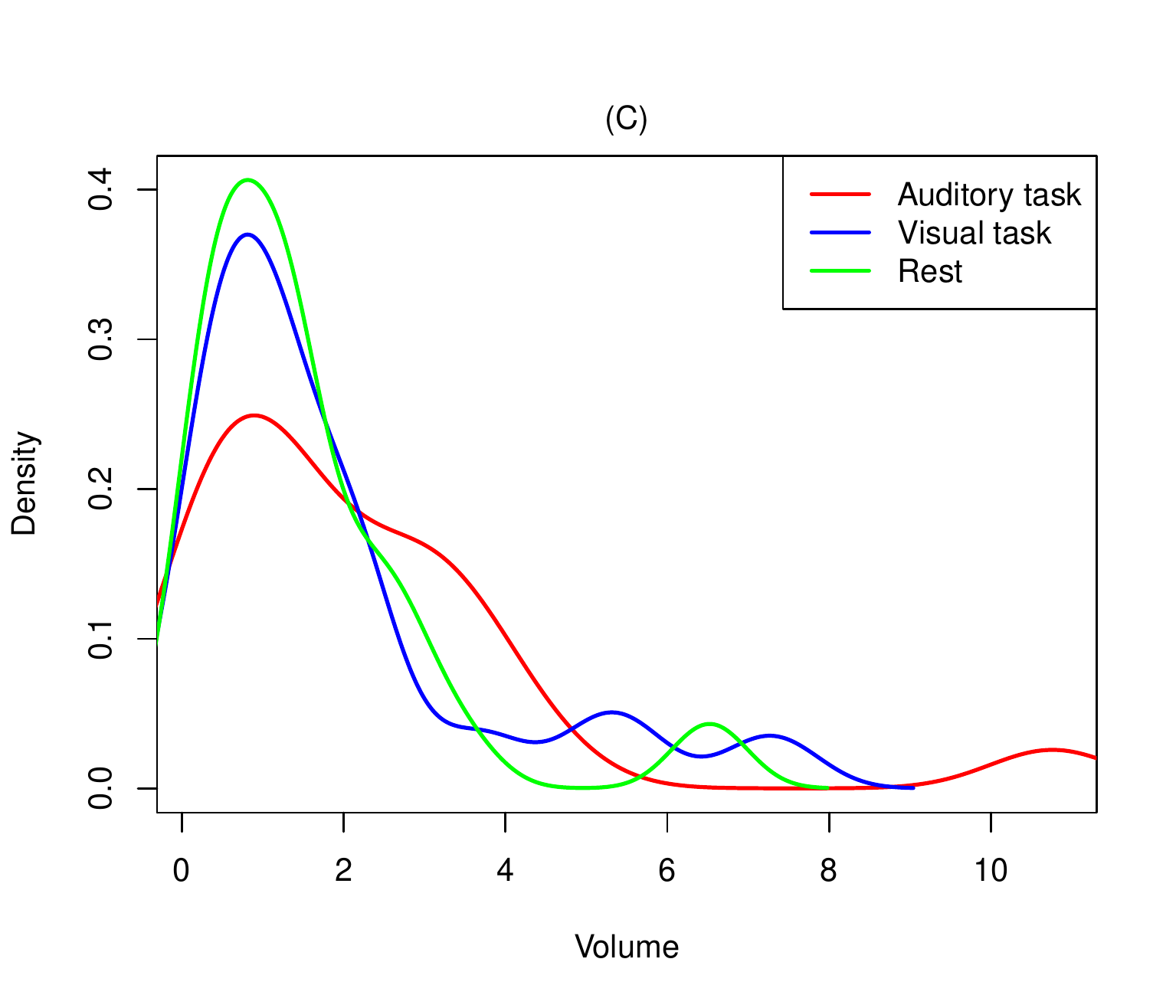} 
    \includegraphics[scale=0.5]{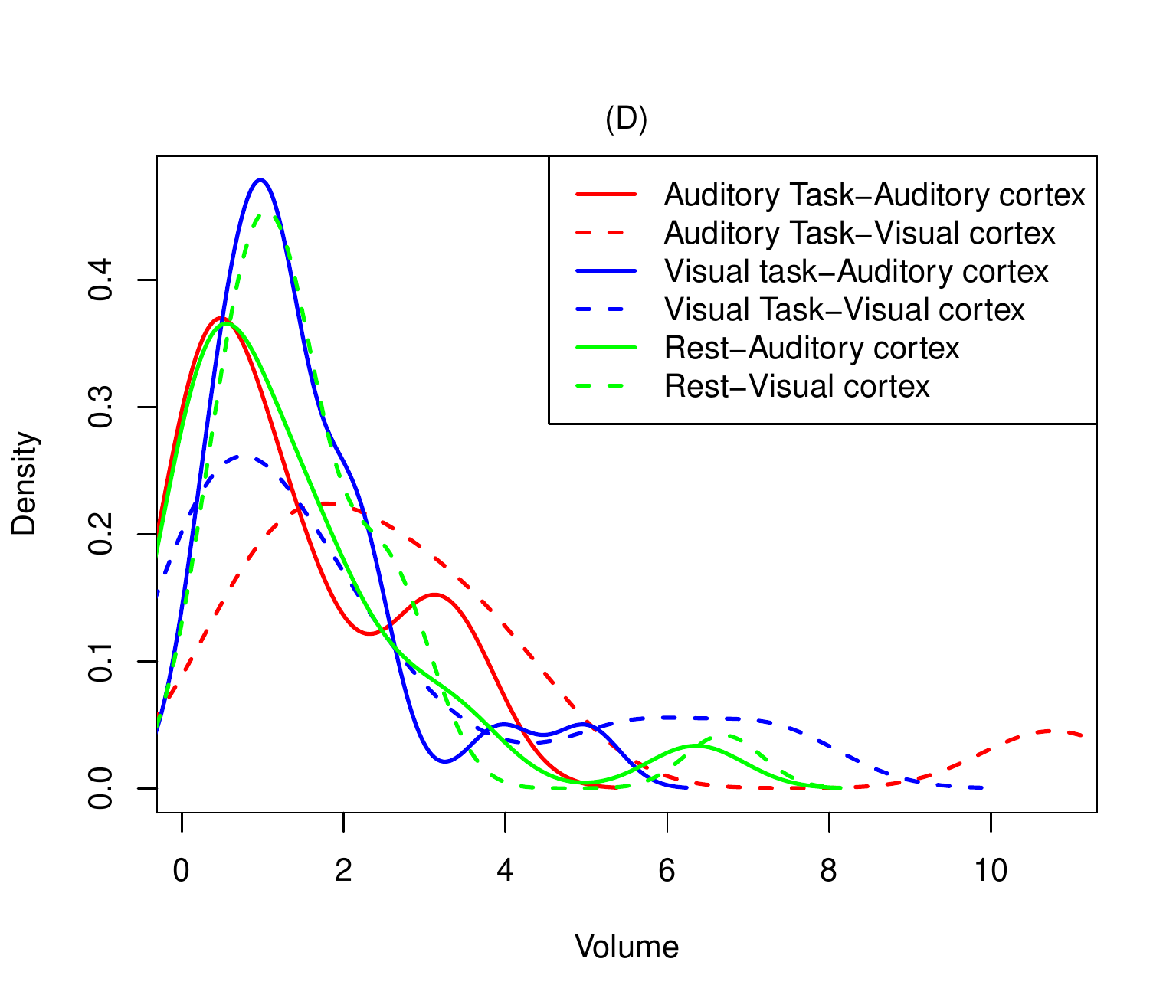}

   \caption{Density plots for the volume of CGS obtained from individual times series.}
   \label{figComp}
\end{figure}

\begin{figure}[H] 
   \centering
    \includegraphics[scale=0.5]{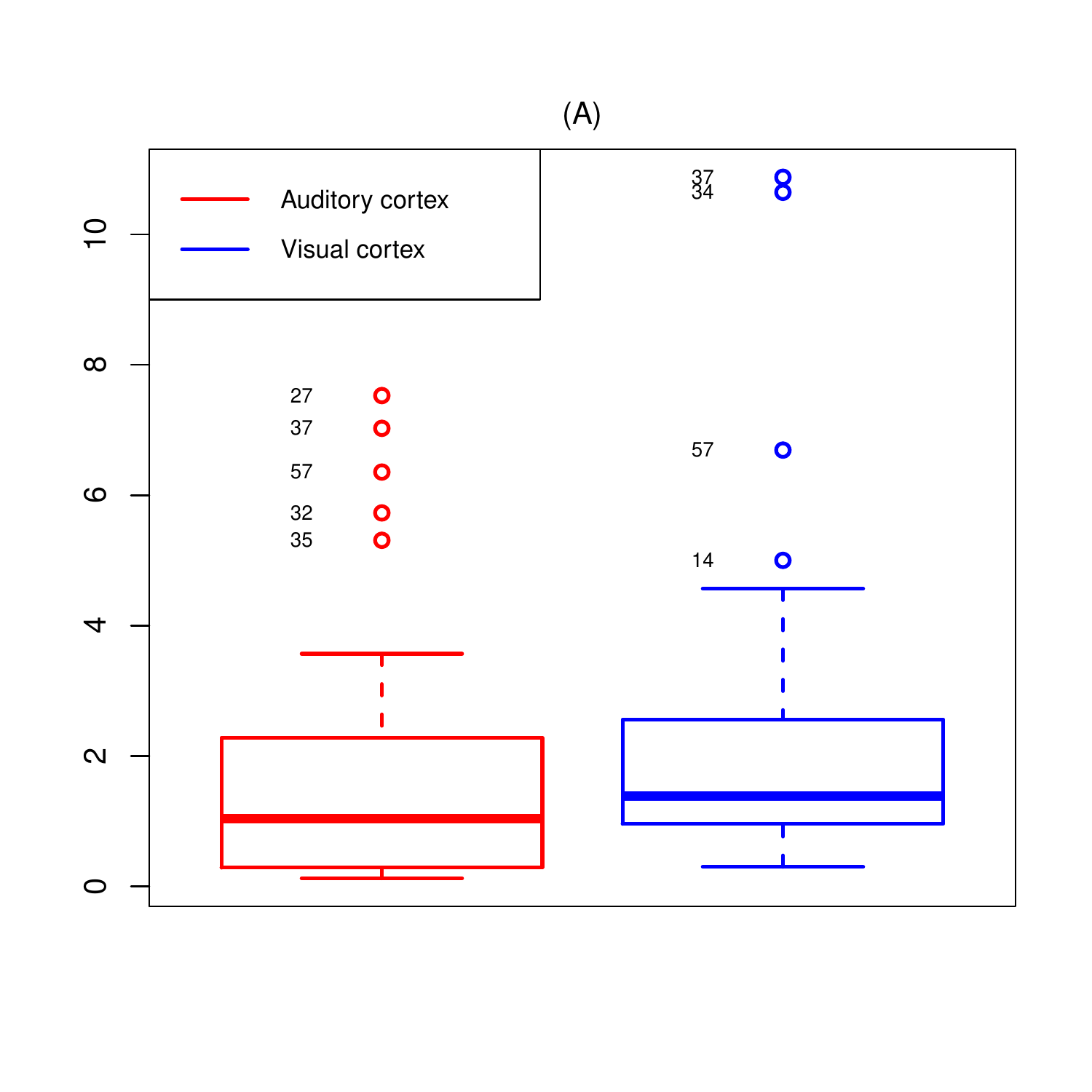}  
       \includegraphics[scale=0.5]{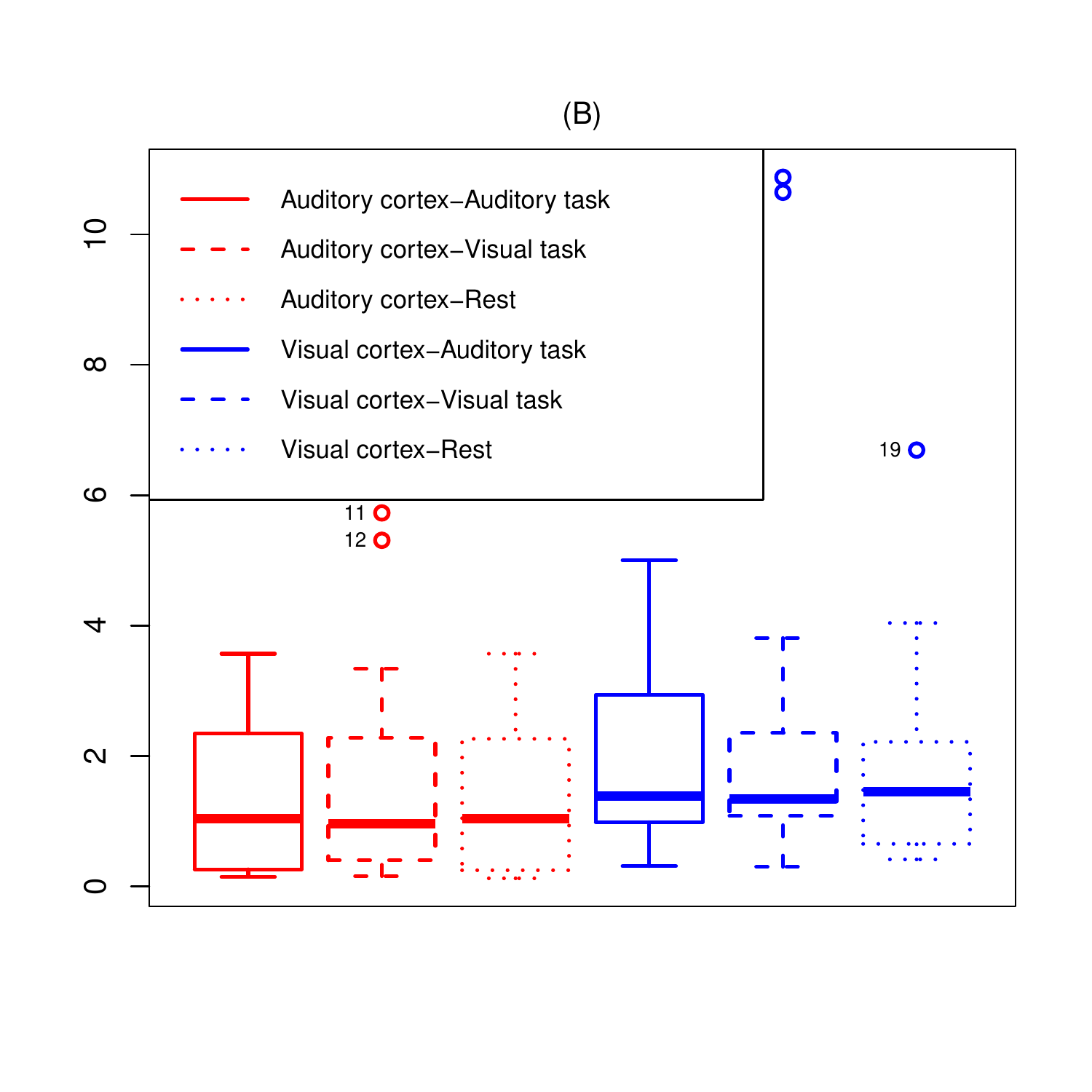}  
        \includegraphics[scale=.5]{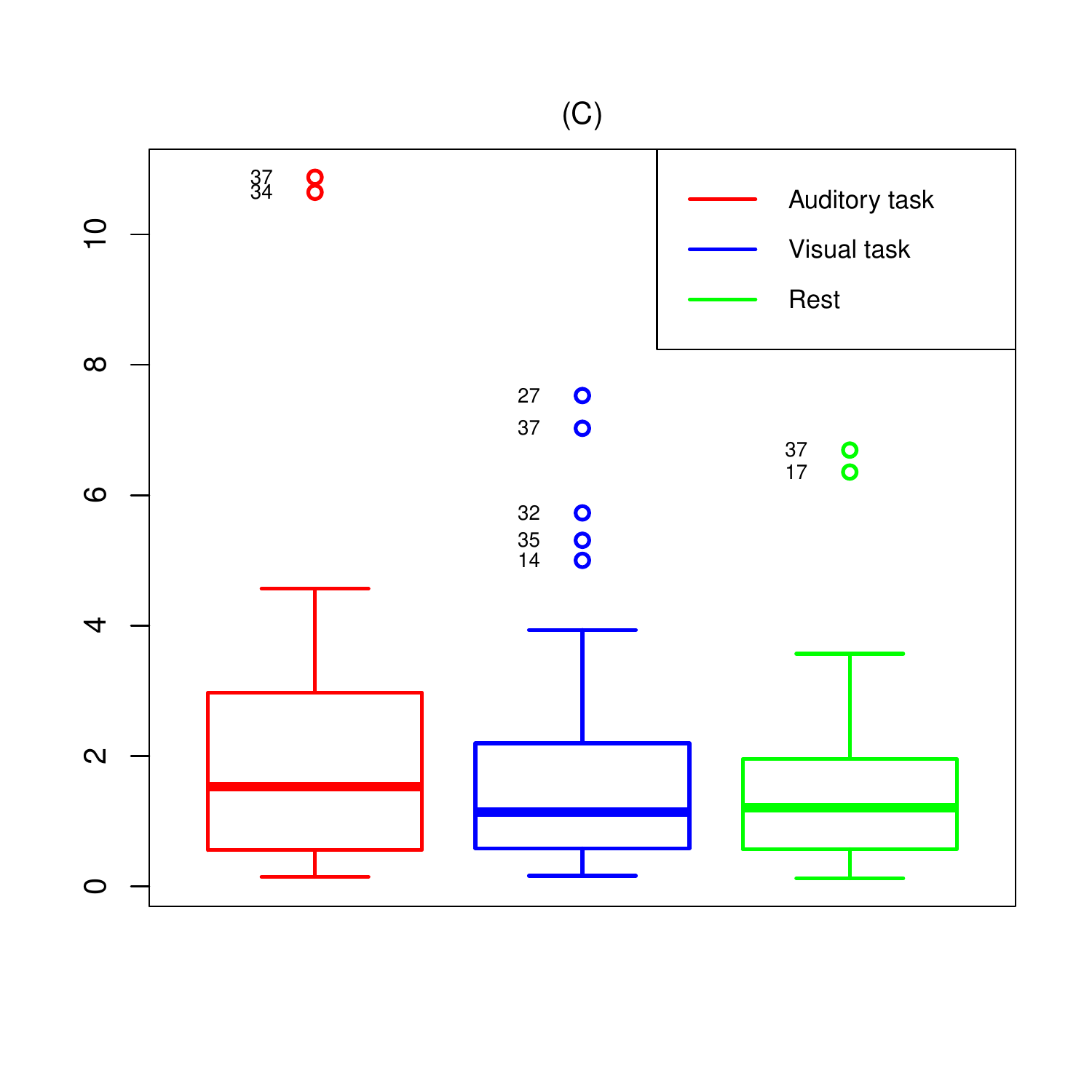}  
 \includegraphics[scale=0.5]{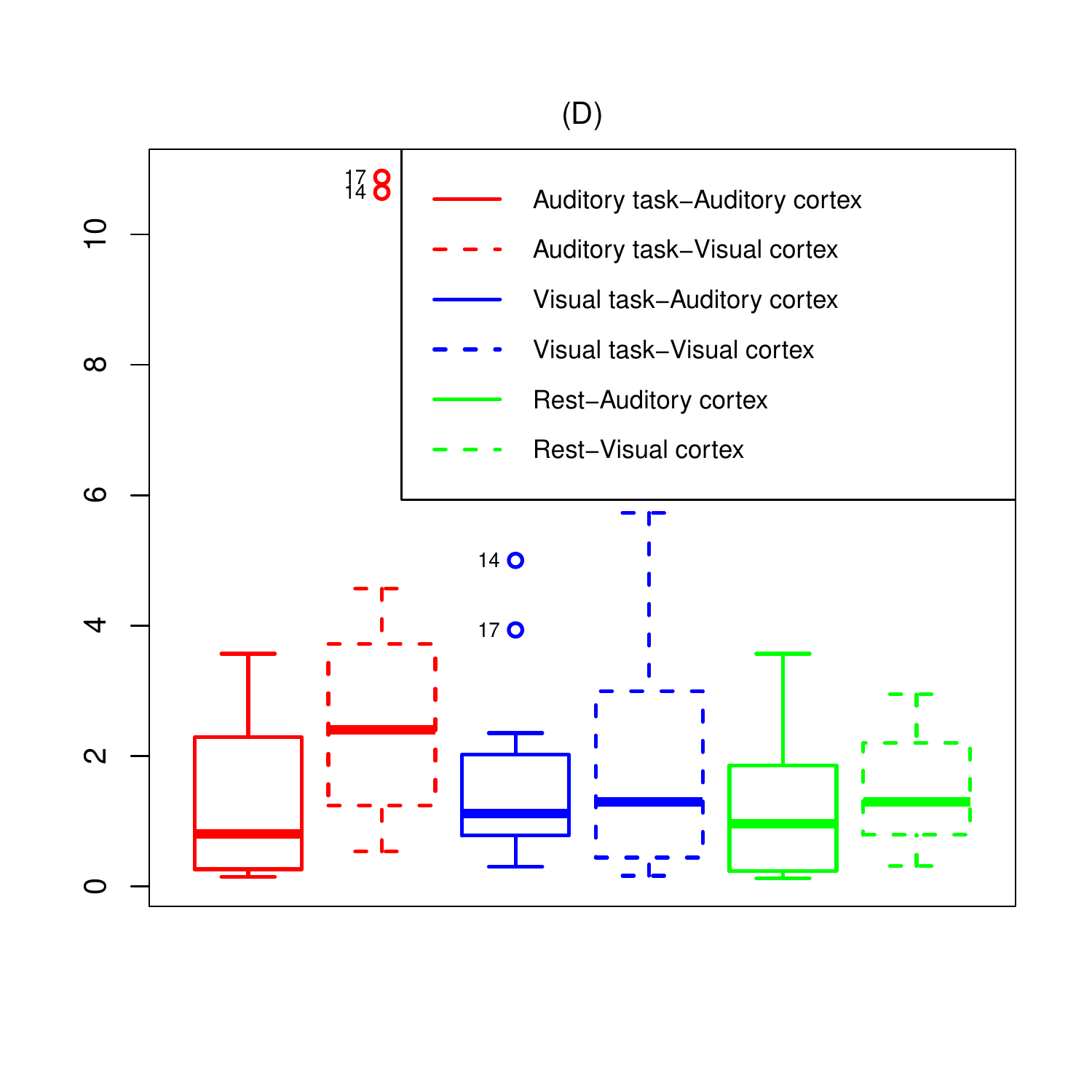}  
  
   \caption{Boxplots  for the volume of CGS obtained from individual times series.}
   \label{figBoxPlotComp}
\end{figure}

\noindent We note there are 20 patients and three  tasks per cortex  for a total of 120 observations. When grouping by cortex, we labelled the observations 1 through 60 per cortex. When grouping by task, we labelled the observations 1 through 40, 1 through 20 when grouping by task within cortex. The numbers shown in the boxplots in Figure \ref{figBoxPlotComp} represent the labels for the corresponding outlying observations. Boxplots (A) does not suggest a  difference between the auditory and visual cortexes, since the notches do overlap for the most part. It shows that there are two common outliers (37 and 57) in both cortexes. This means that there are two patients whose brain activities are more pronounced than others in both cortexes, causing wide excursions in the phase space and therefore large volumes of their CGS. Boxplots (B) and (C) suggest a similarity between the tasks since all the notches do overlap.  Boxplots (D) suggest that the difference observed in Boxplot (A) is due to the difference between the auditory tasks in both cortexes.   Boxplots (B) basically confirm the observations in Boxplots (A)
that all the tasks are similar. A more in-depth analysis is needed to make more meaningful conclusions.

In the tables below, we report the intrinsic discrepancy, that is, the minimum between $d_{KL}(F,G)$ and $d_{KL}(G,F)$, where $d_{KL}(F,G)$ is defined as the Kullback-Leibler (KL) directed divergences between the probability distributions F and G (see Appendix Section \ref{sect:KL}). So the smaller the intrinsic discrepancy, the bigger the difference between $F$ and $G$.

 \begin{table}[H]
\resizebox{1\textwidth}{!}{\begin{minipage}{1\textwidth}
\centering
\begin{tabular}{|c|c|c|c|}
\hline
& Auditory cortex & Visual cortex  & P-value \\ \hline
Auditory cortex & 0 & 0.160 & 0.021\\
\hline
\end{tabular}
\end{minipage}}
\caption{Statistical analysis of the  intrinsic discrepancy between the  cortexes.}
\label{table2}
\end{table} 

 \begin{table}[H]
\resizebox{1\textwidth}{!}{\begin{minipage}{1\textwidth}
\centering
\begin{tabular}{|c|c|c|c|c|}
\hline
& Auditory task & Visual task & Rest&P-value \\ \hline

 \multirow{2}{*}{Auditory task} & 0.000 &  0.104 &  0.029& \\
&-- &   (0.98) & (0.98) &   \multirow{2}{*}{0.451}\\
\cline{1-4}

 \multirow{2}{*}{Visual task} & 0.104 & 0.000 & 0.080 & \\ 
&(0.98) & -- & (0.98) & \\ \hline
\end{tabular}
\end{minipage}}
\caption{Statistical analysis of the  intrinsic discrepancy between the tasks.}
\label{table3}
\end{table}

\noindent To make a meaningful use of  Table \ref{table2} and \ref{table3}, a reference point is needed. However, the densities above allows us to make the hypothesis that the volumes are not normally distributed. A nonparametric test such as a Wilcoxon-Mann-Whitney would be necessary to compare for example the volumes of the two cortexes, controlling for the tasks performed. Such a test  yields a p-value of 0.021, which is small enough to suggest a statistically significant difference between the CGS volumes for Auditory and  Visual cortex. It also suggests that the  intrinsic discrepancy 0.160 obtained above is a sign of a statistical significant difference  between the two cortexes.  Likewise, the nonparametric  Kruskall-Wallis test is used  to compare the tasks and it yields a p-value of 0.451, which is large,  suggesting that  the volumes of the CGS are not significantly different.   The p-values for the pairwise Wilcoxon rank sum test (in parenthesis) with Bonferroni correction are also large. This could be further interpreted as the brain activities of individuals during the auditory task are not significantly different from their brain activities during the visual task. Now, whether these differences or lack thereof are corroborated at the biological level remains to be proved. 

\subsubsection*{Comparison with a Cross-Correlation Function}
\begin{figure}[H] 
\resizebox{1\textwidth}{!}{\begin{minipage}{1.3\textwidth}
\centering
\begin{tabular}{cc}
\bf (A) & \bf (B) \\
          \includegraphics[scale=0.5]{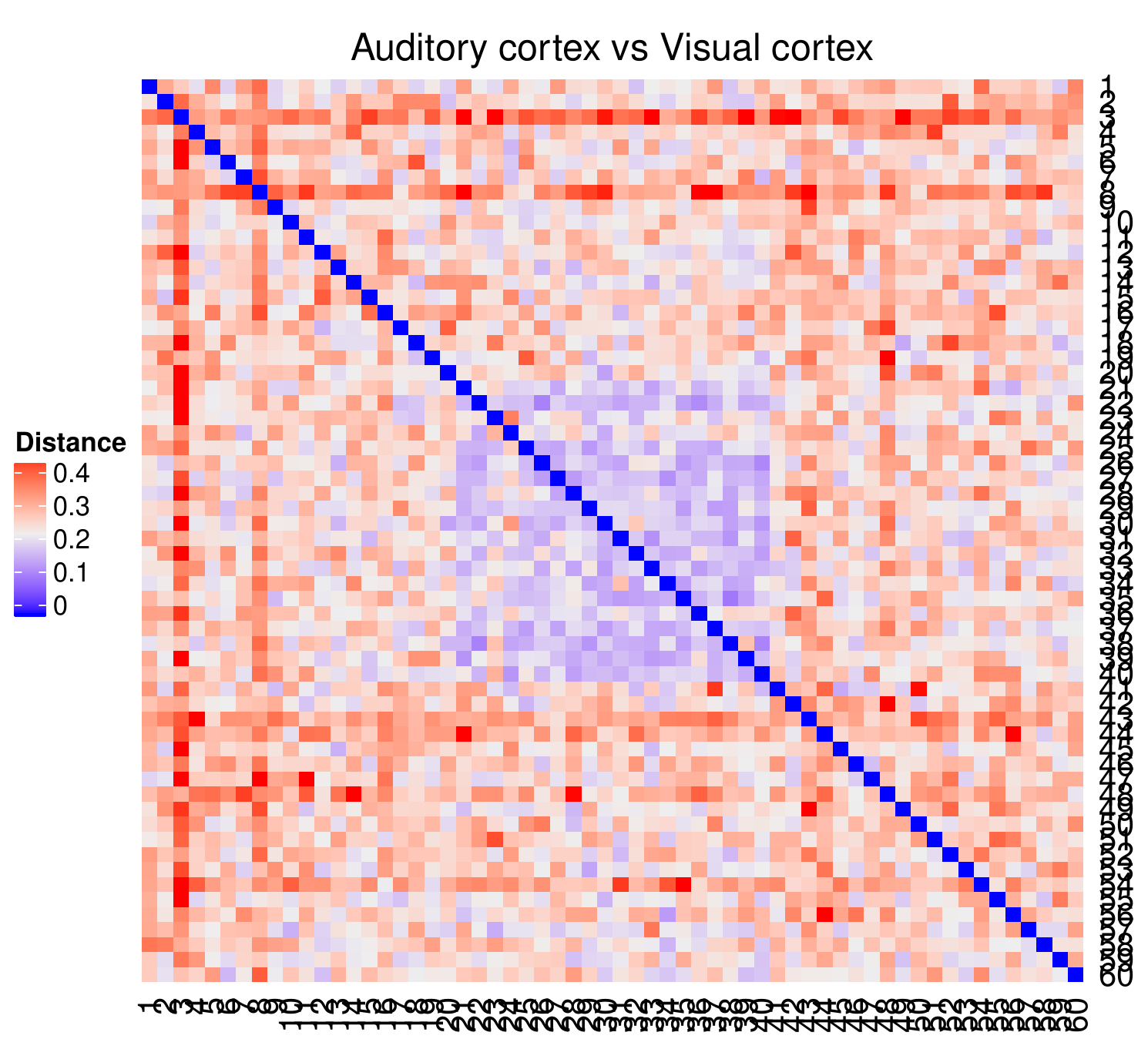}   &
    \includegraphics[scale=0.5]{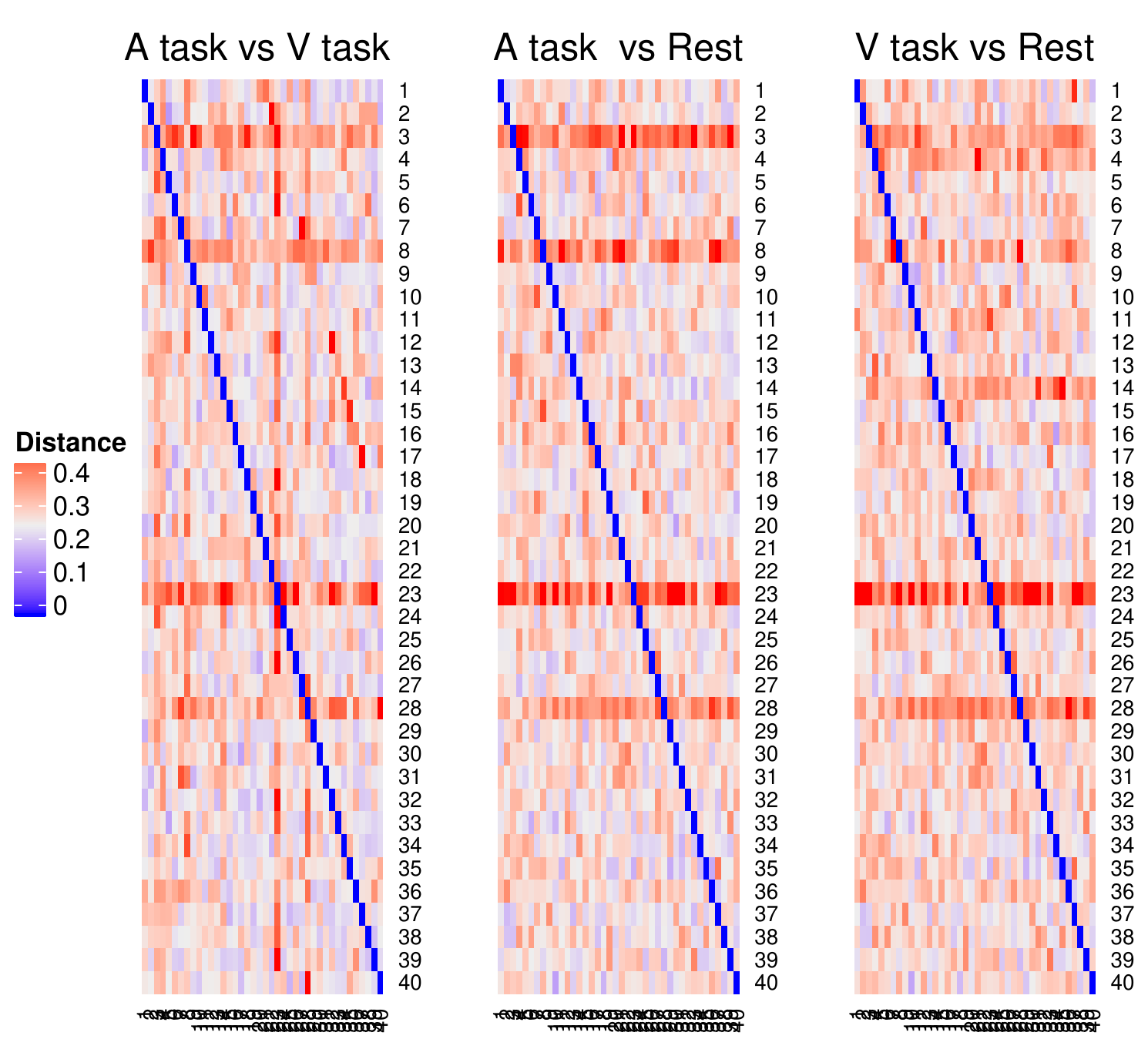}    \\
    \bf (C) & \bf (D)\\
    \includegraphics[scale=0.5]{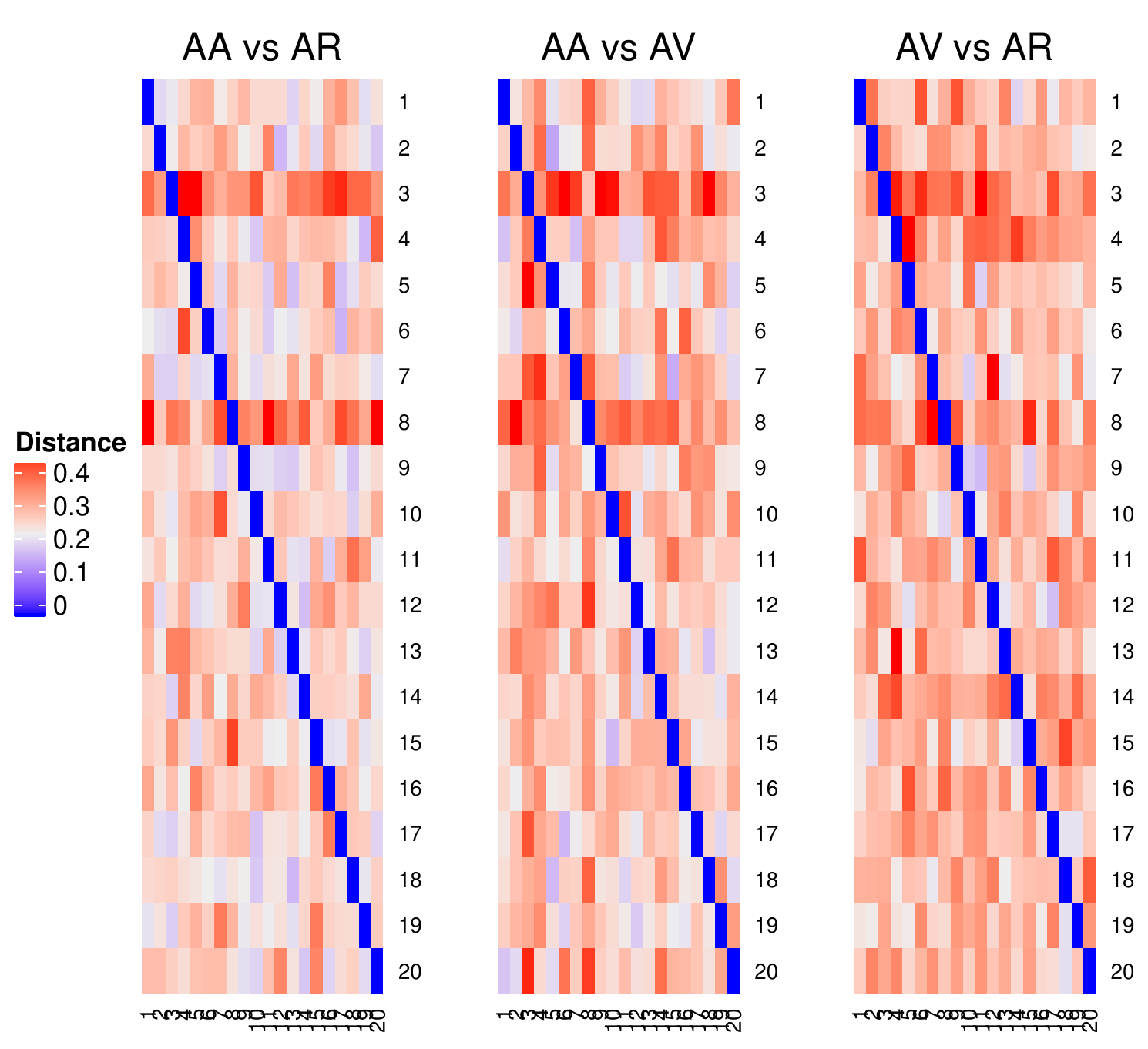}     &
        \includegraphics[scale=0.5]{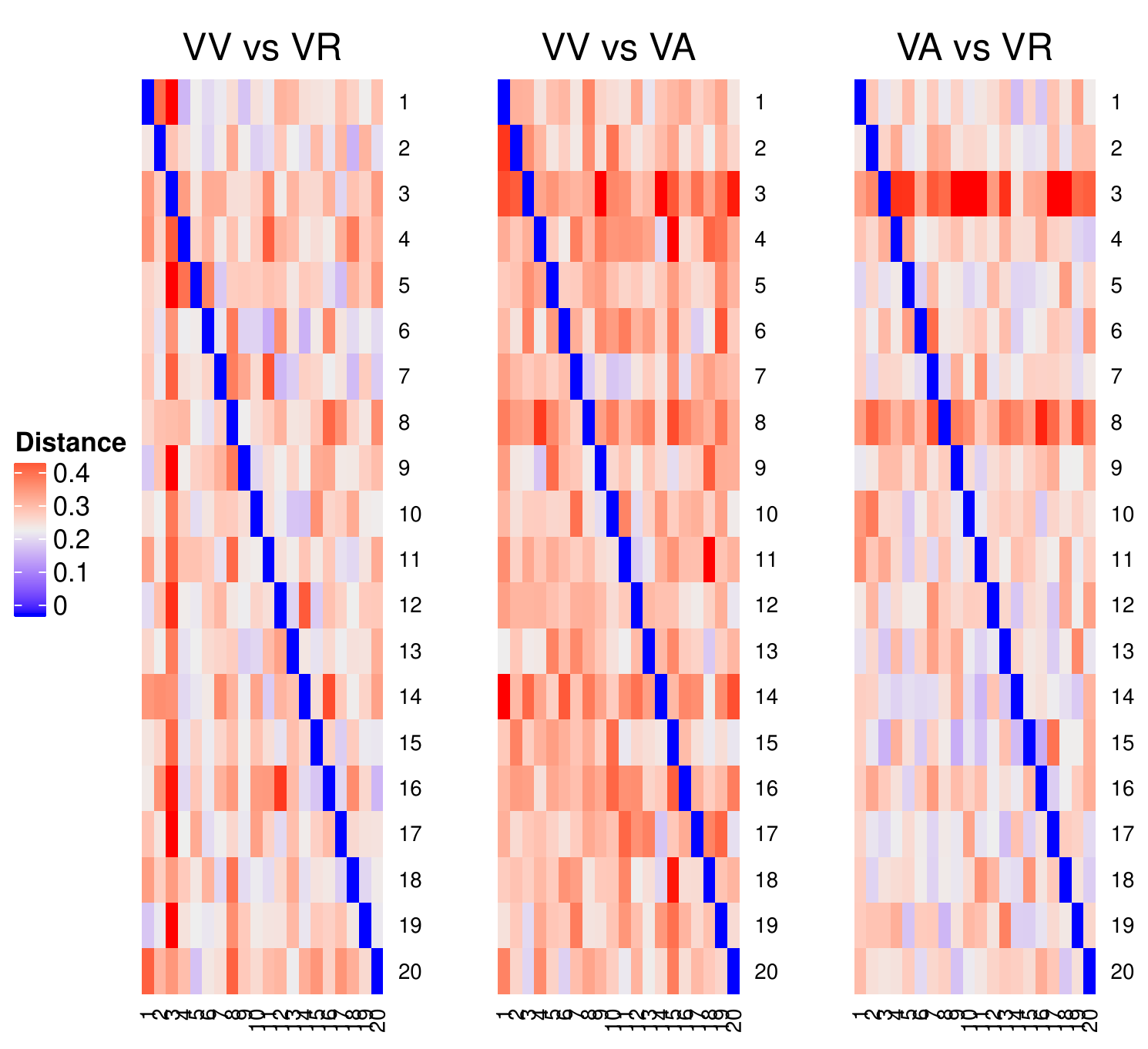}  
        \end{tabular}
        \end{minipage}}
    
   \caption{Plot of $\rho(X,Y)$ between the auditory and visual cortexes.  }
   \label{FigureHeatMapAV}
\end{figure}

In this section, we will compare our method with the  cross-correlation function (CCF), see Appendix, Section 7.5. In the heatmaps below, represented are distances $\rho(X,Y)$ where $X$ and $Y$ are the EEG data (times series) collected on the 20 different patients in different regions of the brain, when subjected to the three tasks above. The heat gradient represents the values of $\rho(X,Y)$.
Figure \ref{FigureHeatMapAV} (A) below represents the comparison between auditory and visual cortexes. Its symmetric nature is indicative of the similarity between the two cortexes. This is in agreement with Figure \ref{figBoxPlotComp} (A).
From  Figure \ref{FigureHeatMapAV} (B) below, we observe that the heatmaps are very similar. In particular, there are four areas that are redder than the majority and indicative of the strongest correlation. A classification by a k-means algorithm  confirms the existence of these two clusters, see Figure \ref{FigureHeatMapAVKmeans} (B) in the Appendix.  This is similar to saying that there is  no significant  difference between the tasks. The same conclusion can be made with  boxplots  in  Figure \ref{figBoxPlotComp} (C), where the outliers in the volume of CGS match precisely the regions with  high correlation. Looking at the first three heatmaps in \ref{FigureHeatMapAV} (C) and (D) , there is  not a huge   difference between the heatmaps, confirming that there is no significant difference between the tasks within the auditory cortex,  an observation also made from  Figure \ref{figBoxPlotComp} (B). The same observation can be made about the last three maps in the visual cortex

\section{Discussion}\label{Discussion}
We offer some items for discussion as a result of our findings:\\
\noindent 1. The method we are proposing has the potential to discriminate brain activities in different parts of the brain. Further, changes in volume could be an indication of changes of activity level in the part of the brain under study. This predictive potential could be used to predict or monitor potential brain disorder. 

\noindent 2. This method has the potential  to be extended to experiments in which EEG data are suitable such as epilepsy,  Alzheimers, attention deficit disorder, learning disabilities, anxiety disorders, fetal alcohol syndrome, autism, chronic pain, insomnia, dyslexia, etc, see for instance \cite{Kumar2012} and the references therein.

\noindent 3. Preliminary analysis of Local Field Potential (LFP) data suggests that this method can be extended beyond EEG. In fact LFP differs from EEG in that the electrodes are inserted in the brain tissue rather than at the surface of the scalp. However, we did not obtain the authorization to release the results on the LFP data at this time.  

\noindent 4. There have been studies suggesting that group averaging of neuroimaging data are not clinically relevant. One possible explanation for this issue is  the lack of focus on individuals. The method we propose can remedy that by providing a measure that is individual-focused.

\noindent 5. We have not addressed the weak stationarity of the EEG data used here and how much stationarity or lack thereof plays a role in the method we are proposing. The data set EDATA was checked for weak stationarity but the data  collected for the Brain Core Initiative was too small for stationarity. This overall is an issue in multiple studies where samples tend to be small.

\noindent 6. One of the drawbacks of the ``distance" above is that its definition may change and therefore this could affect the interpretation of results. In fact,  given two times series $X$ and $Y$,  the distance could also be defined as $\displaystyle \rho(X,Y)=1-\max_{1\leq \tau\leq m_l}\; \{\abs{CCF(X,Y,\tau)}\}$ or by $\rho(X,Y)=\mbox{mean}\{CCF(X,Y,\tau)\}$  with  $CCF(X,Y,\tau)=\frac{E[(X_{t+\tau}-\overline{X})\cdot (Y_{t}-\overline{Y})]}{\sqrt{E[(X_{t}-\overline{X})^2]E[(Y_{t}-\overline{Y})^2]}}$, where $X_{t+\tau}$ is the time-shifted version of $X_t$, and $\tau$ is the time lag separating the two times series $X$ and $Y$, $\overline{X}, \overline{Y}$ are the respective means of the times series $X$ and $Y$. The reader can refer to \cite{Arbabshirani2012} for more information.


\section{Conclusion}
In this paper, we have proposed a method called complex geometric structurization to help analyze EEG signals. The method is based on the embedding theory of dynamical systems and shape analysis. The method works well on epilepsy data. The method has the advantage to discriminate individual EEG and also discriminates groups of  EEG signals. More importantly,  the method performs  better when compared to Cross-Correlation Function. It is important to note that the results and the method, though empirical, offer an important proof of concept relating time series and the volume of their reconstructed complex in three dimension that need to be explored further by validating them  with more solid mathematical concepts. If successful,  this method could be an important   addition  to  the literature of EEG signal analysis and can  be used to explore other brain pathologies such as sleep disorder, schizophrenia, or Alzheimers. Its discrimination potential also can be used to improve our understanding of functional brain connectivity in general.

\section{Appendix}

\subsection{$\alpha$-convex and $\alpha$-shape}

\begin{defn}
A set $A\subseteq \mathbb{R}^m$ is said to be $\alpha$-convex, for $\alpha>0$ if 
\begin{equation}\label{eqn:alphaconvexity}
A=C_{\alpha}(A)=\underset{B}\cup B_{\alpha}(x)\;,
\end{equation}
where $B=\left\{ B_{\alpha}(x): A\cap B_{\alpha}(x) \neq 0\right\}, B_{\alpha}(x)=\{y: \norm{y-x}\geq \alpha\}$, and $\norm{\cdot}$ is a  norm in $\R^m$.
\end{defn}
\noindent The quantity $C_{\alpha}(A)$ in  equation \eqref{eqn:alphaconvexity} is called the $\alpha$-convex hull of $A$. Now suppose we have a sample $S_{n}=\{X_1, \cdots, X_n\}$ obtained from an object $S$ in $\R^m$, then an estimator of $S$ is $C_{\alpha}(S_n)$ if $S$ is assumed $\alpha$-convex. Estimators of  $\alpha$-convex objects   are constructed from arc of circles in two-dimension (2D) and spherical caps in three-dimension (3D). 
\begin{defn} Let $A \subseteq \mathbb{R}^m$ be  $\alpha$-convex. An $\alpha$-shaped estimator of $A$ is an estimator of its $\alpha$-convex hull $C_{\alpha}(A)$ obtained by  approximating  arc of circles  with  polygonal curves in 2D and spherical caps with polyhedral surfaces in 3D.
\end{defn}

\subsection{Estimating the time delay and embedding dimension}\label{sect:est}
 {\bf Estimating the time delay :} There are two popular methods for estimating the time delay $\rho$: the autocorrelation function (ACF) and the average mutual information (AMI). Indeed, consider $N$ measurements of a time series $s(t)$. Then the sample ACF is defined as 
 \begin{equation}
 \ds \rho(t)=\frac{\ds \sum_{n=1}^n\left(s(n+t)-\overline{s}\right)\left(s(n)-\overline{s}\right)}{\ds \sum_{n=1}^n (s(n)-\overline{s})^2}, ~~\mbox{with $\ds \overline{s}=N^{-1}\sum _{n=1}^N s(n)$}\;.
 \end{equation}
The time delay is chosen as the $\rho=\underset{t>0}\min\;\{\rho(t)<0\}$\;.\mbox{}\\
Now define the  AMI as 
\begin{equation}
I(t)=\frac{1}{N}\sum_{n=1}^N \mathbb{P}_r\left(s(n),s(n+t)\right) \log_2\left( \frac{\mathbb{P}_r\left(s(n),s(n+t)\right)}{\mathbb{P}_r\left(s(n)\right)\mathbb{P}_r\left(s(n+t)\right)}\right)\;,
\end{equation}
where  $\mathbb{P}_r\left(s(n)\right)$ and $\mathbb{P}_r\left(s(n),s(n+t)\right)$ are respectively the probability of observing $s(n)$ and the probability of observing $s(n)$ and $s(n+t)$. The time delay is estimated as the first local minima of $I(t)$.\mbox{}\\

\subsection{Estimating the embedding dimension} The most popular method for estimating the  embedding dimension $m$ is the so-called False Nearest Neighbors technique, see for example \cite{Kennel1992}. 

\subsection{Kullback-Leibler divergence function} \label{sect:KL} Given two distributions $F$ and $G$ of a continuous random variable, with respective density functions $f$ and $g$, the  Kullback-Leibler divergence function is defined as 
\begin{equation}
d_{KL}(F,G)=\int_{-\infty}^{\infty} f(x)\log\left(\frac{f(x)}{g(x)}\right)dx\;.
\end{equation}
This quantity measures the degree to which $F$ diverges from $G$. We will use it to  assess the difference between the densities of the different cortex per task  and thus the difference between their  brain activities, see Section 5.2

\subsection{Cross-correlation function}
The cross-correlation function (CCF) is defined as : given a time lag $\tau$, and two time-series $X=\{X_t\}$ and $Y=\{Y_t\}$, the CCF is defined as 
\[CCF(X,Y,\tau)=\frac{\mathbb{E}[(X_{t-\tau}-\overline{X})(Y_{t}-\overline{Y})]}{\sqrt{\mathbb{E}[(X_{t-\tau}-\overline{X})^2]\mathbb{E}[(Y_{t-\tau}-\overline{Y})^2]}}\;,\] 
where $\overline{X}$ and $\overline{Y}$ are the respective mean of the time series $X$ and $Y$.

The CCF is then calculated over range of temporal lags  and a   ``distance" $\rho(\cdot,\cdot)$ is defined as the   maximum absolute CCF over the interval $[1,m_l]$ ( for a  maximum lag value of $m_l$ to be selected) 
\[\rho(X,Y)=\max_{1\leq \tau\leq m_l}\{\abs{CCF(X,Y,\tau)}\}\;.\]
Note that difference defines the similarity between the two time series.

\begin{figure}[H] 
\resizebox{1\textwidth}{!}{\begin{minipage}{1.3\textwidth}
\centering
\begin{tabular}{cc}
\bf (A) & \bf (B) \\
    \includegraphics[scale=0.5]{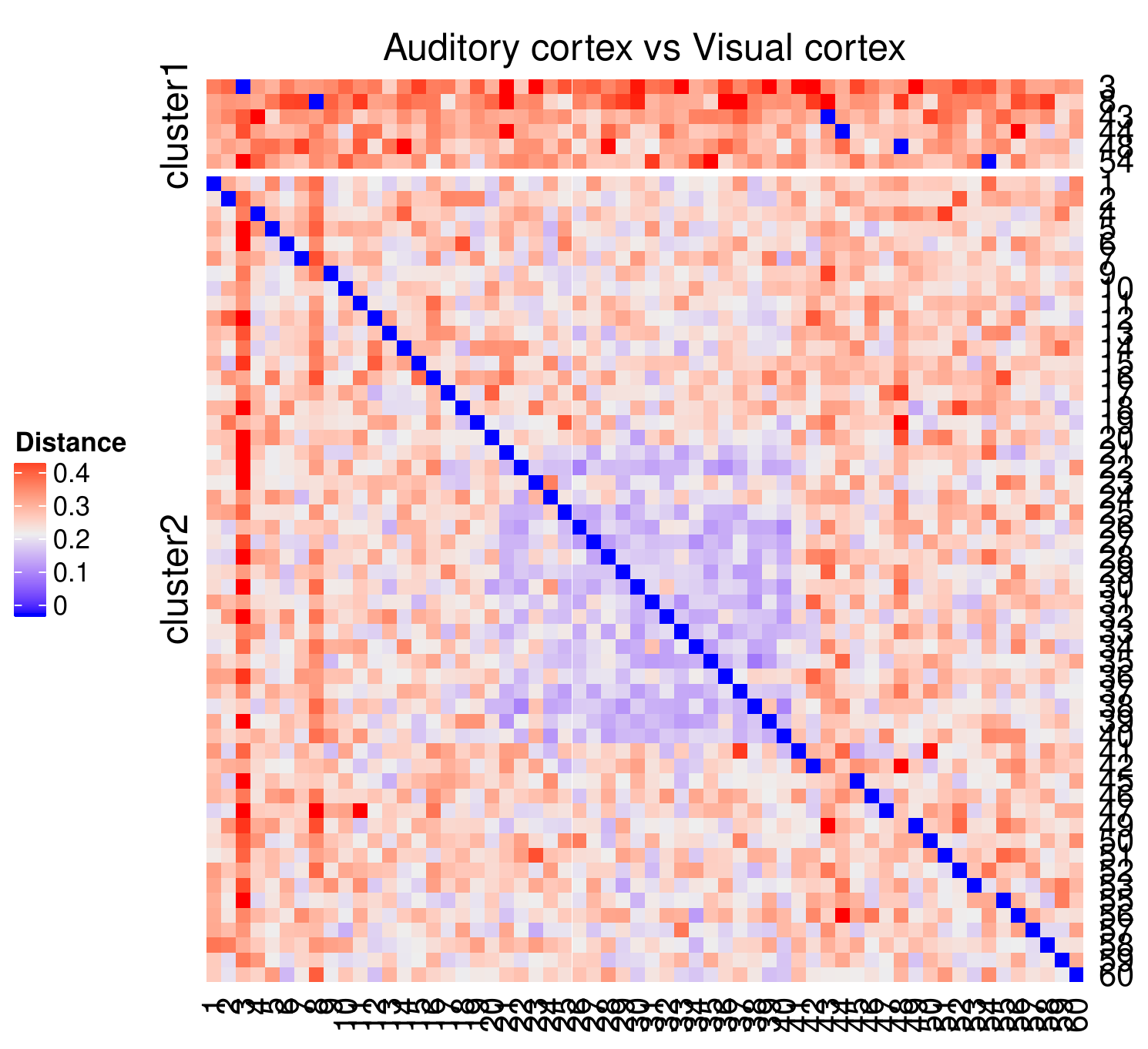}    &
        \includegraphics[scale=0.5]{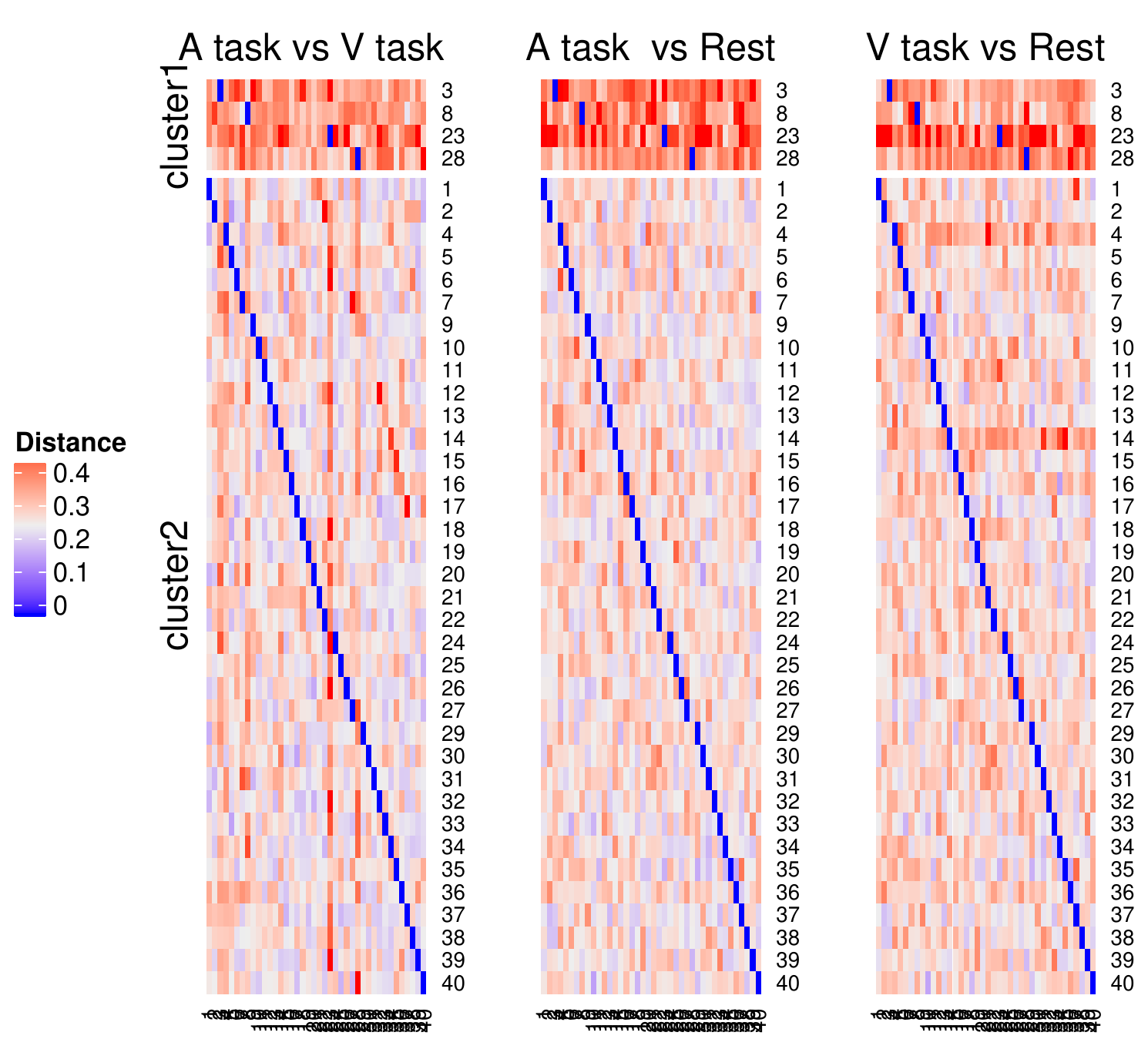} \\
        \bf (C) & \bf (D) \\
            \includegraphics[scale=0.5]{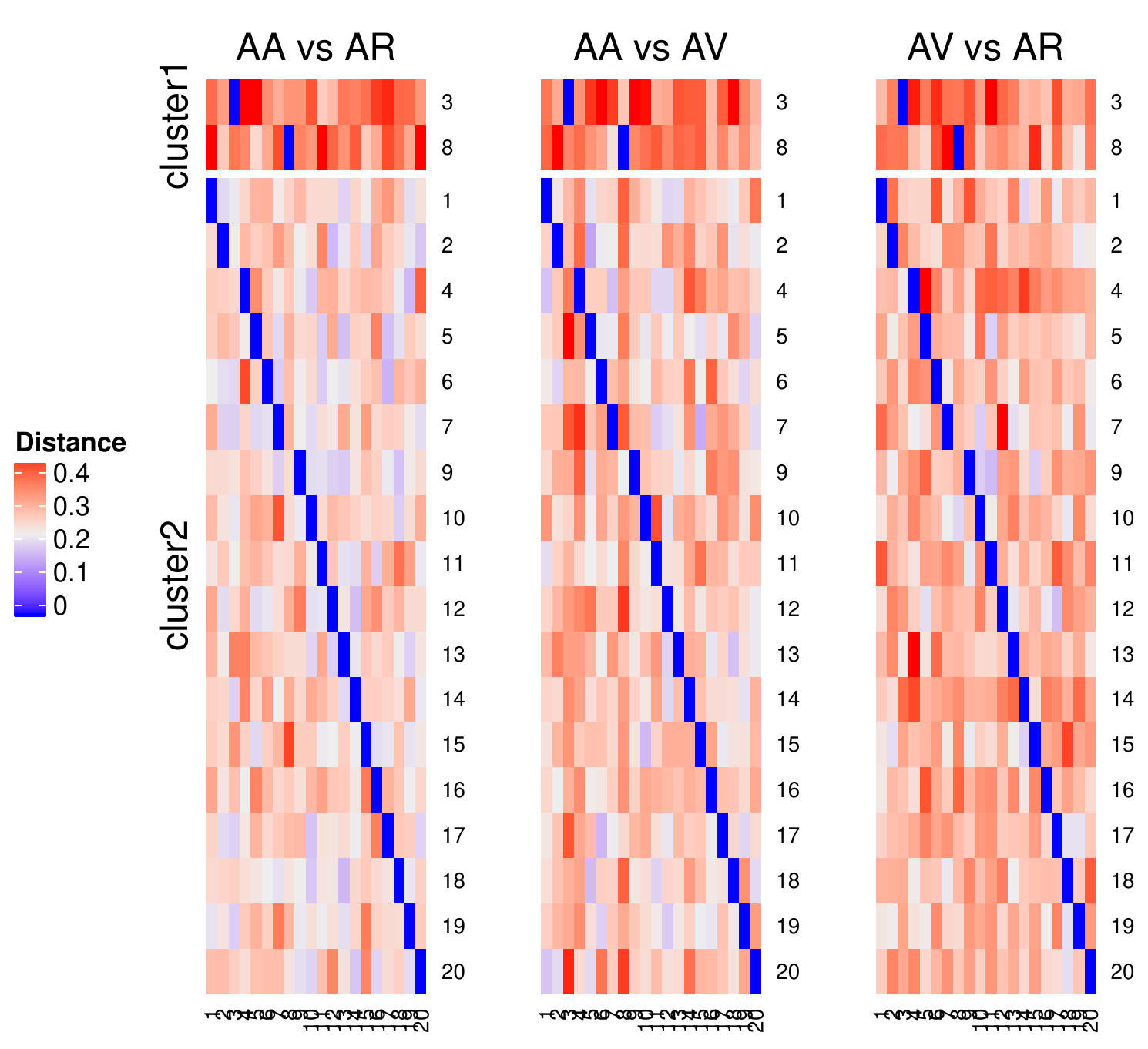}   &
      \includegraphics[scale=0.5]{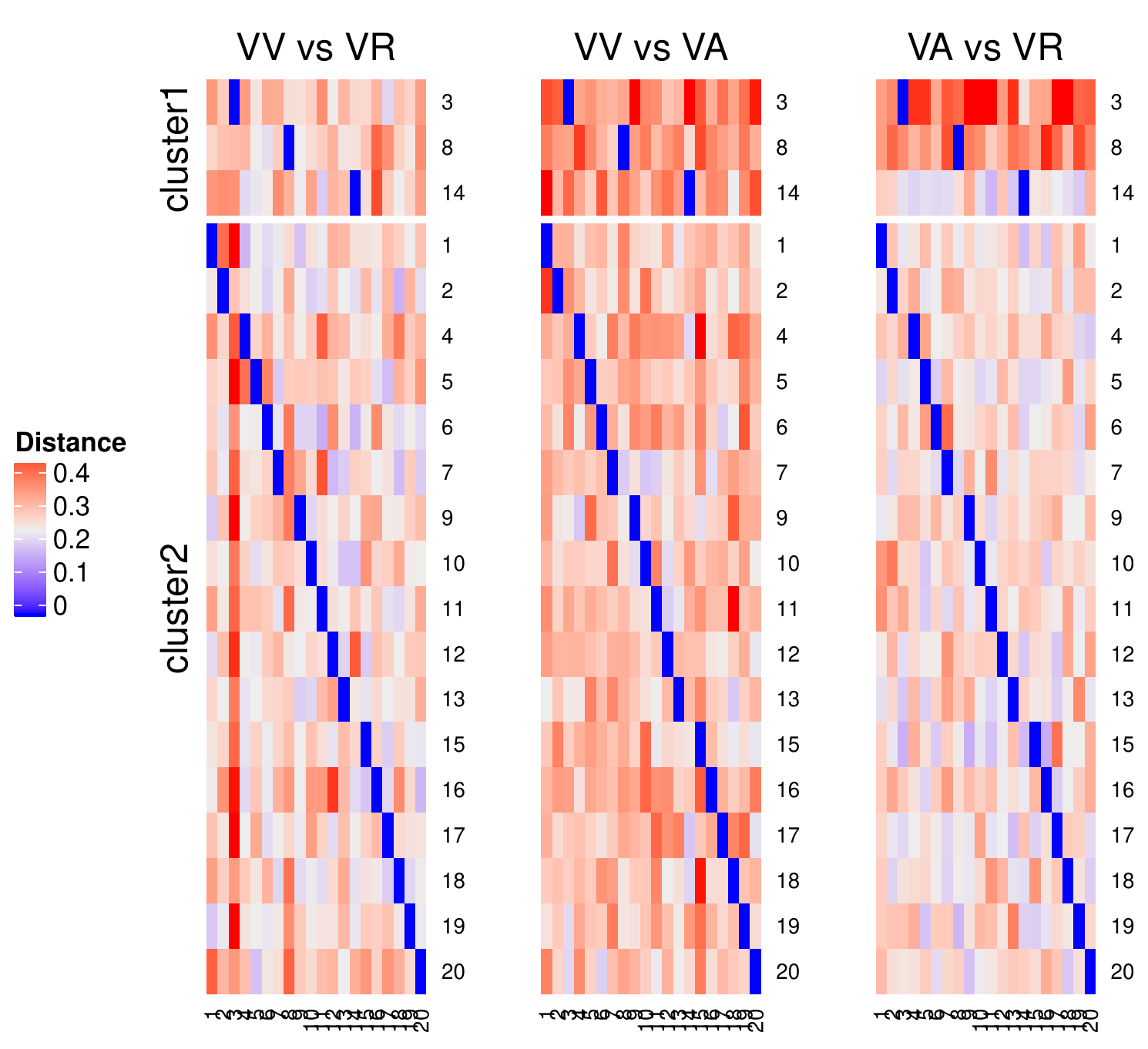} 
      \end{tabular}
      \end{minipage}}                  
   \caption{Plot of $\rho(X,Y)$ between the Auditory and Visual cortexes by cluster. The heat gradient represents the values of $\rho(X,Y)$. }
   \label{FigureHeatMapAVKmeans}
\end{figure}

\bibliography{EEG_CA}

\begin{thebibliography}{26}
\providecommand{\natexlab}[1]{#1}
\providecommand{\url}[1]{\texttt{#1}}
\expandafter\ifx\csname urlstyle\endcsname\relax
  \providecommand{\doi}[1]{doi: #1}\else
  \providecommand{\doi}{doi: \begingroup \urlstyle{rm}\Url}\fi

\bibitem[Andrzejak et~al.(2001)Andrzejak, Lehnertz, Mormann, Rieke, David, and
  Elger]{Andrzejak2001}
R.~G. Andrzejak, K.~Lehnertz, F.~Mormann, C.~Rieke, P.~David, and C.~E Elger.
\newblock Indications of nonlinear deterministic and finite-dimensional
  structures in time seriesof brain electrical activity: Dependence on
  recording region and brain state.
\newblock \emph{Physical Review E}, 2001.
\newblock \doi{10.1103/PhysRevE.64.061907}.

\bibitem[Arbabshirani et~al.(2012)Arbabshirani, Havlicek, Kiehl, Pearlson, and
  Calhoun]{Arbabshirani2012}
M.~R. Arbabshirani, M.~Havlicek, K.~A. Kiehl, G.~D. Pearlson, and V.~D.
  Calhoun.
\newblock Functional network connectivity during rest and task conditions: a
  comparative study.
\newblock \emph{Human brain mapping}, 34\penalty0 (11):\penalty0 2959--2971,
  2012.

\bibitem[Berger(1929)]{Hberger}
H.~Berger.
\newblock Über das elektroenkephalogramm des menschen.
\newblock \emph{Arch Psychiatr Nervenkr}, 87\penalty0 (1):\penalty0 527--570,
  1929.

\bibitem[Bruckner et~al.(2008)Bruckner, Andrews-Hanna, and
  Schacter]{Bruckner2008}
R.~L. Bruckner, J.~R. Andrews-Hanna, and D.~L. Schacter.
\newblock The brain's default network: Anatomy, function, and relevance to
  disease.
\newblock \emph{Ann. N.Y. Acad. Sci}, 1124:\penalty0 1--38, 2008.

\bibitem[Carney et~al.(2011)Carney, Myers, and Geyer]{Carney2011a}
P.~R. Carney, S.~Myers, and J.~D. Geyer.
\newblock Seizure prediction: methods.
\newblock \emph{Epilepsy \& behavior}, 22:\penalty0 S94--S101, 2011.

\bibitem[Celso et~al.(1984)Celso, Ott., Pelikan, and Yorke]{Celso1984}
G.~Celso, E.~Ott., S.~Pelikan, and J.~A. Yorke.
\newblock Strange attractors that are not chaotic.
\newblock \emph{Physica D: Nonlinear Phenomena. Elsevier BV}, 13\penalty0
  (1--2):\penalty0 261--268, 1984.

\bibitem[Celso et~al.(1987)Celso, Ott., and Yorke]{Celso1987}
G.~Celso, E.~Ott., and J.~A. Yorke.
\newblock Chaos, strange attractors, and fractal basin boundaries in nonlinear
  dynamics.
\newblock \emph{Science}, 238\penalty0 (4827):\penalty0 632--638), 1987.

\bibitem[Chen(2011)]{Chen}
P.~Chen.
\newblock Principles of biological science.
\newblock (Last accessed \today), 2011.
\newblock URL \url{http://bio1152.nicerweb.com/}.

\bibitem[David et~al.(2020)David, Machado, I\'{n}acio, and Valentin]{David2020}
S.~A. David, J.A.T Machado, C.M.C. I\'{n}acio, and C.A. Valentin.
\newblock A combined measure to differentiate eeg signals using fractal
  dimension and mfdfa-hurst.
\newblock \emph{Communications in Nonlinear Science and Numerical Simulation},
  84:\penalty0 1, 2020.

\bibitem[Destexhe(1992)]{Destexhe1992}
A~Destexhe.
\newblock \emph{Nonlinear Dynamics of the RhythmicalActivity of the Brain:}.
\newblock PhD thesis, Université Libre de Bruxelles, Brussels, Belgium, 1992.

\bibitem[Destexhe and Babloyantz(1986)]{Destexhe1986}
A.~Destexhe and A.~Babloyantz.
\newblock Low-dimensional chaos in an instance of epilepsy.
\newblock \emph{Proc. Natl. Acad. Sci}, 83:\penalty0 3513--3517, 1986.

\bibitem[Destexhe et~al.(1998)Destexhe, Sepulchre, and
  Babloyantz]{Destexhe1998}
A.~Destexhe, J.~A. Sepulchre, and A.~Babloyantz.
\newblock A comparative study of the experimental quantification of
  deterministic chaos.
\newblock \emph{Physics Letters}, A 132:\penalty0 101--106, 1998.

\bibitem[Edelsbrunner and M\"{u}cke(1994)]{Edelsbrunner1994}
H.~Edelsbrunner and E.P. M\"{u}cke.
\newblock Three-dimensional alpha shapes.
\newblock \emph{ACM Transactions on Graphics}, 13\penalty0 (1):\penalty0
  43--72, 1994.

\bibitem[Fisher et~al.(2009)Fisher, Talathi, Cadotte, and Carney]{Fisher2009a}
N.~K. Fisher, S.~S. Talathi, A.~Cadotte, and P.~R. Carney.
\newblock Epilepsy detecting and monitoring.
\newblock In S.~Tong and N.~V. Thakor, editors, \emph{Quantitative EEG analysis
  methods and clinical applications}, chapter~6, pages 141--165. Artech House,
  2009.

\bibitem[Gardiner et~al.(2018)Gardiner, Behnsen, and Brassey]{Gardiner2018}
J.~D. Gardiner, J.~Behnsen, and C.~A. Brassey.
\newblock Alpha shapes: determining 3d shape complexity across morphologically
  diverse structures.
\newblock \emph{BMC Evolutionary Biology}, 18\penalty0 (184), 2018.
\newblock \doi{10.1186/s12862-018-1305-z}.

\bibitem[Grassberger and Procaccia(1983)]{Grassberger1983}
P.~Grassberger and I.~Procaccia.
\newblock Measuring the strangeness of strange attractors.
\newblock \emph{Physica D. Nonlinear Phenomena}, 9\penalty0 (1-2):\penalty0
  189--208, 1983.

\bibitem[Kennel et~al.(1992)Kennel, Brown, and Abarbanel]{Kennel1992}
M.~Kennel, R.~Brown, and H.~Abarbanel.
\newblock Determining embedding dimension for phase-space reconstruction using
  a geometrical construction.
\newblock \emph{Physical Review A}, 45\penalty0 (6):\penalty0 3403--3411, 1992.

\bibitem[Kwessi and Edwards(2020)]{Kwessi2020}
E.~A. Kwessi and L.~J. Edwards.
\newblock Artificial neural networks with a signed-rank objective function and
  applications.
\newblock \emph{Communication in Statistics-Simulations and Computations},
  2020.
\newblock \doi{10.1080/03610918.2020.1714659}.

\bibitem[Lafarge et~al.(2014)Lafarge, Pateiro-Lopez, Possolo, and
  Dunkers]{Lafarge2014}
T.~Lafarge, B.~Pateiro-Lopez, A.~Possolo, and J.~P. Dunkers.
\newblock R implementation of a polyhedral approximation to a 3d set of points
  using the $\alpha$-shape.
\newblock \emph{J. Stat. Software}, 54\penalty0 (4):\penalty0 1--19, 2014.

\bibitem[Lehnertz and Elger(1998)]{Lehnertz1998}
K.~Lehnertz and C.~E. Elger.
\newblock Can epileptic seizures be predicted? evidence from nonlinear time
  series analysis of brain electrical activity.
\newblock \emph{Physical Review Letters}, 80:\penalty0 5019--5022, 1998.

\bibitem[Lorenz(1963)]{Lorenz1963}
E.~N. Lorenz.
\newblock Deterministic nonperiodic flow.
\newblock \emph{Journal of the Atmospheric Sciences}, 20\penalty0 (2):\penalty0
  130--141, 1963.

\bibitem[Paladin and Vulpiani(1987)]{Paladin1987}
G.~Paladin and A.~Vulpiani.
\newblock Anomalous scaling laws in multifractals objects.
\newblock \emph{Physics Reports}, 156\penalty0 (4):\penalty0 147--225, 1987.

\bibitem[Rickles et~al.(2007)Rickles, Hawe, and Shiell]{Rickles2007}
D.~Rickles, P.~Hawe, and A.~Shiell.
\newblock A simple guide to chaos and complexity.
\newblock \emph{Journal of epidemiology and community health}, 61\penalty0
  (11):\penalty0 933--937, 2007.

\bibitem[Satheesh~Kumar and Bhuvaneswari(2012)]{Kumar2012}
J.~Satheesh~Kumar and P.~Bhuvaneswari.
\newblock Analysis of electroencephalography (eeg) signals and its
  categorization-a study.
\newblock \emph{Procedia Engineering}, 38:\penalty0 2525--2536, 2012.

\bibitem[Takens(1981)]{Takens1981a}
F.~Takens.
\newblock Detecting strange attractors in turbulence dynamical systems and
  turbulence (lecture notes in mathematics), vol. 898, 1981.

\bibitem[Zheng et~al.(2019)Zheng, Fushing, and Ge]{Zheng2019}
J.~Zheng, H.~Fushing, and L.~Ge.
\newblock A data-driven approach to predict and classify epileptic seizures
  from brain-wide calcium imaging video data.
\newblock \emph{IEEE/ACM Transactions on Computational Biology and
  Bioinformatics}, 2019.

\end{thebibliography}

\end{document}